\documentclass[aip, amsmath,amssymb,reprint, longbibliography, final]{revtex4-2}
\usepackage[colorlinks,bookmarks=false,citecolor=blue,linkcolor=red,urlcolor=blue]{hyperref}

\usepackage[usenames,dvipsnames]{color}
\usepackage{soul}

\usepackage{amsfonts}
\usepackage{amsmath}
\let\vec\mathbf
\usepackage{graphicx}
\usepackage{dcolumn}
\usepackage{bm}

\usepackage[utf8]{inputenc}
\usepackage[T1]{fontenc}
\usepackage{mathptmx}

\usepackage[dvipsnames,table]{xcolor}

\usepackage{siunitx}
\mathchardef\mhyphen="2D
\usepackage{floatrow}
\usepackage{multirow}
\usepackage{upgreek}
\usepackage{makecell}
\usepackage{tabularx}
\usepackage{physics}

\newcommand{\suppl}[1]{Supplementary}

\newcommand{\figrefL}[1]{Figure~\ref{fig:#1}}

\newcommand{\subfigref}[2]{Fig.~\hyperref[fig:#1]{\ref*{fig:#1}#2}}
\newcommand{\Supplsubfigref}[2]{\suppl{}~Fig.~\hyperref[fig:#1]{\ref*{fig:#1}#2}}
\newcommand{\SupplsubfigrefL}[2]{\suppl{}~Figure~\hyperref[fig:#1]{\ref*{fig:#1}#2}}
\newcommand{\SupplsubfigrefS}[2]{Fig.~\hyperref[fig:#1]{\ref*{fig:#1}#2}}
\newcommand{\subfigrefL}[2]{Figure~\hyperref[fig:#1]{\ref*{fig:#1}#2}}
\newcommand{\subfigsref}[3]{Figs.~\hyperref[fig:#1]{\ref*{fig:#1}#2}-\hyperref[fig:#1]{\ref*{fig:#1}#3}}
\newcommand{\Supplsubfigsref}[3]{\suppl{}~Figs.~\hyperref[fig:#1]{\ref*{fig:#1}#2}-\hyperref[fig:#1]{\ref*{fig:#1}#3}}
\newcommand{\subfigsrefL}[3]{Figures~\hyperref[fig:#1]{\ref*{fig:#1}#2}-\hyperref[fig:#1]{\ref*{fig:#1}#3}}

\newcommand*{\smallR}[1]{$R_0 = \SI{364}{\nano\meter}$}
\newcommand*{\largeR}[1]{$R_0 = \SI{648}{\nano\meter}$}

\newcommand*{\fp}[1]{$F_{\mathrm{P}}$}
\newcommand*{\lowT}[1]{$T = \SI{4.2}{\kelvin}$}
\newcommand*{\methods}[1]{Methods}
\newcommand{\gtwo}[1]{g^{(2)}({#1})}
\DeclareSIUnit\ML{ML}
\DeclareSIUnit\px{px}
\newcommand*{\SM}[1]{\textbf{\suppl{}~Information}}

\usepackage{array}
\usepackage{tabularx}
\usepackage{arraycols}

\usepackage[flushleft]{threeparttable}

\usepackage{booktabs}

\usepackage{etoolbox}
\robustify\bfseries
\robustify\itshape

\newcolumntype{E}{>{\centering\arraybackslash}X}

\usepackage{microtype}
\usepackage{rotating}

\newcommand*{\oldtext}[2][Old text: ]{\textit{#1}{\textcolor{Red}{#2}}}

\newcommand*{\SMnote}[1]{\suppl{}~Note~#1}

\makeatletter
\let\@fnsymbol\@fnsymbol@latex
\@booleanfalse\altaffilletter@sw
\makeatother

\begin{document}
\preprint{AIP/123-QED}
\newcommand*{\papertitle}{High-throughput quantum photonic devices emitting indistinguishable photons in the telecom C-band}

\author{Pawe\l{}~Holewa}
\email{pawel.holewa@pwr.edu.pl}
\affiliation{Department of Experimental Physics, Faculty of Fundamental Problems of Technology, Wroc\l{}aw University of Science and Technology, Wyb. Wyspia\'{n}skiego 27, 50-370 Wroc\l{}aw, Poland}
\affiliation{DTU Electro, Department of Electrical and Photonics Engineering, Technical University of Denmark, Ørsteds Plads 343, DK-2800 Kongens Lyngby, Denmark}
\affiliation{NanoPhoton - Center for Nanophotonics, Technical University of Denmark, Ørsteds Plads 345A, DK-2800 Kongens Lyngby, Denmark}

\author{Daniel~A.~Vajner}
\affiliation{Institute of Solid State Physics, Technische Universität Berlin, 10623 Berlin, Germany}

\author{Emilia~Zi\k{e}ba-Ost\'{o}j}
\affiliation{Department of Experimental Physics, Faculty of Fundamental Problems of Technology, Wroc\l{}aw University of Science and Technology, Wyb. Wyspia\'{n}skiego 27, 50-370 Wroc\l{}aw, Poland}

\author{Maja~Wasiluk}
\affiliation{Department of Experimental Physics, Faculty of Fundamental Problems of Technology, Wroc\l{}aw University of Science and Technology, Wyb. Wyspia\'{n}skiego 27, 50-370 Wroc\l{}aw, Poland}

\author{Benedek Ga\'{a}l}
\affiliation{DTU Electro, Department of Electrical and Photonics Engineering, Technical University of Denmark, Ørsteds Plads 343, DK-2800 Kongens Lyngby, Denmark}

\author{Aurimas~Sakanas}
\affiliation{DTU Electro, Department of Electrical and Photonics Engineering, Technical University of Denmark, Ørsteds Plads 343, DK-2800 Kongens Lyngby, Denmark}

\author{Marek~Burakowski}
\affiliation{Department of Experimental Physics, Faculty of Fundamental Problems of Technology, Wroc\l{}aw University of Science and Technology, Wyb. Wyspia\'{n}skiego 27, 50-370 Wroc\l{}aw, Poland}

\author{Pawe\l{}~Mrowi\'{n}ski}
\affiliation{Department of Experimental Physics, Faculty of Fundamental Problems of Technology, Wroc\l{}aw University of Science and Technology, Wyb. Wyspia\'{n}skiego 27, 50-370 Wroc\l{}aw, Poland}

\author{Bartosz~Krajnik}
\affiliation{Department of Experimental Physics, Faculty of Fundamental Problems of Technology, Wroc\l{}aw University of Science and Technology, Wyb. Wyspia\'{n}skiego 27, 50-370 Wroc\l{}aw, Poland}

\author{Meng~Xiong}
\affiliation{DTU Electro, Department of Electrical and Photonics Engineering, Technical University of Denmark, Ørsteds Plads 343, DK-2800 Kongens Lyngby, Denmark}
\affiliation{NanoPhoton - Center for Nanophotonics, Technical University of Denmark, Ørsteds Plads 345A, DK-2800 Kongens Lyngby, Denmark}

\author{Kresten~Yvind}
\affiliation{DTU Electro, Department of Electrical and Photonics Engineering, Technical University of Denmark, Ørsteds Plads 343, DK-2800 Kongens Lyngby, Denmark}
\affiliation{NanoPhoton - Center for Nanophotonics, Technical University of Denmark, Ørsteds Plads 345A, DK-2800 Kongens Lyngby, Denmark}

\author{Niels~Gregersen}
\affiliation{DTU Electro, Department of Electrical and Photonics Engineering, Technical University of Denmark, Ørsteds Plads 343, DK-2800 Kongens Lyngby, Denmark}

\author{Anna~Musia\l{}}
\affiliation{Department of Experimental Physics, Faculty of Fundamental Problems of Technology, Wroc\l{}aw University of Science and Technology, Wyb. Wyspia\'{n}skiego 27, 50-370 Wroc\l{}aw, Poland}

\author{Alexander~Huck}
\affiliation{Center for Macroscopic Quantum States (bigQ), Department of Physics, Technical University of Denmark, DK-2800 Kongens Lyngby, Denmark}

\author{Tobias~Heindel}
\affiliation{Institute of Solid State Physics, Technische Universität Berlin, 10623 Berlin, Germany}

\author{Marcin~Syperek}
\email{marcin.syperek@pwr.edu.pl}
\affiliation{Department of Experimental Physics, Faculty of Fundamental Problems of Technology, Wroc\l{}aw University of Science and Technology, Wyb. Wyspia\'{n}skiego 27, 50-370 Wroc\l{}aw, Poland}

\author{Elizaveta~Semenova}
\email{esem@fotonik.dtu.dk}
\affiliation{DTU Electro, Department of Electrical and Photonics Engineering, Technical University of Denmark, Ørsteds Plads 343, DK-2800 Kongens Lyngby, Denmark}
\affiliation{NanoPhoton - Center for Nanophotonics, Technical University of Denmark, Ørsteds Plads 345A, DK-2800 Kongens Lyngby, Denmark}

\graphicspath{ {./Figures/} }


\begin{abstract}
Single indistinguishable photons at telecom C-band wavelengths are essential for quantum networks and the future quantum internet.
However, high-throughput technology for single-photon generation at $\SI{1550}{\nano\meter}$ remained a missing building block to overcome present limitations in quantum communication and information technologies.
Here, we demonstrate the high-throughput fabrication of quantum-photonic integrated devices operating at C-band wavelengths based on epitaxial semiconductor quantum dots.
Our technique enables the deterministic integration of single pre-selected quantum emitters into microcavities based on circular Bragg gratings.
Respective devices feature the triggered generation of single photons with ultra-high purity and record-high photon indistinguishability.
Further improvements in yield and coherence properties will pave the way for implementing single-photon non-linear devices and advanced quantum networks at telecom wavelengths.
\end{abstract}

\title{\papertitle{}}
\maketitle

\section*{Introduction}

A quantum network~\cite{Kimble2008} based on remote nodes interconnected via fiber-optical links and capable of transferring quantum information using flying qubits will provide the backbone for the implementation of protocols for secure communication~\cite{Gisin2002,Vajner2022} and distributed quantum computing~\cite{Serafini2006}.
Notably, the network can rely on the existing silica-fiber-based infrastructure, utilizing a low-loss channel for the transmission of photons with a wavelength in the telecom C-band around $\SI{1550}{\nano\meter}$~\cite{Cao2022}.
These quantum network architectures can benefit from existing components and classical signal management protocols, hence making it feasible to transfer quantum information over large distances~\cite{Cao2022}.

In recent years, the technology for the epitaxial growth of self-assembled quantum dots (QDs) has rapidly advanced, resulting in the demonstration of QD-based single-photon sources (SPSs) with excellent characteristics.
These include high photon extraction efficiencies ($\sim\SI{79}{\percent}$)~\cite{Gazzano2013}, high single photon generation rates ($\sim\SI{1}{\giga\hertz}$)~\cite{Tomm2021}, and near unity photon indistinguishability ($>\SI{96}{\percent}$)~\cite{Tomm2021,Wang2016}, however, all achieved outside the telecom-relevant C-band.
Besides the extraordinary material quality, these characteristics are achieved owing to efficient light-matter coupling between a QD and a suitable photonic element. 
For efficient coupling, spectral and spatial matching is required between the quantum emitter and the engineered photonic mode, which is challenging due to the spatial and spectral distribution of epitaxial QDs. Until now, the QD coupling to photonic cavities operating around $\SI{1550}{\nano\meter}$ has only been realized using non-deterministic fabrication processes, limiting device yield and scalability~\cite{Nawrath2023}.

In this article, we report on the high-throughput fabrication of nanophotonic elements around pre-selected individual QDs emitting single and indistinguishable photons in the telecom C-band.
For this purpose, we develop a near-infrared (NIR) imaging technique for self-assembled InAs/InP QDs utilizing a hybrid sample geometry with enhanced out-of-plane emission from single QDs~\cite{Holewa2022Mirror} and a thermo-electrically cooled InGaAs camera in a wide-field imaging configuration.
In combination with two electron-beam lithography (EBL) steps, our method enables an overall positioning accuracy of $\SI{90}{\nano\meter}$ of the QD with respect to the photonic element and allows for rapid data collection as compared to competing techniques based on scanning in-situ imaging~\cite{Liu2021ReviewPositioning}.
We apply our technique for the deterministic integration of pre-selected QDs into circular Bragg grating (CBG) cavities.
The proposed technological workflow allows us to greatly enhance the device fabrication yield reaching $\sim\SI{30}{\percent}$, which is a significant improvement compared to $<\SI{1}{\percent}$ that would typically be achieved with a random placement approach.
The QD-CBG coupling is evidenced by a Purcell factor $\sim 5$, and our devices demonstrate a state-of-the-art photon extraction efficiency of $\eta=\SI{16.6(27)}{\percent}$ into the first lens with a numerical aperture (NA) of $0.4$, a high single-photon purity associated with $\gtwo{0} =\SI{3.2(6)E-3}{}$, and a record-high photon-indistinguishability of $V = \SI{19.3(26)}{\percent}$ for QD-based SPSs at C-band wavelengths.

\section*{Results}
\subsection*{Design of a QD structure for wide-field imaging}

	\begin{figure*}[htb] %
		\begin{center} %
    	\includegraphics[width=1\columnwidth]{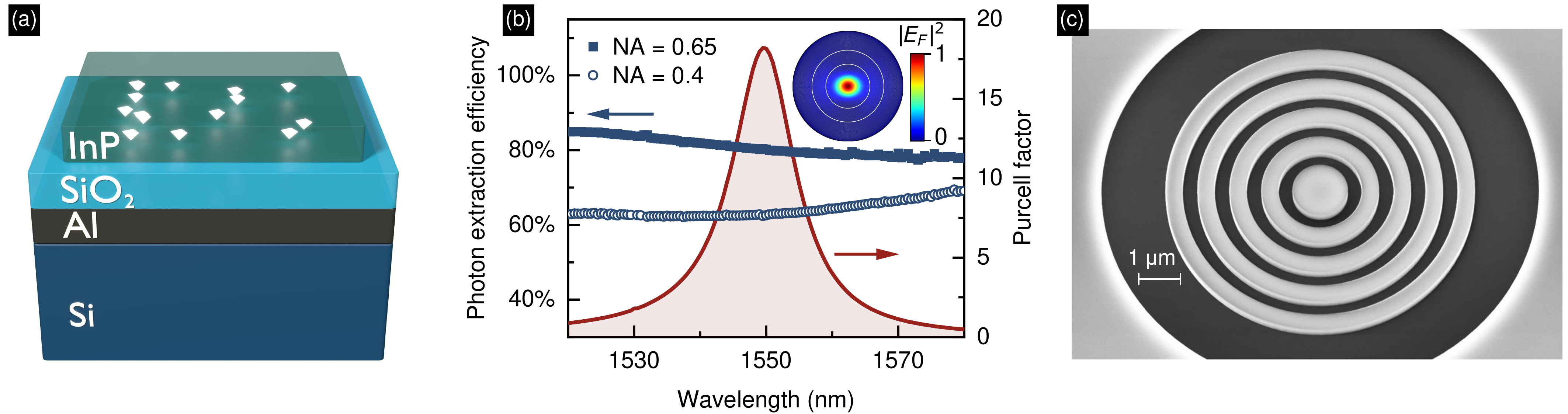} 
	    \end{center}	 %
		\caption{\label{fig:Structure}
		Structure with quantum dots (QDs) for imaging and optimized circular Bragg gratings (CBGs).
		a,~The layer stack for efficient localization of single InAs/InP QDs emitting at C-band,
		b,~calculated CBG Purcell factor (dark red line, right axis) and photon extraction efficiency (numerical aperture $\mathrm{NA}=0.65$, full squares, and $\mathrm{NA}=0.4$, empty circles, left axis).
        Inset: far-field emission pattern with rings marking $\mathrm{NA}=0.4$ and $\mathrm{NA}=0.65$.
        $E_F$ -- normalized electric field amplitude.
		c,~Scanning electron microscope image of a CBG cavity etched in InP on top of SiO$_2$.}
	\end{figure*}

Imaging at very low light levels at wavelengths $\SI{>1}{\micro\meter}$ is challenging due to the high level of electronic noise of respective camera systems.
Although cameras based on InGaAs achieve quantum efficiencies $>80\%$, they are characterized by a factor of $>10^4$ higher dark currents compared to Si-based devices.
The photon emission rate from the sample is therefore of key importance for the ability to image and localize single QDs.
Following our previous work~\cite{Holewa2022Mirror}, we have designed a planar sample geometry that significantly enhances the photon extraction efficiency, allowing to localize single QDs and the subsequent fabrication of photonic elements.

The planar QD structure consists of an epitaxially grown $\SI{312}{\nano\meter}$-thick InP slab containing a single layer of InAs QDs.
The InP slab is atop a $\SI{359}{\nano\meter}$-thick SiO$_2$ layer with a bottom Al mirror bonded to a Si wafer carrier (\subfigref{Structure}{a}, see \methods{}).
Overall, this geometry enhances the QD emission in the out-of-plane direction by a factor $>7$ as compared to bulk InP samples, reaching a total photon extraction efficiency of $>\SI{10}{\percent}$ from a single QD for $\mathrm{NA}=0.4$~\cite{Holewa2022Mirror}.
In fact, this design turned out to be crucial for the imaging step, as the SNR of the QD emission was insufficient for investigated structures without a backside mirror (see \SMnote{3}).
For QD localization later in the experiment, we structure the top InP layer in a mesh with fields of size $(50\times50)~\si{\micro\meter\squared}$ separated by $\SI{10}{\micro\meter}$-wide ridges (see \methods{}), where the field edges are used as alignment marks (AMs) for imaging.
The fields are furthermore organized in blocks accompanied by InP crosses that allow us to align the electron beam to specific target QDs during the EBL process (see \SMnote{2} for the optical microscope image of the sample surface with fabricated cavities).

For the self-assembled Stranski-Krastanov QD epitaxy, we employed the near-critical growth regime in metalorganic vapor-phase epitaxy (MOVPE)~\cite{Berdnikov2023} (see \methods{}), and obtained a QD surface density of $\SI{3.1e8}{\per\square\centi\meter}$ corresponding to an average QD separation of $\SI{1.5}{\micro\meter}$. 
Since QDs exhibit a size, shape, and strain distribution, only a fraction of the QDs have their ground-state optical transition in the C-band.
With a~$\SI{1550(8)}{\nano\meter}$ bandpass filter, we find on average $N_{\mathrm{F}}=10$ QDs per field, which translates to an effective QD density of $\SI{4E5}{\per\square\centi\meter}$ and an average QD separation of $\sim\SI{16}{\micro\meter}$. 

\subsection*{Design of circular Bragg grating cavities}
The CBG geometry is optimized using the modal method (see \SMnote{1}) to enhance the cavity figures of merit at $\SI{1550}{\nano\meter}$, namely the collection efficiency at the first lens and the Purcell factor (\fp{}).
As opposed to other implementations~\cite{Nawrath2023}, we consider a simplified CBG geometry consisting of a central mesa and only four external rings.
According to our calculations, this number is sufficient for high $\eta$ and \fp{}, providing a smaller footprint and less complexity in the fabrication process.
The in-plane cavity dimensions include the central mesa radius of \largeR{}, the grating period of $\SI{747}{\nano\meter}$, and the separation between InP rings (air gap) of $\SI{346}{\nano\meter}$.
For these geometrical parameters, \subfigref{Structure}{b} shows the calculated broadband $\eta$ that amounts to nearly $\SI{62}{\percent}$ and $\SI{82}{\percent}$ at $\SI{1550}{\nano\meter}$ for a NA of $0.4$ and $0.65$, respectively, which is similar to other state-of-the-art CBG designs~\cite{Barbiero2022-1p55um,Bremer2022,Rickert2023}.
The wavelength dependence of the Purcell factor, presented in \subfigref{Structure}{b}, mimics the CBG cavity mode centered at $\SI{1550}{\nano\meter}$ and reaches a maximum value $F_{\mathrm{P}}=18.1$ with a quality factor of $110$.
The influence of the cavity geometry on the dispersion of $\eta$ and \fp{} suggests that the cavity grating together with the Al mirror creates a photonic bandgap that governs the $\eta$-dependence and enhances the emission directionality, while the InP membrane thickness and central mesa diameter crucially affect the center wavelength of the \fp{}-dependence.
A scanning electron microscopy (SEM) image of a fabricated CBG cavity is shown in \subfigref{Structure}{c}.

\subsection*{Optical imaging and QD localization}
	%
	\begin{figure*}[htb] %
		\begin{center} %
    	\includegraphics[width=0.85\columnwidth]{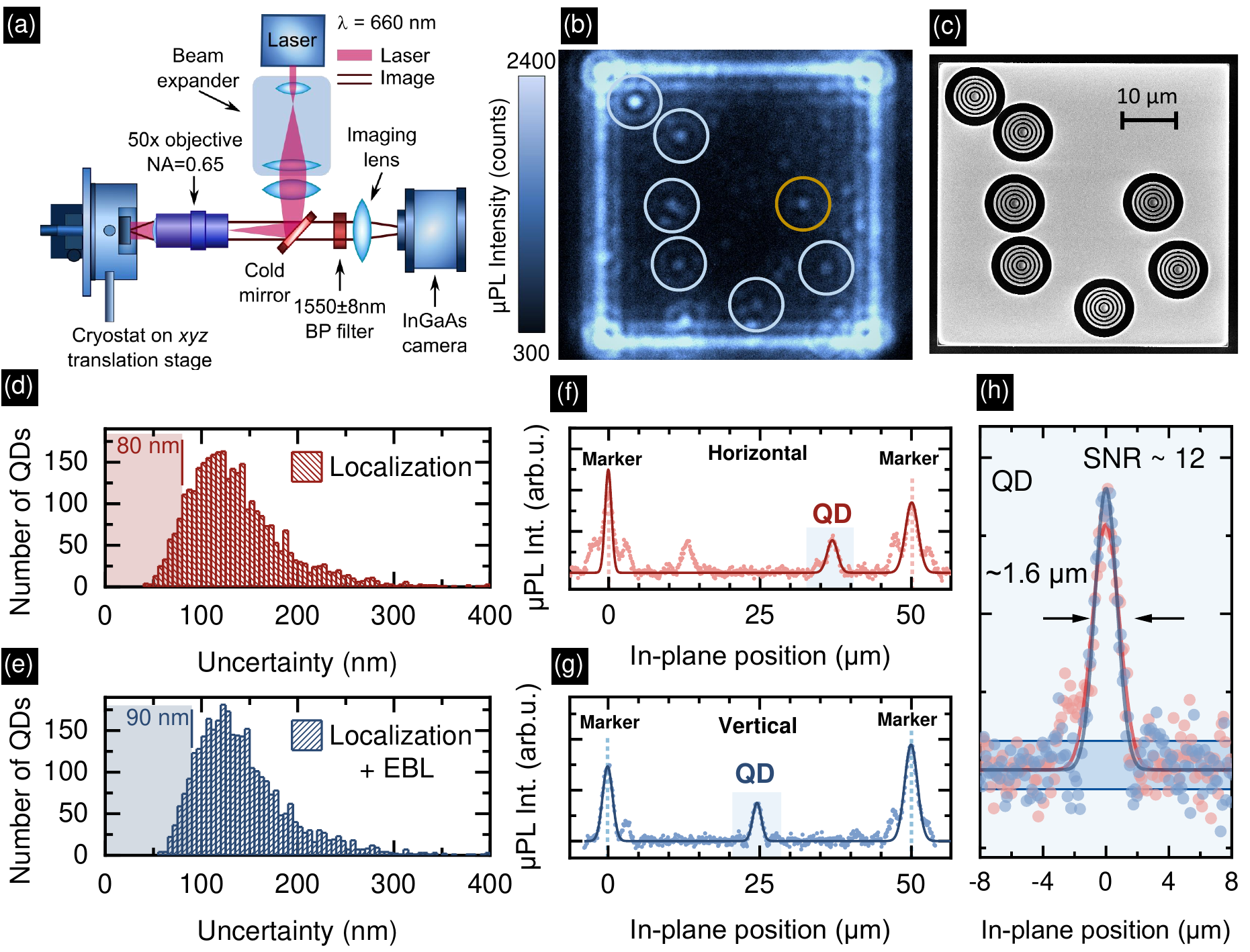} 
	    \end{center}	 %
		\caption{\label{fig:Imaging}
        Microphotoluminescence imaging of quantum dots (QDs).
		a,~The optical setup used for imaging, BP -- bandpass filter, NA -- numerical aperture,
		b,~a microphotoluminescence map of a $(50\times50)~\si{\micro\meter\squared}$ InP field containing seven localized InAs QDs emitting at C-band,
		c,~scanning electron microscope image of CBGs fabricated atop the localized and preselected QDs,
        d, e,~histograms of QD localization accuracy (d) and overall cavity placement accuracy (e) for all detected spots, with markers for $10^\mathrm{th}$ distribution percentiles,
		f,~g,~exemplary microphotoluminescence map cross-sections showing the signal of the QD labeled with the orange circle in b and alignment marks together with Gaussian fits used for the QD localization,
		h,~close-up of the QD signal from e, centered to $\SI{0}{\micro\meter}$.
        The full width at half maximum of the fitted Gaussian profiles is $\SI{\sim1.6}{\micro\meter}$ and the signal-to-noise ratio (SNR) is $12$.}
	\end{figure*}

The NIR imaging setup utilizes a wide-field bright microscope configuration as shown in \subfigref{Imaging}{a}.
The structure with QDs above the Al reflector is mounted in an optical cryostat at \lowT{} movable by an x-y-z stage for targeting fabricated fields that are imaged consecutively.
For sample illumination, we use a $\SI{660}{\nano\meter}$ continuous-wave (CW) semiconductor laser diode, spatially shaped with a beam expander, and focused on the backside of a commercially-available microscope objective ($\mathrm{NA}=0.65$) with $50\times$ magnification, $\SI{57}{\percent}$ transmission in the NIR, and $\SI{10}{\milli\meter}$ working distance.
This configuration provides nearly homogeneous surface illumination across a $(50\times50)~\si{\micro\meter\squared}$ field and high photon collection efficiency.
The spatially-distributed QD microphotoluminescence ($\upmu$PL) and scattered light from the field edges (here used as AMs) are collected by the same objective and pass through a cold mirror cutting off the laser light.
Finally, the emission is projected onto a thermo-electrically cooled InGaAs-based camera with a 
$(12.8\times10.24)~\si{\milli\meter\squared}$ chip and a pixel size of $(20\times20)~\si{\micro\meter\squared}$. 
With the $4\times$ magnification lens in front of the camera, the setup has a $200\times$ magnification, enabling the optimal filling of the entire camera chip with a single field (\subfigref{Imaging}{b}).
The $\SI{1550}{\nano\meter}$ band-pass filter with $\SI{8}{\nano\meter}$ full-width half-maximum (FWHM) placed in front of the imaging lens selects QDs with emission in the C-band.

\subfigrefL{Imaging}{b} shows a representative image of a field recorded with a camera integration time of $\SI{2.5}{\second}$.
The QDs can clearly be recognized as individual bright spots with FWHM $\approx\SI{1.6}{\micro\meter}$ (see \subfigref{Imaging}{h} and \SMnote{3}) and Airy rings around.
The square-shaped outline of the field scatters light and is used as AM for QD localization.

The localization of QDs is performed by taking vertical and horizontal cross-sections both crossing at a QD emission spot in the $\upmu$PL intensity map.
Each cross-section thus contains the position of the target QD relative to two AMs (\subfigsref{Imaging}{f}{g}).
Gaussian profiles fitted to the QD and the AM signals are subsequently used to determine the QD peak position relative to the AMs.
The average signal-to-noise ratio for QD emission spots is $10.6$, emphasizing the importance of the $7\times$ emission enhancement in the planar structure as compared to bulk InP.
We find that for the brightest $\SI{10}{\percent}$ of all QDs with SNR$\,>15.5$, the position is fitted with an uncertainty of $<\SI{54}{\nano\meter}$ in 1D and with an uncertainty of the AM position of $<\SI{36}{\nano\meter}$, resulting in a total uncertainty of QD position in 1D of $<\SI{62}{\nano\meter}$.
In 2D, this translates to $\SI{80}{\nano\meter}$ accuracy for the QD localization.
Finally, taking into account the EBL alignment accuracy of $\SI{40}{\nano\meter}$ as measured in our previous work~\cite{Sakanas2019}, we estimate the overall accuracy of 2D CBG placement to $\Delta R=\SI{90}{\nano\meter}$.
The histograms of QD localization accuracy and overall cavity placement accuracy for all detected spots are shown in \subfigref{Imaging}{f} and \subfigref{Imaging}{g} respectively, and the $\SI{80}{\nano\meter}$ and $\SI{90}{\nano\meter}$ levels are marked for reference.
Medians for the distributions are slightly larger, $\SI{127}{\nano\meter}$ and $\SI{133}{\nano\meter}$, due to processing of all detected spots, irrespective of their brightness and expected cavity fabrication precision.
Details on the localization algorithm, derivation and discussion of the uncertainties, and data on the accuracy of cavity positioning are given in the \SMnote{3}.
Following the localization of suitable QDs, CBGs are fabricated using EBL with proximity error correction and an optimized inductively coupled plasma-reactive ion etching (ICP-RIE) process (see \methods{}).
The SEM image presented in \subfigref{Imaging}{c} is taken from the same field after fabricating the CBG cavities around pre-selected QDs indicated by the circles in \subfigref{Imaging}{b}.

%
	\begin{figure*}[htb]%
		\begin{center} %
    	\includegraphics[width=0.999\columnwidth]{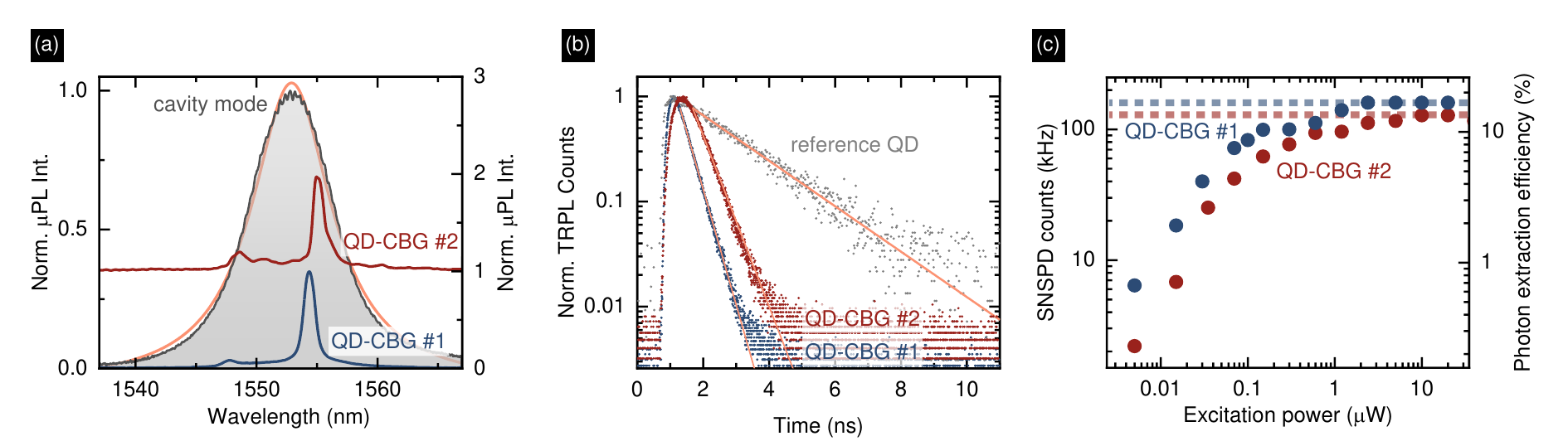} 
	    \end{center}
		\caption{\label{fig:QD-in-cavity}
		Characteristics of exemplary fabricated quantum dot-circular Bragg grating (QD-CBGs) devices \#1 and \#2.
		a,~microphotoluminescence ($\upmu$PL) spectra for QDs in devices \#1 and \#2 overlaid on the cavity mode of device \#2 (grey) and fitted with a Lorentzian profile (orange), stacked for clarity,
		b,~time-resolved $\upmu$PL (TRPL) traces for these QDs with the reference QD decay,
		c,~power-dependent count rates registered on the superconducting nanowire single-photon detector (SNSPD).
		Horizontal lines mark the line $\upmu$PL signal saturation level used for the determination of photon extraction efficiency.}
	\end{figure*}
\subsection*{Deterministic process yield}
We use a $\upmu$PL setup to quantify the process yield that we define as the ratio between the number of QD-CBG devices with QD emission spectra matching the CBG mode and the number of all CBGs investigated, and we obtain $Y=\SI{30}{\percent}$.
This value should be compared with the yield that would be obtained in a statistical QD-CBG fabrication process.
As we estimate the average number of QDs per field of size $F = \SI{50}{\micro\meter}$ to $N_{\mathrm{F}}=10$, the probability of covering one of them with the central mesa of diameter $2R_0$ is $Y_{\mathrm{random}}\sim N_{\mathrm{F}}\times\left(2R_0/F\right)^2=\SI{0.67}{\percent}\ll Y$.
Some of the QD (inside a CBG) emission spectra are significantly broadened (median linewidth of $\SI{0.76}{\nano\meter}$, see \SMnote{6}) as compared to the narrowest recorded linewidth of $\SI{0.14}{\nano\meter}$ (identical to the spectrometer resolution).
We attribute the broadening to the impact of surface states and point defects caused by the cavity fabrication, effectively resulting in the spectral wandering of the QD emission line~\cite{Liu2018}.
Such defects can also introduce non-radiative recombination centers in the close vicinity of or even into the QD, quench the optical emission, and effectively reduce the process yield.
Using the temperature-dependent $\upmu$PL studies, we make sure that even the broadened emission lines follow the expected Varshni trend, ensuring that these spectral lines can indeed be associated with the QD emission as the temperature dependence of the cavity mode energy is much weaker.

\subsection*{Purcell enhancement}
In the following, we discuss the optical properties of two exemplary devices, QD-CBG \#1 and QD-CBG \#2, each containing a single pre-selected QD coupled to the CBG cavity mode (see \SMnote{6} for the properties of a third device QD-CBG \#3).
\subfigrefL{QD-in-cavity}{a} shows the narrow QD emission lines overlaid on the cavity mode with $Q=194$, the latter obtained under high power cavity excitation, evidencing good spectral overlap between the cavity mode and the QD emission.
We interpret the dominant QD emission lines in both devices as trions (CX), due to their linear intensity dependence on excitation power, and the lack of fine-structure splitting.
This is in line with typical spectra for our InAs/InP QDs with preferential CX recombination where the average CX binding energy was measured to be $\SI{4.7}{\milli\electronvolt}$~\cite{Holewa2022Mirror}.

The coupling between the QD and the CBG cavity is evidenced by the observation of a reduced emission decay time as compared to the decay of QDs in the planar reference structure.
For the CX line in QD-CBG \#1 and \#2 we record decay times of $\tau_{\mathrm{\#1}} = \SI{0.40(1)}{\nano\second}$ and $\tau_{\mathrm{\#2}} =\SI{0.53(1)}{\nano\second}$, respectively (\subfigref{QD-in-cavity}{b}).
To take statistical QD-to-QD fluctuations for the reference decay time into account, we estimate the average decay time of 8 QD CX lines observed from dots located
outside of the cavities, i.e. without Purcell-induced modification of the radiative lifetime,
and obtain $\tau_{\mathrm{ref}} = \SI{1.99(16)}{\nano\second}$, while the single reference shown in \subfigref{QD-in-cavity}{b} has a decay time of $\SI{2.01(2)}{\nano\second}$.
Therefore, the measured Purcell factor for QD-CBG \#1 is $F_{\mathrm{P}} = \SI{5.0(4)}{}$ and $F_{\mathrm{P}} = \SI{3.8(3)}{}$ for QD-CBG \#2.
Although the obtained Purcell factors are comparable with $F_{\mathrm{P}} = \SI{3}{}$ obtained in the non-deterministic fabrication approach~\cite{Nawrath2023}, we expect it to be much higher if the QD would perfectly match the cavity mode both spectrally and spatially.
However, the expected Purcell factor decays rapidly with the dipole displacement $\vec{r}_0$ from the cavity center (decreases by half for $|\vec{r}_0|=\SI{100}{\nano\meter}$, see \SMnote{1}), as the fabricated CBG is optimized for a higher-order mode that exhibits electric field minima along the cavity radial direction (see \SMnote{1}).
Hence, the relatively large total positioning uncertainty for QD-CBGs \#1--\#3 ($\Delta R\sim\SIrange{140}{150}{\nano\meter}$) and the non-ideal spectral emitter-cavity overlap explain the reduced $F_{\mathrm{P}}$ as compared to the model.

\subsection*{Photon extraction efficiency}
We evaluate the photon extraction efficiency by recording the power-dependent $\upmu$PL signal with a superconducting nanowire single-photon detector (SNSPD) in a calibrated optical setup (see \subfigref{QD-in-cavity}{c}).
The setup has a total transmission of $\SI{1.1(2)}{\percent}$ (see \SMnote{5}).
The measured $\eta$ values are corrected by the factor $\sqrt{1-\gtwo{0}}$, to account for the detection of secondary photons due to the refilling of QD states~\cite{Yang2020,Kumano2016}.
Here, the $\gtwo{0}$ value is obtained under the excitation power $P_{\mathrm{sat}}$ corresponding to saturation of the CX line.
Evaluating the CX emission, we obtain an extraction efficiency $\eta_{\mathrm{\#1}}=\SI{16.6(27)}{\percent}$ for QD-CBG \#1 and $\eta_{\mathrm{\#2}}=\SI{13.3(22)}{\percent}$ for QD-CBG \#2 using an objective with $\mathrm{NA}=0.4$.

\subsection*{Single-photon emission purity}
The photon statistics of a quantum light source is of fundamental importance for applications in photonic quantum technologies.
In the following, we investigate the single-photon purity of the emission from QD-CBG \#2 by analyzing the photon autocorrelation function $\gtwo{\tau}$ (cf. \SMnote{7} for details on the data analysis and complementary $\gtwo{\tau}$ measurements).

\begin{figure*}[htb] %
    \begin{center} %
    \includegraphics[width=.6\columnwidth]{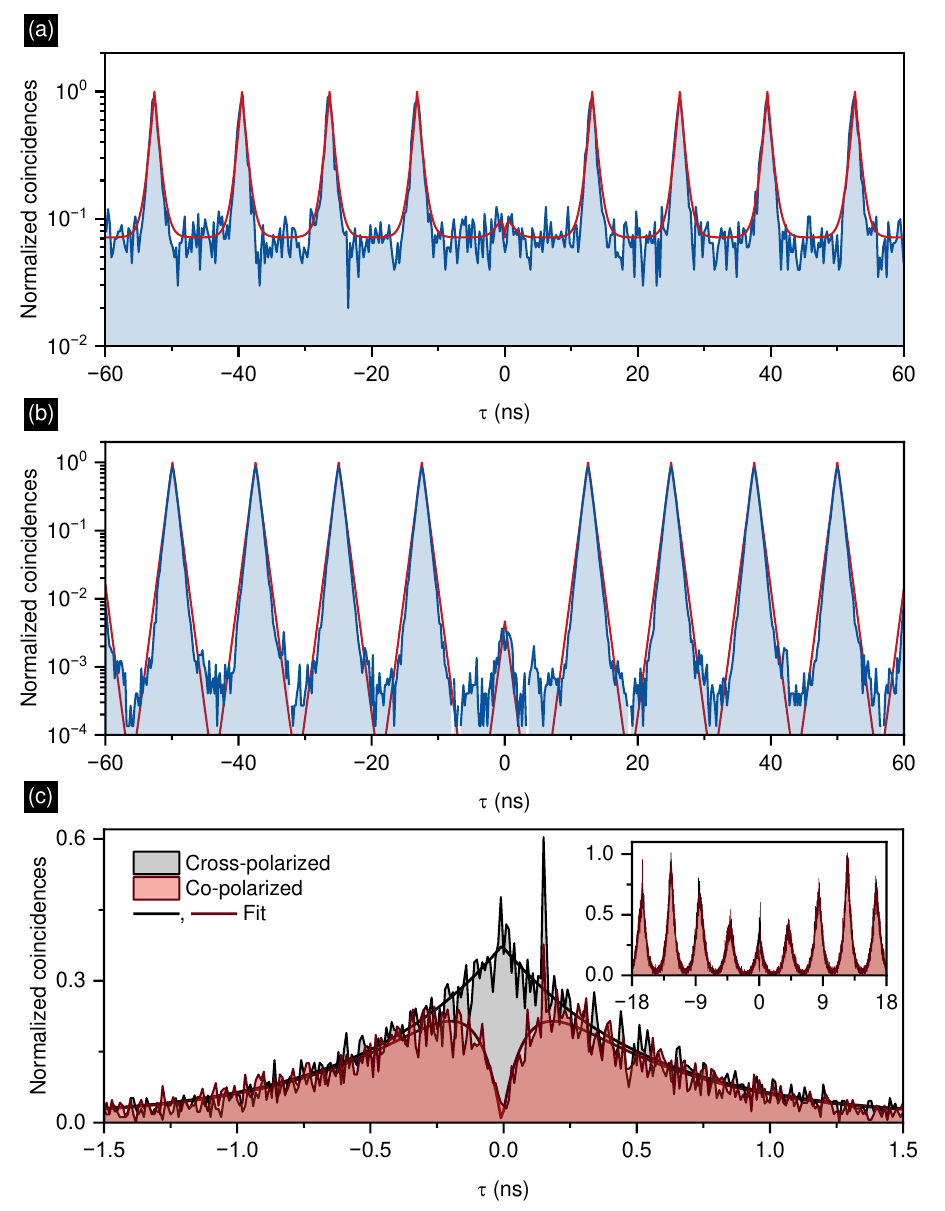} 
    \end{center}	 %
    \caption{\label{fig:g2-and-HOM}
    Quantum optical experiments on device \#2.
    a, b,~The second-order autocorrelation function $\gtwo{\tau}$ of the triggered photons for the (a)
    above-band excitation, and (b) LO-phonon-assisted quasi-resonant excitation,
    c,~Hong-Ou-Mandel histogram for cross- and co-polarized photons evidencing the indistinguishability by the two-photon interference visibility of $V = \SI{19.3(26)}{\percent}$, and the post-selected value of $V_{\mathrm{PS}} = 99.8^{+0.2}_{-2.6}\si{\percent}$.
    Inset: Data for larger delay $\tau$ range.}
\end{figure*}

\figrefL{g2-and-HOM} depicts the measured $\gtwo{\tau}$ histograms obtained under pulsed off-resonant excitation at a power $0.5\times P_{\mathrm{sat}}$ (\subfigref{g2-and-HOM}{a}) and LO-phonon-assisted, quasi-resonant excitation at $0.04\times P_{\mathrm{sat}}$ (\subfigref{g2-and-HOM}{b}).
Under off-resonant excitation, the single-photon purity is limited by recapture processes resulting in $\gtwo{0}_{\mathrm{fit}} = \SI{0.05(2)}{}$, where the uncertainty is mainly determined by the background level $B$.
From the fit, we determine a decay time of $\tau_{\mathrm{dec}}=\SI{0.67(3)}{\nano\second}$, in good agreement with the spontaneous emission decay time observed in \subfigref{QD-in-cavity}{b}, $\tau_{\mathrm{\#2}} =\SI{0.53(1)}{\nano\second}$.

Under weak quasi-resonant excitation at $P\ll P_{\mathrm{sat}}$, the probability for charge-carrier recapture is strongly reduced, resulting in almost negligible background contributions ($B=0$) (\subfigref{g2-and-HOM}{b}) and a fitted value of $\gtwo{0}_{\mathrm{fit}} = \SI{4.7(26)E-3}{}$ at $P=0.04 \times P_{\mathrm{sat}}$.
Additionally, we evaluated the raw antibunching value by integrating the raw coincidences around $\tau=0$ over a full repetition period normalized by the Poisson level of the side peaks.
This results in $\gtwo{0}_{\mathrm{raw}} = \SI{3.2(6)E-3}{}$, with the error deduced from the standard deviation of the distribution of counts in the side peaks.
As discussed later, this result compares favorably with previous reports on non-deterministically fabricated QD-CBGs.

	%
\subsection*{Photon indistinguishability}

Finally, we explore the photon indistinguishability of QD-CBG \#2 by Hong-Ou-Mandel (HOM)-type two-photon interference (TPI) experiments~\cite{Hong1987} (see \methods{} and \SMnote{7} for details on the experimental setup, data analysis, and complementary TPI measurements).
The HOM histograms recorded for co- and cross-polarized measurement configurations are presented in \subfigref{g2-and-HOM}{c}, and were obtained under pulsed quasi-resonant excitation with identical experimental conditions as the $\gtwo{\tau}$ measurement presented in the previous section (at $0.04 \times P_{\mathrm{sat}}$).
The data shown in \subfigref{g2-and-HOM}{c} is not corrected for multi-photon events or contributions from residual laser light.

The HOM histograms feature a characteristic pattern that we analyze following the methodology described in Ref.~\onlinecite{Thoma2016}.
The reduced area of the central peak in the co-polarized measurement, as compared to the maximally distinguishable cross-polarized measurement, is a distinct signature of the two-photon coalescence due to a significant degree of photon indistinguishability.
The visibility of the TPI is calculated from the ratio of the fitted central peak area in the co- and cross-polarized measurements as $V = 1-A_{\mathrm{Co}}/A_{\mathrm{Cross}}$.
We obtain a TPI visibility of $V = \SI{19.3(26)}{\percent}$ with the accuracy being the propagated fitting errors reflecting the statistics of the experimental data. The temporally post-selected visibility at zero delay time ($\tau=0$) is $V_{\mathrm{PS}} = 99.8^{+0.2}_{-2.6}\si{\percent}$, limited only by the system temporal response.

\subsection*{Photon coherence time}
From the width of the central dip, we extract a photon coherence time of $T_2=\SI{103(13)}{\pico\second}$, while the highest extracted coherence time is $T_2=\SI{176(9)}{\pico\second}$, measured for the lowest excitation power (see \SMnote{7}).
Given the Purcell-reduced lifetime of $T_1=\SI{400}{\pico\second}$, this results in a $T_2/T_1$-ratio of $0.44$, comparing favorably with previous reports for QDs emitting in the telecom C-band~\cite{Anderson2021,Nawrath2019}.
The still relatively short coherence time observed in our work is mainly attributed to fluctuating charges in the QD environment, suggested by the observed time-dependent spectral diffusion of emission lines, which most probably limits the observed indistinguishability.
The coherence properties may be further improved by implementing electrical charge stabilization via electric gates or using droplet epitaxy as an alternative growth technique~\cite{Anderson2021}.

To gain further insights into the coherence properties, we performed direct measurements of the $T_2$ time using an all-fiber-based Michelson interferometer (MI; see \methods{}).
We extract a coherence time of up to \SI{62(3)}{\pico\second} for the QD-CBG~\#2 under weak CW above-band excitation (see \SMnote{7}).
Note that while this value is lower than the $T_2$ time extracted from the dip in the HOM experiment in \subfigref{g2-and-HOM}{c}, a direct comparison is not possible due to the different excitation schemes applied.
An analysis of MI data from a total of three different CBG devices yields $T_2$ values between \SI{18}{\pico\second} and \SI{60}{\pico\second}.
These numbers compare favorably with MI-measured coherence times of $\SIrange{6}{30}{\pico\second}$ obtained for three different QDs in planar regions on the same sample.
The observed coherence times are comparable with reports in the literature for SK InAs/InP QDs \cite{Anderson2021}.
While the direct comparison of the $T_2$ values measured for QDs with and without CBG should be treated with care due to the relatively low statistics, these results indicate that the microcavity integration does not degrade the optical coherence of the emitted photons.
Future work in this direction may include a more elaborate study allowing for deeper insights into the limiting dephasing mechanisms and their timescales, e.g., by applying photon-correlation Fourier spectroscopy~\cite{Brokmann2006}.
\section*{Discussion}
The scalable fabrication of active quantum photonic devices operating in the telecom C-band has been a long-standing challenge.
This is mainly due to the random size and strain distribution of epitaxially grown QDs causing an inhomogeneous broadening of the emission and difficulties in localizing suitable QDs due to the high electronic noise level of detector arrays sensitive around $\SI{1.55}{\micro\meter}$ wavelength.
In this work, we present a solution to this problem based on a hybrid sample design.
We fabricate an InP layer containing epitaxial QDs on top of a Al reflector placed on a Si wafer carrier.
This geometry significantly enhances the photon extraction efficiency from the QDs by a factor $>7$ enabling the localization of single QDs in a wide-field imaging setup with a thermo-electrically cooled InGaAs camera.
For the $\SI{10}{\percent}$ brightest QDs our setup achieves an imaging SNR$\,>15.5$ and a localization uncertainty of $\sim\SI{80}{\nano\meter}$ with respect to alignment marks.
After final EBL processing, we achieve an overall uncertainty of $\sim\SI{90}{\nano\meter}$ for fabricating a nano-photonic device around pre-selected QDs.
The localization accuracy of our setup is comparable to setups operating in the $\SIrange{770}{950}{\nano\meter}$ range ($\SI{48}{\nano\meter}$~\cite{Kojima2013}, $\SI{30}{\nano\meter}$~\cite{Sapienza2015}) where Si-based sensors with four orders of magnitude lower electronic noise can be used.
The accuracy in our setup can be further improved by using higher-NA objectives inside the cryostat together with an overall increased microscope magnification, which was reported to reduce the localization accuracy down to $\SI{5}{\nano\meter}$~\cite{Liu2017}.
Alternatively, in-situ EBL~\cite{Gschrey2015} or photolithography~\cite{Dousse2008} scanning techniques provide a similar accuracy down to $\sim\SI{35}{\nano\meter}$, but are comparably slow, require cathodoluminescence signal for QD localization or, in case of photolithography, are not suitable to define reliably sub-$\si{\micro\meter}$ features.

To exemplify our approach, we fabricate CBG cavities with a resonance wavelength of $\SI{1.55}{\micro\meter}$ around some of the pre-selected QDs.
The low QD density of $\SI{3.1e8}{\per\square\centi\meter}$ guarantees their spatial isolation sufficient for the imaging procedure while maximizing the number of devices that can be fabricated per field.
The QD-to-CBG coupling is evidenced by a Purcell factor of $F_{\mathrm{P}} = \SI{5.0(4)}{}$, further increasing the single-photon emission rate and final source brightness.
Using our approach, we obtain a total process yield of $\SI{30}{\percent}$ for finding a pre-selected QD spectrally matching the CBG cavity, which is a significant improvement compared to the yield achievable with a random placement approach ($\sim\SI{0.7}{\percent}$).
For our QD-CBG device, we measure a photon extraction efficiency of $\SI{16.6(27)}{\percent}$ with a $\mathrm{NA} = 0.4$ objective, which is comparable to previously reported devices fabricated probabilistically and operating at C-band wavelengths~\cite{Nawrath2023}, as well as deterministically fabricated CBGs at O-band with In(Ga)As/GaAs QDs~\cite{Xu2022}.
However, no cavity positioning accuracy, fabrication yield, and HOM visibility data were provided in Ref.~\onlinecite{Xu2022}.
The discrepancy between the simulated and measured photon extraction efficiency is attributed to the residual defects in the epitaxial material and possible material damage due to the dry etching.
The fabrication can be optimized to eliminate both effects.

Our QD-CBG devices feature excellent single-photon emission purities with raw values down to $\gtwo{0}=\SI{3.2(6)E-3}{}$, beating previous records for non-deterministically~\cite{Nawrath2023} and deterministically~\cite{Xu2022} fabricated QD-CBGs, as well as most QDs operating in the C-band~\cite{Takemoto2015,Zeuner2021}, while being not yet competitive with the state-of-the-art~\cite{Miyazawa2016}.

Importantly, we report triggered TPI experiments for InP-based cavity-coupled QDs with emission wavelengths in the telecom C-band, which is crucial for applications in quantum information processing (QIP).
We generate indistinguishable photons with a TPI visibility up to $V = \SI{19.3(26)}{\percent}$ and a post-selected value of $V_{\mathrm{PS}} = 99.8^{+0.2}_{-2.6}\si{\percent}$ at zero time delay.
Previous reports on C-band QD-SPSs by other groups were based either on droplet epitaxy InAs/InP QDs in planar structures~\cite{Anderson2021,Wells2023} or InAs QDs grown on GaAs followed by an InGaAs metamorphic buffer, also located in planar structures~\cite{Nawrath2019,Nawrath2021} or embedded in randomly placed CBGs~\cite{Nawrath2023}.
The photon coherence time measured using a Michelson interferometer is up to \SI{62(3)}{\pico\second} for the QD-CBG~\#2 under weak CW above-band excitation.

Further improvement in the photon indistinguishability is of utmost importance for applications in QIP. 
This is challenged by the strong coupling of QDs to their semiconductor environment via charge and spin noise, both causing QD decoherence~\cite{Kuhlmann2013ChargeNoise,Thoma2016}.
It is thus important to stabilize the QD environment by removing the excess charge carriers from the vicinity of QDs, e.\,g., by integrating them into a p-i-n junction, which is expected to increase the photon coherence and indistinguishability substantially.
On the other hand, tuning the QD emission energy using strain~\cite{Joens2011,Zeuner2018}, or quantum-confined Stark effect~\cite{Zhai2020,Zhai2022}, would address the challenge of QD ensemble inhomogeneous broadening by fine-tuning the QD energy to match the cavity mode.
The QD tuning is feasible using the reported approach but requires a different cavity design~\cite{Buchinger2023,Shih2023}.
Implementing coherent optical pumping schemes, such as two-photon resonant excitation \cite{Vajner2024}, also for scalably fabricated devices, while avoiding the excess charge carriers that could originate, e.g., from the electrical QD excitation, is a crucial next step to further improve the photon coherence time and hence indistinguishability~\cite{Reigue2019}.

Moreover, the InP material system used in our work appears to be advantageous for QD-based quantum photonic devices operating in the C-band and compared to GaAs-based devices.
Despite the careful strain engineering involved in the epitaxy of QDs on GaAs~\cite{Sittig2022}, the metamorphic buffer complicates the device engineering and QD growth.
In contrast, an unstrained InP system is free from threading dislocations that would be a source of dangling bonds causing non-radiative recombination, thus lowering the efficiency~\cite{Seravalli2011}.

In conclusion, our work opens the route for the high-throughput fabrication of telecom C-band wavelength quantum photonic devices with QDs delivering flying qubits, i.e., single or entangled photons~\cite{Lu2021}, or acting as a non-linear element for QIP~\cite{Javadi2015}.
Improvements in our optical imaging setup will further increase the device yield and positioning accuracy, while the electric control and coherent excitation of QD emitters will further push the achievable photon indistinguishability.\\

\begin{scriptsize}
\noindent{\bf \normalsize Methods}\\
\noindent{\bf Epitaxial growth and fabrication of planar structure with QDs--}
The structures were grown on epi-ready (001)-oriented InP substrates by the low-pressure metalorganic vapor-phase epitaxy (MOVPE) TurboDisc\textregistered\ reactor using arsine (AsH$_3$), phosphine (PH$_3$), tertiarybutylphosphine (TBP), trimethylgalium (TMG) and trimethylindium (TMIn) precursors with H$_2$ as a carrier gas.
We grow the $\SI{0.5}{\micro\meter}$-thick InP buffer followed by $\SI{200}{\nano\meter}$-thick In$_{0.53}$Ga$_{0.47}$As lattice-matched to InP etch-stop layer and a $\SI{156}{\nano\meter}$-thick InP layer at $\SI{610}{\degreeCelsius}$.
Then, the temperature is decreased to $\SI{493}{\degreeCelsius}$, stabilized under TBP for $\SI{180}{\second}$ and AsH$_3$ for $\SI{27}{\second}$.
The nucleation of QDs occurs in the near-critical regime of Stranski-Krastanov growth mode after deposition of nominally $\SI{1.22}{\ML}$-thick InAs at growth rate $\SI{.53}{\ML\per\second}$ under TMIn and AsH$_3$ flow rates of $\SI{11.8}{\micro\mole\per\minute}$ and $\SI{590}{\micro\mole\per\minute}$, respectively (V/III ratio of $50$). 
Nucleated QDs are annealed for $\SI{3.5}{\second}$ at the growth temperature in AsH$_3$ ambient before the deposition of a $\SI{156}{\nano\meter}$-thick InP capping layer ($\SI{12}{\nano\meter}$ at $\SI{493}{\degreeCelsius}$, and the remaining $\SI{144}{\nano\meter}$ after increasing the temperature up to $\SI{610}{\degreeCelsius}$) what finishes the growth sequence.\\

\noindent{\bf Fabrication of the sample for $\upmu$PL imaging--} 
After the QD epitaxy, SiO$_2$ is deposited in plasma-enhanced chemical vapor deposition (PECVD).
This layer is intended to be $\SI{358.6}{\nano\meter}$-thick, and it covered with a $\SI{120}{\nano\meter}$-thick Al layer deposited via electron-beam evaporation.
The flipped structure is bonded to the Si chip carrier utilizing spin-coated AP3000 adhesion promoter and benzocyclobutene (BCB) on Si and AP3000 on the InP wafer.
The bonding is done by applying the force of $\sim\SI{2}{\kilo\newton}$ in vacuum at $\SI{250}{\degreeCelsius}$.
The substrate removal step is done by $\sim\SI{60}{\minute}$ dip in HCl and the InGaAs etch-stop layer is subsequently removed in H$_2$SO$_4$:H$_2$O$_2$:H$_2$O=1:8:80 mixture.
Next, by employing electron-beam lithography (EBL) followed by inductively coupled plasma-reactive ion etching (ICP-RIE) to etch InP down to SiO$_2$, the square imaging fields were fabricated with $\SI{50}{\micro\meter}$ side-length and AMs dedicated to EBL outside the fields.
Therefore, there are different AMs for optical imaging (edges of the imaging fields) and for EBL alignment marks detection (InP crosses).
This approach is justified by the simplification of the fabrication flow by avoiding the deposition of metallic AMs, relying instead on the outline of the field visible due to the $\upmu$PL signal scattering from its edges.
The material contrast between InP and SiO$_2$/Al regions is sufficient for the AM detection during the EBL alignment step.\\

\noindent{\bf Modeling of the CBG--} The QD is modeled as a classical dipole~\cite{Novotny2012}, and the numerical simulations of the CBG geometry are performed using a modal method employing a true open boundary condition~\cite{Guer2021}.
See the \SMnote{1} for further details. \\

\noindent{\bf Deterministic fabrication of QD-CBG devices--}
After determining the positions of QDs with respect to the AMs, the CBG pattern is defined in the CSAR e-beam resist using e-beam lithography using high-precision alignment based on the InP mark detection in JEOL JBX-9500FSZ e-beam writer.
The pattern is transferred into the PECVD-deposited $\SI{110}{\nano\meter}$-thick SiN$_x$ layer using ICP-RIE with SF$_6$-based etch recipe.
Residual CSAR is stripped in Remover 1165 followed by $\SI{10}{\minute}$ descum in the barrel-type plasma asher.
Subsequently, the pattern is transferred into the InP layer in ICP-RIE by HBr-based etch.
The calculated design is first scaled and fabricated using a nominally identical heterostructure to investigate the mode energy vs. size dependence and to account for the fabrication imperfections.
Additionally, we experimentally determine the $\SI{\sim15}{\nano\meter}$ temperature-induced blueshift of the mode energy between a room and low temperature resulting from the contraction of the structure (introducing size and strain changes), as well as from the change of the refractive indices of the layers.\\

\noindent{\bf Optical characterization of devices--}
The structure with QD-CBG devices is held in a helium-flow cryostat allowing for control of the sample temperature in the range of $\SIrange{4.2}{300}{\kelvin}$. 
For our standard $\upmu$PL studies, the structures are optically excited through a microscope objective with $\mathrm{NA}=0.4$ or $0.65$ and $20\times$ magnification using $\SI{660}{\nano\meter}$ or $\SI{805}{\nano\meter}$ light generated with semiconductor laser diodes.
The same objective is used to collect the $\upmu$PL signal and to direct it for spectral analysis into a $\SI{1}{\meter}$-focal-length monochromator equipped with a liquid-nitrogen-cooled InGaAs multichannel array detector, providing spatial and spectral resolution of $\SI{\sim2}{\micro\meter}$ and $\SI{\sim25}{\micro\electronvolt}$, respectively.

The photon extraction efficiency and time-resolved $\upmu$PL are measured in the same setup.
Here, QDs are excited by $\SI{\sim50}{\pico\second}$-long pulses with a repetition rate of $\SI{80}{\mega\hertz}$ and a central wavelength of $\SI{805}{\nano\meter}$.
At the same time, the second monochromator output port is equipped with the fiber coupling system, transmitting the signal to an NbN-based SNSPD (Scontel) with $\SI{\sim87}{\percent}$ quantum efficiency in the range $\SIrange{1.5}{1.6}{\micro\meter}$ and $\sim200$ dark counts per second.
A multichannel picosecond event timer (PicoHarp~300 by PicoQuant GmbH) analyzes the single photon counts as a time-to-amplitude converter.
The overall time resolution of the setup is $\SI{\sim80}{\pico\second}$.
Experimental setups are shown in \SMnote{4}, and data on the setup transmission efficiency used for determining the photon extraction efficiency is given in \SMnote{5}.\\

\noindent{\bf Photon autocorrelation measurements--}
For the photon-autocorrelation measurements, QD-CBG devices were optically excited using a Ti:Sapphire (Ti:Sa) laser (Coherent Mira-HP) or a widely tunable ps-pulsed laser system based on an optical parametric oscillator (OPO) (picoEmerald by APE GmbH) with repetition rates of $\SI{76}{\mega\hertz}$ and $\SI{80}{\mega\hertz}$, respectively.
We use a fiber-coupled bandpass filter ($\mathrm{FWHM}=\SI{\sim0.4}{\nano\meter}$) for spectrally selecting the QD emission, followed by a 50:50 fiber beam splitter.
For the off-resonant excitation, we use a microscope objective with $\mathrm{NA}=0.7$ and $100\times$ magnification and excite the QD emission with $\SI{\sim2}{\pico\second}$-long pulses at $\SI{830}{\nano\meter}$ from the Ti:Sa.
The signal is detected with a pair of SNSPDs with $\SI{\sim87}{\percent}$ and $\SI{\sim92}{\percent}$ quantum efficiency at $\SI{1550}{\nano\meter}$.
For the quasi-resonant excitation, we use an aspheric lens ($\mathrm{NA}=0.6$) mounted inside a low-vibration closed-cycle cryostat (attoDRY800 by Attocube Systems AG) cooled to $\SI{4.5}{\kelvin}$.
Here, the OPO-laser is used and adjusted to a pulse length of $\SI{5}{\pico\second}$.
Single photons are detected via SNSPDs with $\SI{80}{\percent}$ detection efficiency at $\SI{1550}{\nano\meter}$ and $\SI{57}{\pico\second}$ timing jitter (complete system temporal response).
The excitation energy was determined in photoluminescence excitation experiments to be $\SI{0.83537}{\electronvolt}$ ($\SI{37.57}{\milli\electronvolt}$ above the QD emission energy, cf. \SMnote{7}), which was also used for following TPI experiments.\\

\noindent{\bf Photon-indistinguishability measurements--} 
In the TPI experiments, an additional $\SI{4}{\nano\second}$ delay was introduced between consecutive laser pulses by adding an imbalanced free-space Mach-Zehnder interferometer (MZI) in the excitation path, which was compensated in the HOM setup on the detection side.
Hence, the excitation sequence is composed of pairs of pulses separated by $\SI{4}{\nano\second}$, every $\SI{12.5}{\nano\second}$ corresponding to $\SI{80}{\mega\hertz}$ laser repetition rate.
Free-space waveplates were used to match the polarization for the TPI inside the fiber beam splitter.
The exact polarization was set by using a polarimeter at the beam splitter output in combination with a laser tuned to the QD emission wavelength.
Fine-tuning the relative delay between both MZI arms was used to match precisely the detection and excitation delay, respectively.
The contrast of classical Michelson interference of the laser with itself was used for optimization.
See \SMnote{7} for the details of the HOM data analysis and the scheme of the experimental setup.\\

\noindent{\bf Coherence measurements--}
For measurements of the $T_2$ time, an all-fiber-based Michelson interferometer (MI) was implemented~\cite{Anderson2021}, consisting of a $2\times2$-port $50:50$ fiber beam splitter with both exit ports terminated by a Faraday mirror, reflecting the light with $\ang{90}$ polarization rotation.
The necessary coarse and fine temporal delay is controlled by a variable optical delay stage and a piezo-driven fiber stretcher in the two MI arms, respectively.
The single-photon signal is coupled to one input port of the MI and detected at the second input port using a SNSPD. 
The MI setup in its configuration features $\SI{80}{\percent}$ overall transmission (excluding the BS) and allows for the measurement of coherence times of up to $\SI{1}{\nano\second}$.
The maximally achievable interference contrast was measured with a CW laser at $\SI{1550}{\nano\meter}$ to be $\SI{98}{\percent}$, limited only by the slight intensity mismatch due to the reduced transmission through the optical delay line.
For each temporal delay adjusted via the coarse variable delay line, a fine temporal scan is performed via the fiber stretcher, resulting in interference fringes with an amplitude depending on the overall delay.
The interference fringes are evaluated by subtracting a constant amount of dark counts and evaluating the interference contrast via $v = (I_{\text{max}}-I_{\text{min}})/(I_{\text{max}}+I_{\text{min}})$.
Finally, the $T_2$ time is extracted by fitting a two-sided exponential decay to the interference visibility $v$ data as a function of the coarse delay set in the MI with the uncertainty representing the fit accuracy.\\

\noindent{\bf \normalsize Data availability}\\
The source data for plots in Figs. 1--4 and 50 representative $\upmu$PL maps with full results of data analysis leading to QD localization have been deposited in the Figshare database under accession code \href{https://doi.org/10.6084/m9.figshare.24530050}{10.6084/m9.figshare.24530050}~\cite{Holewa2024dataset}.\\

\noindent{\bf \normalsize Code availability}\\
The codes of this study are available from the corresponding authors upon reasonable request.\\

\end{scriptsize}
\noindent{\bf \normalsize References}
%

\begin{scriptsize}
 \,\\
\noindent{\bf \normalsize Acknowledgements}\\
The authors acknowledge financial support from the Danish National Research Foundation through NanoPhoton - Center for Nanophotonics, grant number DNRF147, and the Center for Macroscopic Quantum States (bigQ), grant number DNRF142.
P.\,H., M.\,B., P.\,M., A.\,M, and B.\,K. acknowledge financial support from the Polish National Science Center (Grants~No.~2020/36/T/ST5/00511, 2020/39/D/ST5/02952, 2020/39/D/ST5/03359).
D.\,A.\,V. and T.\,H. acknowledge financial support by the German Federal Ministry of Education and Research (BMBF) via the project “QuSecure” (Grant~No.~13N14876) within the funding program Photonic Research Germany, the BMBF joint project “tubLAN Q.0” (Grant~No.~16KISQ087K), and by the Einstein Foundation via the Einstein Research Unit “Quantum Devices”.
N.\,G. acknowledges support from the European Research Council (ERC-CoG “UNITY”, Grant No. 865230), and from the Independent Research Fund Denmark (Grant No.~DFF-9041-00046B).\\

Danish National Research Foundation: DNRF147, E. S. and K. Y.; DNRF142, A. H.;
Polish National Science Center: 2020/36/T/ST5/00511, P. H.; 2020/39/D/ST5/02952, P. M.; 2020/39/D/ST5/03359, B. K.;
European Research Council: 865230, N. G.;
Independent Research Fund Denmark: DFF-9041-00046B, N. G.;
German Federal Ministry of Education and Research: 13N14876, T. H.; 16KISQ087K, T. H.;
Einstein Foundation: Einstein Research Unit “Quantum Devices”, T. H.\\

This version of the article has been accepted for publication after peer review but is not the Version of Record and does not reflect post-acceptance improvements or any corrections.
The Version of Record is available online at: \href{http://dx.doi.org/10.1038/s41467-024-47551-7}{http://dx.doi.org/10.1038/s41467-024-47551-7}.\\

\noindent{\bf \normalsize Author Contributions Statement}\\
E. S.	conceived the project.
M. S. provided the concept of optical imaging and designed spectroscopic experiments.
P. H. and E. S.	optimized and performed the QD growth.
P. M., P. H., B. K., and M. S.	constructed the $\upmu$PL imaging system.
B. G. and N.~G.	designed the CBG.
P. H with the help of A. S., M. X., and K. Y.	developed the nanofabrication process.
P. H and E. Z.-O., with the help of M. B. and P. M. and the advice of M. S.	carried out imaging experiments and device characterization, including above-band excitation photon autocorrelation.
P. H.	localized the QDs and fabricated the devices.
D. A. V. performed the photon autocorrelation and two-photon interference experiments with the help of M. W., E. Z.-O., A. M., and under the supervision of T. H.
P.~H., E. Z.-O., D. A. V., M. W., A. H., with the advice of A. M., M. S., and T. H.	analyzed and interpreted the data.
P. H., with the help of A. H. and D. A. V., and with the advice of T.~H., M. S., and E. S. wrote the paper and SI.
These authors jointly supervised this work: M.~S. and E. S.\\

\noindent{\bf \normalsize Competing Interests Statement}\\
The authors declare no competing interests.\\

\end{scriptsize}
\end{document}


\newcommand*{\papertitle}{High-throughput quantum photonic devices emitting indistinguishable photons in the telecom C-band}

\author{Pawe\l{}~Holewa}
\email{pawel.holewa@pwr.edu.pl}
\affiliation{Department of Experimental Physics, Faculty of Fundamental Problems of Technology, Wroc\l{}aw University of Science and Technology, Wyb. Wyspia\'{n}skiego 27, 50-370 Wroc\l{}aw, Poland}
\affiliation{DTU Electro, Department of Electrical and Photonics Engineering, Technical University of Denmark, Ørsteds Plads 343, DK-2800 Kongens Lyngby, Denmark}
\affiliation{NanoPhoton - Center for Nanophotonics, Technical University of Denmark, Ørsteds Plads 345A, DK-2800 Kongens Lyngby, Denmark}

\author{Daniel~A.~Vajner}
\affiliation{Institute of Solid State Physics, Technische Universität Berlin, 10623 Berlin, Germany}

\author{Emilia~Zi\k{e}ba-Ost\'{o}j}
\affiliation{Department of Experimental Physics, Faculty of Fundamental Problems of Technology, Wroc\l{}aw University of Science and Technology, Wyb. Wyspia\'{n}skiego 27, 50-370 Wroc\l{}aw, Poland}

\author{Maja~Wasiluk}
\affiliation{Department of Experimental Physics, Faculty of Fundamental Problems of Technology, Wroc\l{}aw University of Science and Technology, Wyb. Wyspia\'{n}skiego 27, 50-370 Wroc\l{}aw, Poland}

\author{Benedek Ga\'{a}l}
\affiliation{DTU Electro, Department of Electrical and Photonics Engineering, Technical University of Denmark, Ørsteds Plads 343, DK-2800 Kongens Lyngby, Denmark}

\author{Aurimas~Sakanas}
\affiliation{DTU Electro, Department of Electrical and Photonics Engineering, Technical University of Denmark, Ørsteds Plads 343, DK-2800 Kongens Lyngby, Denmark}

\author{Marek~Burakowski}
\affiliation{Department of Experimental Physics, Faculty of Fundamental Problems of Technology, Wroc\l{}aw University of Science and Technology, Wyb. Wyspia\'{n}skiego 27, 50-370 Wroc\l{}aw, Poland}

\author{Pawe\l{}~Mrowi\'{n}ski}
\affiliation{Department of Experimental Physics, Faculty of Fundamental Problems of Technology, Wroc\l{}aw University of Science and Technology, Wyb. Wyspia\'{n}skiego 27, 50-370 Wroc\l{}aw, Poland}

\author{Bartosz~Krajnik}
\affiliation{Department of Experimental Physics, Faculty of Fundamental Problems of Technology, Wroc\l{}aw University of Science and Technology, Wyb. Wyspia\'{n}skiego 27, 50-370 Wroc\l{}aw, Poland}

\author{Meng~Xiong}
\affiliation{DTU Electro, Department of Electrical and Photonics Engineering, Technical University of Denmark, Ørsteds Plads 343, DK-2800 Kongens Lyngby, Denmark}
\affiliation{NanoPhoton - Center for Nanophotonics, Technical University of Denmark, Ørsteds Plads 345A, DK-2800 Kongens Lyngby, Denmark}

\author{Kresten~Yvind}
\affiliation{DTU Electro, Department of Electrical and Photonics Engineering, Technical University of Denmark, Ørsteds Plads 343, DK-2800 Kongens Lyngby, Denmark}
\affiliation{NanoPhoton - Center for Nanophotonics, Technical University of Denmark, Ørsteds Plads 345A, DK-2800 Kongens Lyngby, Denmark}

\author{Niels~Gregersen}
\affiliation{DTU Electro, Department of Electrical and Photonics Engineering, Technical University of Denmark, Ørsteds Plads 343, DK-2800 Kongens Lyngby, Denmark}

\author{Anna~Musia\l{}}
\affiliation{Department of Experimental Physics, Faculty of Fundamental Problems of Technology, Wroc\l{}aw University of Science and Technology, Wyb. Wyspia\'{n}skiego 27, 50-370 Wroc\l{}aw, Poland}

\author{Alexander~Huck}
\affiliation{Center for Macroscopic Quantum States (bigQ), Department of Physics, Technical University of Denmark, DK-2800 Kongens Lyngby, Denmark}

\author{Tobias~Heindel}
\affiliation{Institute of Solid State Physics, Technische Universität Berlin, 10623 Berlin, Germany}

\author{Marcin~Syperek}
\email{marcin.syperek@pwr.edu.pl}
\affiliation{Department of Experimental Physics, Faculty of Fundamental Problems of Technology, Wroc\l{}aw University of Science and Technology, Wyb. Wyspia\'{n}skiego 27, 50-370 Wroc\l{}aw, Poland}

\author{Elizaveta~Semenova}
\email{esem@fotonik.dtu.dk}
\affiliation{DTU Electro, Department of Electrical and Photonics Engineering, Technical University of Denmark, Ørsteds Plads 343, DK-2800 Kongens Lyngby, Denmark}
\affiliation{NanoPhoton - Center for Nanophotonics, Technical University of Denmark, Ørsteds Plads 345A, DK-2800 Kongens Lyngby, Denmark}

\graphicspath{ {./Figures/} }
\title{\textbf{\large \SM\ \\
		---\\
		\papertitle{}
	}}

\maketitle
\tableofcontents
\ra{1.3}
\preprint{Supplementary information}
\vspace*{1cm}
\section{Optical simulations of the cavity geometry}

The CBG geometry is modeled using a modal method employing a true open geometry boundary condition~\cite{Guer2021}.
Here, the geometry is divided into uniform layers along a propagation \textit{z} axis, and the field is expanded in eigenmodes of each uniform layer.
The QD is modeled as a classical dipole emitter using the equivalence principle~\cite{Novotny2012}. 
The eigenmode expansion coefficients in the QD layer are computed using the reciprocity theorem~\cite{Lavrinenko2014}, and the fields are connected at each layer interface using the $S$ matrix formalism~\cite{Lavrinenko2014,Li1996}. 

We model the Purcell factor $F_{\rm P} = P / P_0$ as the power $P$ emitted by the classical dipole relative to the power $P_0$ in a bulk medium.
The power $P$ and the electric near field $\vec{E}(\vec{r})$ generated by a dipole $\vec{d}$ with frequency $\omega_0$ at the position $\vec{r}_0$ can be written in terms of the optical Green's function $\overleftrightarrow{\vec{G}} (\vec{r},\vec{r}')$ as \cite{Novotny2012}
\begin{align}
P(\vec{r}_0) &=  \frac{{\omega _0^3}{\mu _0} \left| \vec{d} \right|^2} {2} \text{Im} \left(\vec{n}_{\rm d}^* \cdot \overleftrightarrow{\vec{G}} ({\vec{r}}_0,{\vec{r}}_0)  \cdot \vec{n}_{\rm d} \right) \label{eq:P_def} \\
{\vec{E}}({\vec{r}}) &= {\omega _0^2}{\mu _0}  \overleftrightarrow{\vec{G}} ({\vec{r}},{\vec{r}}_0) {\vec{d}}, \label{eq:E_GF}
\end{align}
where $\vec{n}_{\rm d} = \vec{d} / |\vec{d}|$ is the dipole orientation.
The corresponding far field $P_{\mathrm{FF}}(\theta,\varphi,\vec{r}_0)$ is then determined from \eqnref{E_GF} using a standard near field to far field transformation~\cite{Balanis2016}.
The total collected power $P_{\mathrm{Lens}}$ detected by the lens with a given numerical aperture (NA) is obtained by integration of $P_{\mathrm{FF}}(\theta,\varphi,\vec{r}_0)$ over the unit solid angle $\Omega$ as
\begin{equation}
	P_{\mathrm{Lens}}(\vec{r}_0) = \int_{\theta<\theta_{\mathrm{NA}}}  P_{\mathrm{FF}}(\theta,\varphi,\vec{r}_0) \dd{\Omega},
\end{equation}
where $\theta_{\mathrm{NA}}$ is defined by the NA of the lens. Finally, the extraction efficiency is defined as $\eta=P_{\mathrm{Lens}}/P$.

The dominant lines in all investigated QD-CBG devices are trions (CX) which emit circularly polarized photons ($\sigma^\pm$).
We thus model the trion state dipole orientation as
\begin{equation}
	\vec{n}_{\rm d} = \vec{n}_{\rm CX} =  \frac{1}{\sqrt{2}}(\vec{r} \pm i\vec{\varphi}), \label{eq:circ_dipole}
\end{equation}
where $\vec{r}$ and $\vec{\varphi}$ are unit vectors of the cylindrical coordinate system.
Inserting \eqnref{circ_dipole} into \eqnref{P_def}, we obtain the power $P_{\rm CX}$ emitted by the trion given by
\begin{equation}
	P_{\rm CX}(\vec{r}_0) = \frac{P_{r}(\vec{r}_0) +  P_{\varphi}(\vec{r}_0)}{2}, \label{eq:Purcell-CX}
\end{equation}
where $P_{\rm r}$ ($P_{\rm \varphi}$) is the power emitted by a dipole at position $\vec{r}_0$ oriented along the $r$ ($\varphi$) axis. 
Similarly, the far field generated by the trion becomes 
\begin{equation}
	P_{\mathrm{FF,CX}}(\theta,\varphi,\vec{r}_0) = \frac{1}{2}\left(P_{\mathrm{FF},r}(\theta,\varphi,\vec{r}_0)+P_{\mathrm{FF},\varphi}(\theta,\varphi,\vec{r}_0)\right),
\end{equation}
where $r$ and $\varphi$ again refer to far fields generated by the two dipole orientations.
Finally, the total photon extraction efficiency for the trion at the position $\vec{r}_0$ becomes
\begin{equation} \label{eq:eta-CX}
	\eta_{\mathrm{CX}}(\vec{r}_0) = \frac{P_{\mathrm{Lens,CX}}(\vec{r}_0)}{P_{\mathrm{CX}}(\vec{r}_0)} = \frac{P_{\mathrm{Lens},r}(\vec{r}_0)+P_{\mathrm{Lens},\varphi}(\vec{r}_0)}{P_{r}(\vec{r}_0)+P_{\varphi}(\vec{r}_0)}.
\end{equation}

\begin{figure}[htb] %
	\begin{center} %
		\includegraphics[width=0.8\columnwidth]{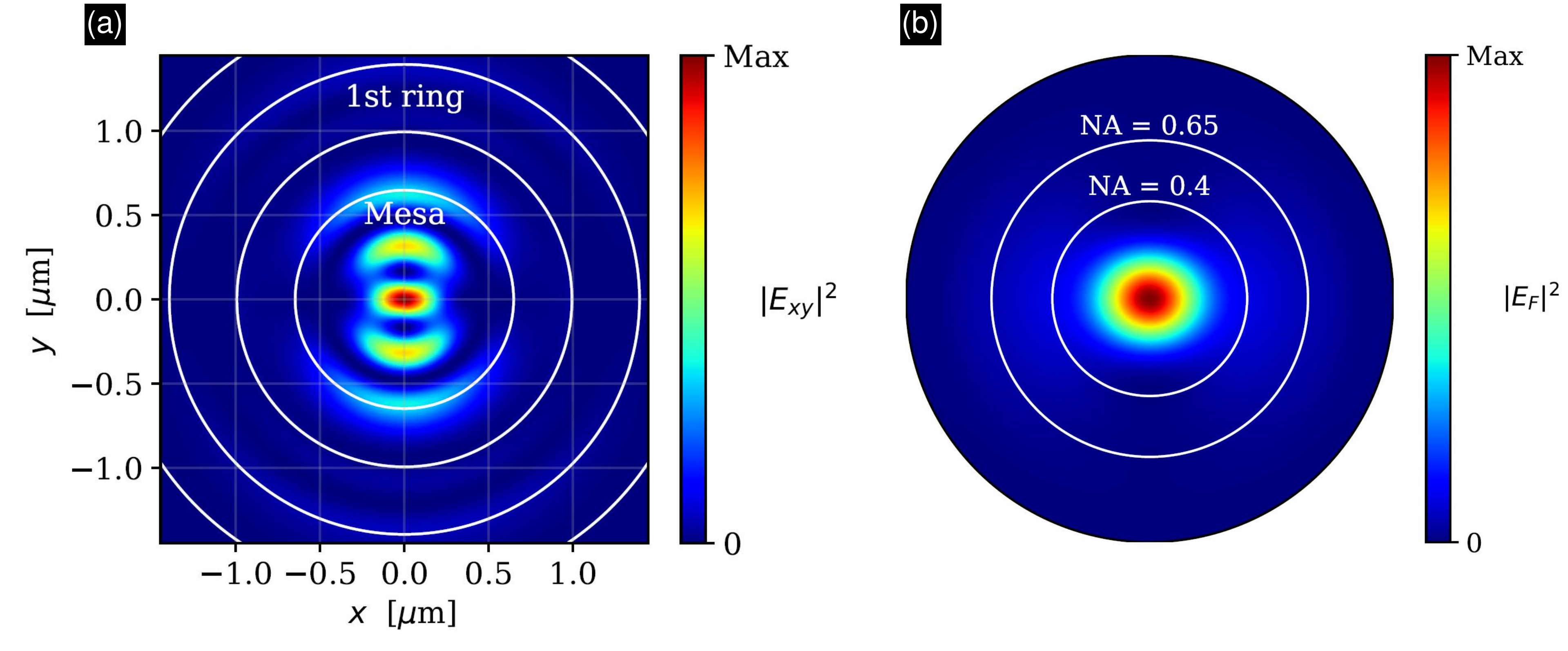} 
	\end{center}
	\caption{\label{fig:mode-profiles}
 Circular Bragg grating cavity mode.
		\textbf{a},~Near field and \textbf{b}, far-field mode profiles of the CBG with 4 rings generated by a linear dipole $\vec{n}_{\rm d} = \vec{r}$.
  $E_F$ -- amplitude of electric field, $E_{XY}$ -- amplitude of the in-plane electric field.
  }
\end{figure}

The near and far fields for the cavity mode are presented in \Supplsubfigref{mode-profiles}. 
The near-field profile shown in \Supplsubfigref{mode-profiles}{a} evidences the higher-order nature of the optimized cavity mode characteristic of the CBG design~\cite{Yao2018}.

\begin{figure}[!htb] %
		\begin{center} %
			\includegraphics[width=0.4\columnwidth]{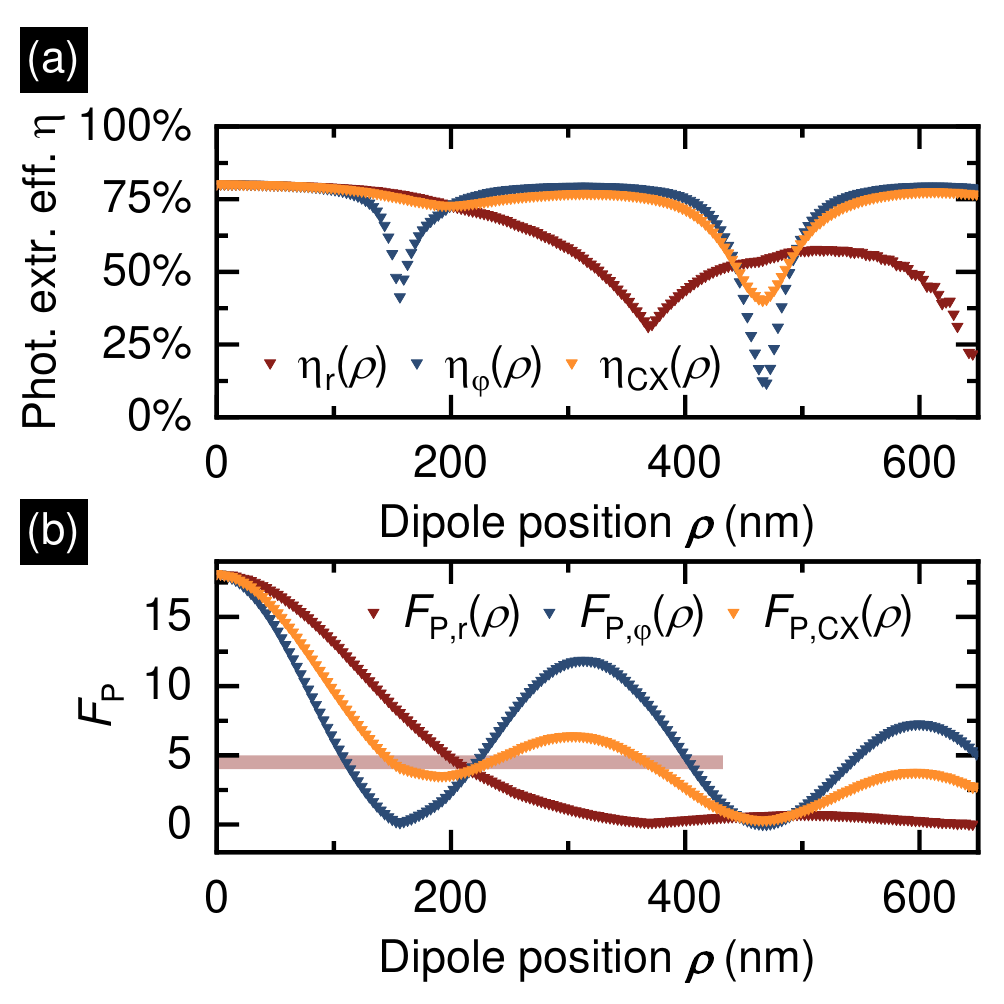} %
		\end{center} %
		\caption{\label{fig:QD-placement-tolerance}%
		Quantum dot displacement tolerance.
  \textbf{a}, the photon extraction efficiency $\eta(\rho)$ and 
  \textbf{b}, Purcell factor $F_{\mathrm{P}}(\rho)$ as a function of the dipole-center separation $\rho$ computed for the $r$, $\varphi$ and trion (CX) dipole orientations.}
	\end{figure}

The extraction efficiency $\eta$ and Purcell factor $F_{\mathrm{P}}$ computed using \eqnref{Purcell-CX} and \eqnref{eta-CX} as a function of spatial misalignment $\rho$ of the QD is presented in \Supplfigref{QD-placement-tolerance}.
Whereas the photon extraction efficiency overall displays robustness towards misalignment, the decay of the Purcell factor with $\rho$ is much more pronounced.
The variations of $F_{\mathrm{P}}$ along the $r$ and $\varphi$ axes are quite different and result from the different variations of the field profile shown in \Supplfigref{mode-profiles} along the $x$ and $y$ axes.

\section{Nanofabrication of the devices}

\subsection{MOVPE growth of InAs/InP quantum dots}

	\begin{figure}[htb] %
		\begin{center} %
    	\includegraphics[width=0.4\columnwidth]{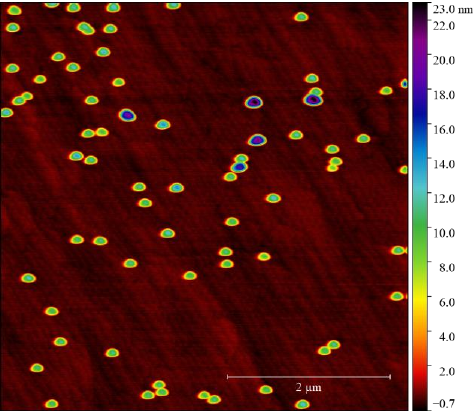} 
	    \end{center}
	    \caption{\label{fig:AFM}
        Atomic force microscopy (AFM) image of the InP surface with InAs quantum dots grown under nominally the same conditions as those used for the imaging.}
	\end{figure}

\SupplfigrefL{AFM} presents the atomic force microscopy (AFM) image of a reference structure that has the same InAs/InP quantum dots (grown under nominally same conditions) without InP capping layer.
We estimate the density of QDs to be $\SI{3.1e8}{\per\square\centi\meter}$ at the center of the wafer where the imaging fields are fabricated.

\subsection{Deterministic fabrication of the cavities}
	\begin{figure}[htb] %
		\begin{center} %
    \includegraphics[width=0.5\columnwidth]{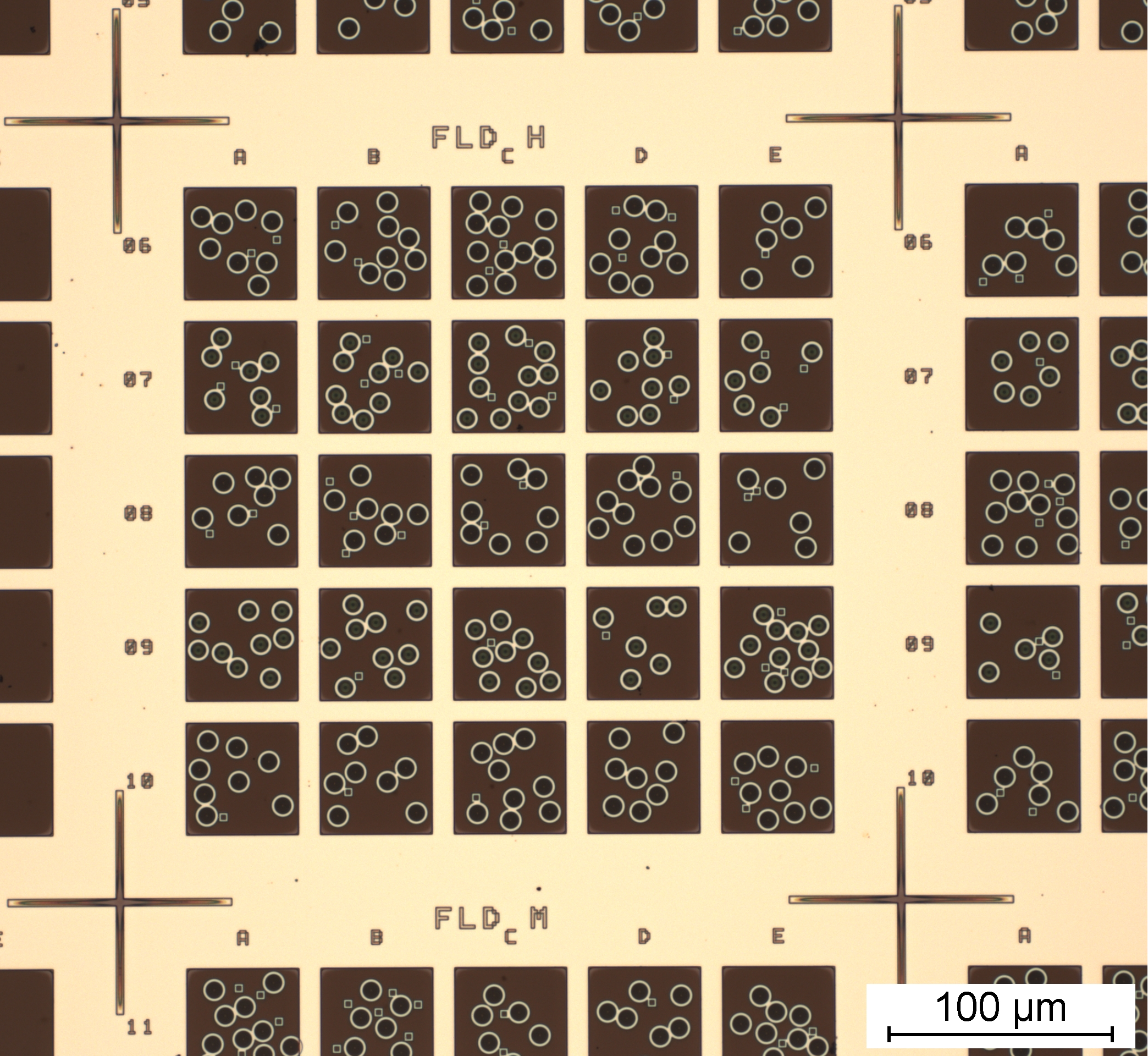} 
	    \end{center}
	    \caption{\label{fig:optical-microscope}
        The optical microscope image of the sample's surface taken at $20\times$ magnification with the cavities transferred to InP.}
	\end{figure}

\Supplfigref{optical-microscope} shows the optical microscope image of the sample's surface taken at $20\times$ magnification with the cavities transferred to InP.
The imaged fragment of the chip shows a $5\times5$ pattern of the imaging fields with four InP crosses at the corners of the pattern.
Also visible are additional, $\SI{2}{\micro\meter}$ side-length square mesas.

\section{Accuracy of determination of QD positions from $\upmu$PL maps and final cavity positioning}
\subsection{Signal-to-noise ratio}
\begin{figure}[!htb] %
		\begin{center} %
			\includegraphics[width=0.4\columnwidth]{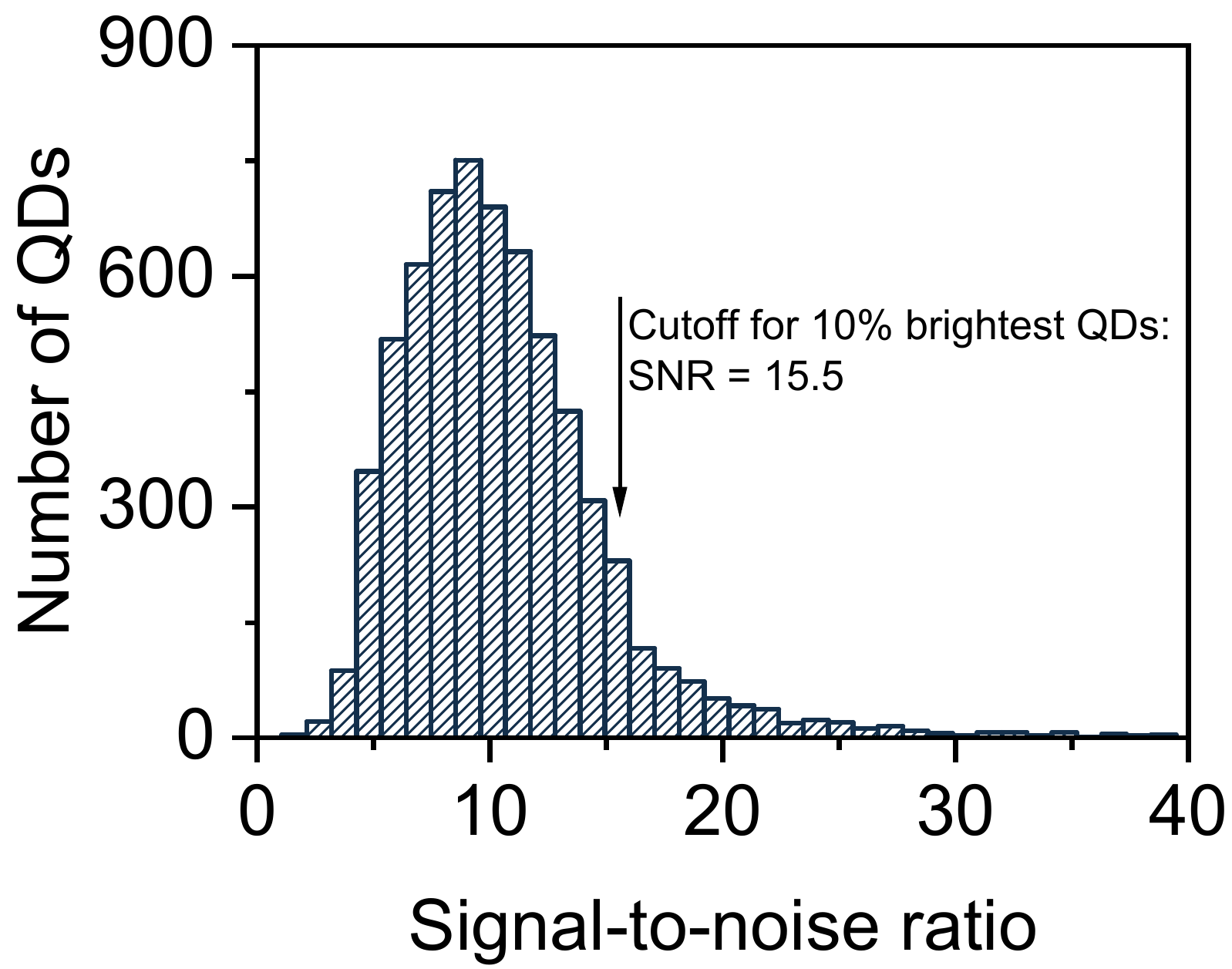} %
		\end{center} %
		\caption{\label{fig:SNR-histogram}%
		Histogram of signal-to-noise ratio (SNR) for the recorded QD spots.}
	\end{figure}

\Supplfigref{SNR-histogram} shows the histogram of signal-to-noise ratio (SNR) for the mapped QDs.
The average SNR is $10.6$ which underlines the crucial role of the $7$-fold enhancement of the signal intensity for the planar structure by application of the metallic mirror~\cite{Holewa2022Mirror}.
Based on this result, we deduce that for a sample without such a mirror, the SNR would be roughly $7$ times lower (SNR of about $1.5$), which would make the optical localization of QDs impossible (compare with the noise level in Figs. 2e and 2f in the main text).

\subsection{Localization algorithm}
Our algorithm calculates the scaling factor $P \left[{\si{\px}}/{\si{\micro\meter}}\right]$ to change the map unit from pixels ($\si{\px}$) to $\si{\micro\meter}$.
For a given $\upmu$PL map, we identify $n$ QD spots and take $2n$ cross-sections of the map (each QD is sectioned in two directions).
The width of each cross-section is taken to equal a few pixels centered at the QD spot maximum to increase the signal-to-noise ratio (SNR).
We found the best SNR for 10 pixels and used this value consistently.

For the $i$-th cross-section, we determine the positions $\mathbb{M}_{i,l} \left[\si{\px}\right]$, $\mathbb{M}_{i,u} \left[\si{\px}\right]$\footnote{For clarity, we use the blackboard-bold font for quantities given in pixels, e.\,g. $\mathbb{M}$, $\mathbb{Q}$. The normal font is used for the same dimensions given in $\si{\micro\meter}$, e.\,g. $Q$.} of the field boundaries (acting as reference/alignment marks, AMs; indices $l$ and $u$ stand for lower and upper field boundaries, the same indices are applied also for left and right field edges) and $\mathbb{Q}_{i} \left[\si{\px}\right]$ for the QD spot position.
All values $\mathbb{M}_{i,l}$, $\mathbb{M}_{i,u}$, and $\mathbb{Q}_{i}$ are the centers of the Gaussian peaks, fitted to respective maxima on the $\upmu$PL map cross-section.
Additionally, we assume that each field is a square of size $F=\SI{50}{\micro\meter}$. 

$P$ is calculated by averaging over all cross-section-related coefficients $P_i$ recorded for a given field ($P=\bar{P_i}$) to ensure its highest accuracy, according to the formula
\begin{equation}\label{eq:P}
    P = \frac{1}{2n} \sum_{i=1}^{2n} P_i
      = \frac{1}{2n} \sum_{i=1}^{2n} \frac{|\mathbb{M}_{i,u} - \mathbb{M}_{i,l}|}{F}.
\end{equation}
Calculating $P$ separately for each map accounts for possible slight changes in the magnification due to defocusing of the sample surface during the cryostat translation, however, we find very low dispersion of $P$ coefficients for different $\upmu$PL maps (see the following section and \Supplfigref{Histogram-deltaP}).

Then, the $i$-th QD position $Q_i~\left[\si{\micro\meter}\right]$ (vertical or horizontal) is calculated as 
\begin{equation}\label{eq:Qi}
    Q_i = \frac{\mathbb{Q}_{i} - \mathbb{M}_{i,l}}{P}.
\end{equation}

\subsection{Uncertainty of scaling the $\upmu$PL maps}
The accuracy of the scaling factor $\Delta P_i$ for a single ($i$-th) $\upmu$PL map cross-section can be calculated by propagating the uncertainties in \eqnref{P}:
\begin{displaymath}
    \Delta P_i = \sqrt{ \left(\frac{\Delta\mathbb{M}_{i,u}}{F}\right)^2 + \left(\frac{\Delta\mathbb{M}_{i,l}}{F}\right)^2 + \left(\frac{\Delta F}{F^2}\right)^2}.
\end{displaymath}
The uncertainty of the field size $\Delta F$ has two contributions, the uncertainty of the electron beam lithography alignment, estimated to $\Delta C=\SI{40}{\nano\meter}$~\cite{Sakanas2019} and the over-etching $\Delta x$ during the ICP-RIE step (estimated to be up to $\Delta x=\SI{40}{\nano\meter}$), potentially influencing $F$ by $2\Delta x = \SI{80}{\nano\meter}$.
$\Delta\mathbb{M}_{i,u}$ and $\Delta\mathbb{M}_{i,l}$ are standard errors of the numerical fitting.

However, to increase the accuracy of the determination of the scaling factor, we average the $P_i$ values for all cross-sections taken in a particular $\upmu$PL map, as can be seen in \eqnref{P}, so as the uncertainty $\Delta P$ we adopt the estimator of the standard error of mean $\hat{\sigma}_{\bar{P_i}}$ for the sample defined as a set of all $2n$ factors $P_i = |\mathbb{M}_{i,u} - \mathbb{M}_{i,l}| / F$ determined for a given $\upmu$PL map:
\begin{equation}\label{eq:deltaP}
    \Delta P = \hat{\sigma}_{\bar{P_i}} = \frac{\sigma_{P_i}}{\sqrt{2n}},
\end{equation}
where $\sigma_{P_i}$ is the sample standard deviation.

\begin{figure}[htb] %
		\begin{center} %
    	\includegraphics[width=0.4\columnwidth]{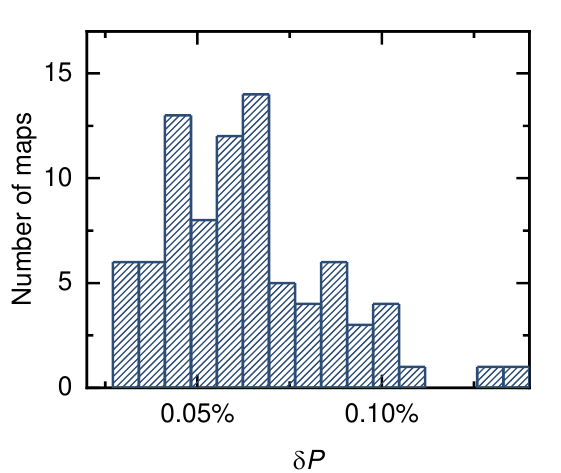} 
	    \end{center}	 
		\caption{\label{fig:Histogram-deltaP}
		The histogram of relative microphotoluminescence map scaling uncertainty $\delta P = \Delta P/P$ for the set of analyzed $m=84$ exemplary microphotoluminescence maps.}
	\end{figure}

According to calculated statistics for $m=84$ exemplary $\upmu$PL maps, we can compare the estimators of the standard error of mean $\hat{\sigma}_{\bar{P_i}}$ determined separately for each $\upmu$PL map and reach the following conclusions:
\begin{enumerate}
\item We find that the coefficients $P_i$ are very close to each other, evidencing no perceptible elongation or distortion of the registered images.
This can be evaluated by the analysis of the set of calculated (for all $m$ maps) uncertainties $\Delta P=\hat{\sigma}_{\bar{P_i}}$ scaled by the determined $P$ factors: $\delta P = \Delta P/P$.
We plot the histogram of $\delta P$ in \Supplfigref{Histogram-deltaP} and find the median value of $\delta P$-distribution of $\SI{0.06}{\percent}$.
In other words, for half of the analyzed maps, the distribution of $P_i$ factors is sufficiently narrow to determine the $P$ factor with relative uncertainty $\delta P<\SI{0.06}{\percent}$.
This value includes the potential elongation of the image (difference in vertically and horizontally determined $P_i$ factors).
\item The stability of the setup and overall repetitivity of the imaging process is high as the image magnification varies very little between maps.
This can be evaluated by the estimator of the standard error of mean $\hat{\sigma}_{\bar{P}}$ calculated not for different cross-sections [averaging $P_i$ values for a given map, as in \eqnref{P}] but for different maps (averaging $P$ values for all maps).
We obtain $\hat{\sigma}_{\bar{P}}/\bar{P} = \SI{0.013}{\percent}$.
We take this value as the estimation of variation of the setup magnification during the imaging process.
\end{enumerate}

Then, the accuracy of cavity positioning $\Delta R$ depends primarily on the accuracy of the QD localization ($\Delta Q$) and EBL alignment uncertainty $\Delta C$.
The $\Delta Q$ itself depends on the fit standard errors ($\Delta \mathbb{M}$, $\Delta \mathbb{Q}$), and scaling factor uncertainty $\Delta P$.

\subsection{Image rotation $\varphi$}
Although we carefully align the image of the field with the horizontal and vertical axes of the detector array, we assume that there can be some indiscernible rotation of the image $\varphi$ on the level of up to a few degrees.
Even if this is the case, the rotation results first in the larger separation between AM peaks in the cross-sections ($|\mathbb{M}_{i,u} - \mathbb{M}_{i,l}|/\cos{\varphi}$ instead of $|\mathbb{M}_{i,u} - \mathbb{M}_{i,l}|$) and this is translated to scaled $P$: $P\to P/\cos{\varphi}$, according to \eqnref{P}.
The distance between lower AM and QD is however also elongated, $\left(\mathbb{Q}_{i} - \mathbb{M}_{i,l}\right)/\cos{\varphi}$ instead of $\left(\mathbb{Q}_{i} - \mathbb{M}_{i,l}\right)$.
According to \eqnref{Qi}, the new QD position is:
\begin{displaymath}
    Q_{i,\mathrm{new}} = \frac{\left(\mathbb{Q}_{i} - \mathbb{M}_{i,l}\right)/\cos{\varphi}}{P/\cos{\varphi}} = Q_i,
\end{displaymath}
so that the slight image rotation has no influence on the determination of QD position $Q_i$ in our approach.

\subsection{One-dimensional QD position $\Delta Q_i$}
The determination of $i$-th QD position $Q_i$ is influenced by the scaling factor uncertainty $\Delta P$, as well as QD ($\Delta\mathbb{Q}_{i}$) and lower AM ($\Delta\mathbb{M}_{i,l}$) fit uncertainties (standard errors of the numerical fitting), and is calculated by propagating the uncertainties, according to \eqnref{Qi}:
\begin{equation}\label{eq:DQi}
    \Delta Q_i = \sqrt{\left(\frac{\Delta\mathbb{Q}_{i}}{P}\right)^2 + \left(\frac{\Delta\mathbb{M}_{i,l}}{P}\right)^2 + \left(\frac{\left(\mathbb{Q}_{i} - \mathbb{M}_{i,l}\right)\Delta P}{P^2} \right)^2}.
\end{equation}

\subsection{Two-dimensional QD position $\Delta Q$}
We combine the one-dimensional QD position uncertainties $\Delta Q_i$ into the two-dimensional uncertainty using the formula
\begin{equation}\label{eq:DQ}
    \Delta Q = \sqrt{\left(\Delta Q_h\right)^2 + \left(\Delta Q_v\right)^2},
\end{equation}
where $\Delta Q_h$, $\Delta Q_v$ are $\Delta Q_i$ values calculated according to \eqnref{DQi} for horizontal and vertical cross-sections.
We use the $\Delta Q$ value to determine the accuracy of our $\upmu$PL imaging method.

\subsection{Distance between QD position and cavity center $R$ -- accuracy of cavity positioning $\Delta R$}
We express the uncertainty $\Delta R$ of the expected distance $R=0$ between the QD position and the cavity center as
\begin{equation}\label{eq:DR}
    \Delta R = \sqrt{\left(\Delta Q\right)^2 + \left(\Delta C\right)^2}.
\end{equation}

\subsection{Example calculations for QD-CBGs \#1--\#3}
Finally, in \Suppltabref{uncertainties-ABC} we show the uncertainties involved in the determination of the QD position $\Delta Q_i$ accordingly to \eqnref{DQi} and of the accuracy of cavity positioning $\Delta R$ accordingly to \eqnref{DR} for three exemplary QD-CBGs \#1--\#3, described in the article (QD-CBG \#1 and QD-CBG \#2) and in the following part of this document (QD-CBG~\#3).

    \begin{table}[htb]
    \caption{\label{tab:uncertainties-ABC} Uncertainties involved in the determination of accuracy of cavity positioning $\Delta R$ for exemplary quantum dot-circular Bragg grating (QD-CBGs) devices \#1--\#3.
    $\Delta \mathbb{Q}_i/P$ -- 1D fit uncertainty for the QD peak center determination,
    $\Delta \mathbb{M}_i/P$ -- 1D fit uncertainty for the alignment mark (AM) center determination,
    $\Delta Q_i$ -- 1D uncertainty of QD localization,
    $\Delta Q$ -- uncertainty of QD localization in 2D.}    
    \begin{tabular}{c c c c c c c c}
    \toprule\toprule
    \thead{Device} & \thead{Orientation} & \thead{$\Delta \mathbb{Q}_i/P$} & \thead{$\Delta \mathbb{M}_i/P$}  & \thead{$\left(\mathbb{Q}_{i} - \mathbb{M}_{i,l}\right) \Delta P/P^2$} & \thead{$\Delta Q_i$} & \thead{$\Delta Q$} & \thead{$\Delta R$} \\
    \midrule
    \mr{QD-CBG \#1} & Vertical   & $\SI{61.0}{\nano\meter}$ & $\SI{14.4}{\nano\meter}$ & $\SI{19.9}{\nano\meter}$ & $\SI{65.8}{\nano\meter}$ & \mr{$\SI{141.9}{\nano\meter}$} & \mr{$\SI{147.4}{\nano\meter}$} \\
                    & Horizontal & $\SI{120.0}{\nano\meter}$& $\SI{29.0}{\nano\meter}$ & $\SI{23.5}{\nano\meter}$ & $\SI{125.7}{\nano\meter}$&                                &                              \\\addlinespace
    \mr{QD-CBG \#2} & Vertical   & $\SI{61.9}{\nano\meter}$ & $\SI{12.7}{\nano\meter}$ & $\SI{20.1}{\nano\meter}$ & $\SI{66.3}{\nano\meter}$ & \mr{$\SI{137.3}{\nano\meter}$} & \mr{$\SI{143.0}{\nano\meter}$} \\
                    & Horizontal & $\SI{115.7}{\nano\meter}$& $\SI{26.1}{\nano\meter}$ & $\SI{20.2}{\nano\meter}$ & $\SI{120.2}{\nano\meter}$&                                &                              \\\addlinespace
    \mr{QD-CBG \#3} & Vertical   & $\SI{112.2}{\nano\meter}$& $\SI{20.7}{\nano\meter}$ & $\SI{12.0}{\nano\meter}$ & $\SI{114.7}{\nano\meter}$& \mr{$\SI{132.9}{\nano\meter}$} & \mr{$\SI{138.8}{\nano\meter}$} \\
                    & Horizontal & $\SI{59.5}{\nano\meter}$ & $\SI{19.5}{\nano\meter}$ & $\SI{24.6}{\nano\meter}$ & $\SI{67.2}{\nano\meter}$ &                                &                              \\\addlinespace
    \bottomrule     
    \end{tabular}
    \end{table}

\subsection{Calculation of the diffraction-limited spot size}
In this and the following subsections, we consider a QD as a point light source and calculate the expected observed width of such a source in our imaging setup shown in Fig.~2a of the main text.
Its emission can be described by the point spread function (PSF) of the setup, which for the approximation of 2D paraxial wide-field fluorescence microscope forms an Airy disc (first-order Bessel function of the first kind)~\cite{Zhang2007}.

Our QD localization algorithm uses the approximation of the Bessel function by the Gaussian profile, which we use for fitting the QD- and AM-related signal in the cross-sections of $\upmu$PL maps.

First, we calculate the standard deviation $\sigma_{\mathrm{diff}}$ of the Gaussian curve that best approximates the PSF of our setup according to the formula~\cite{Zhang2007}:
\begin{equation}
    \sigma_{\mathrm{diff}} \approx 0.21 \frac{\lambda_{\mathrm{QD}}}{\mathrm{NA}} = \SI{0.501}{\micro\meter}
\end{equation}
with employed NA = 0.65 and $\lambda_{\mathrm{QD}}=\SI{1.55}{\micro\meter}$.
Then, we translate $\sigma_{\mathrm{diff}}$ to full width at half-maximum (FWHM) as 
\begin{equation}\label{eq:FHWM-diff}
    \mathrm{FWHM}_{\mathrm{diff}} = 2\sqrt{2\ln2}\times\sigma_{\mathrm{diff}} \approx\SI{1.18}{\micro\meter}.
\end{equation}

\subsection{Statistics on QD spot size and the accuracy of QD positioning}
	\begin{figure}[htb] %
		\begin{center} %
    	\includegraphics[width=1\columnwidth]{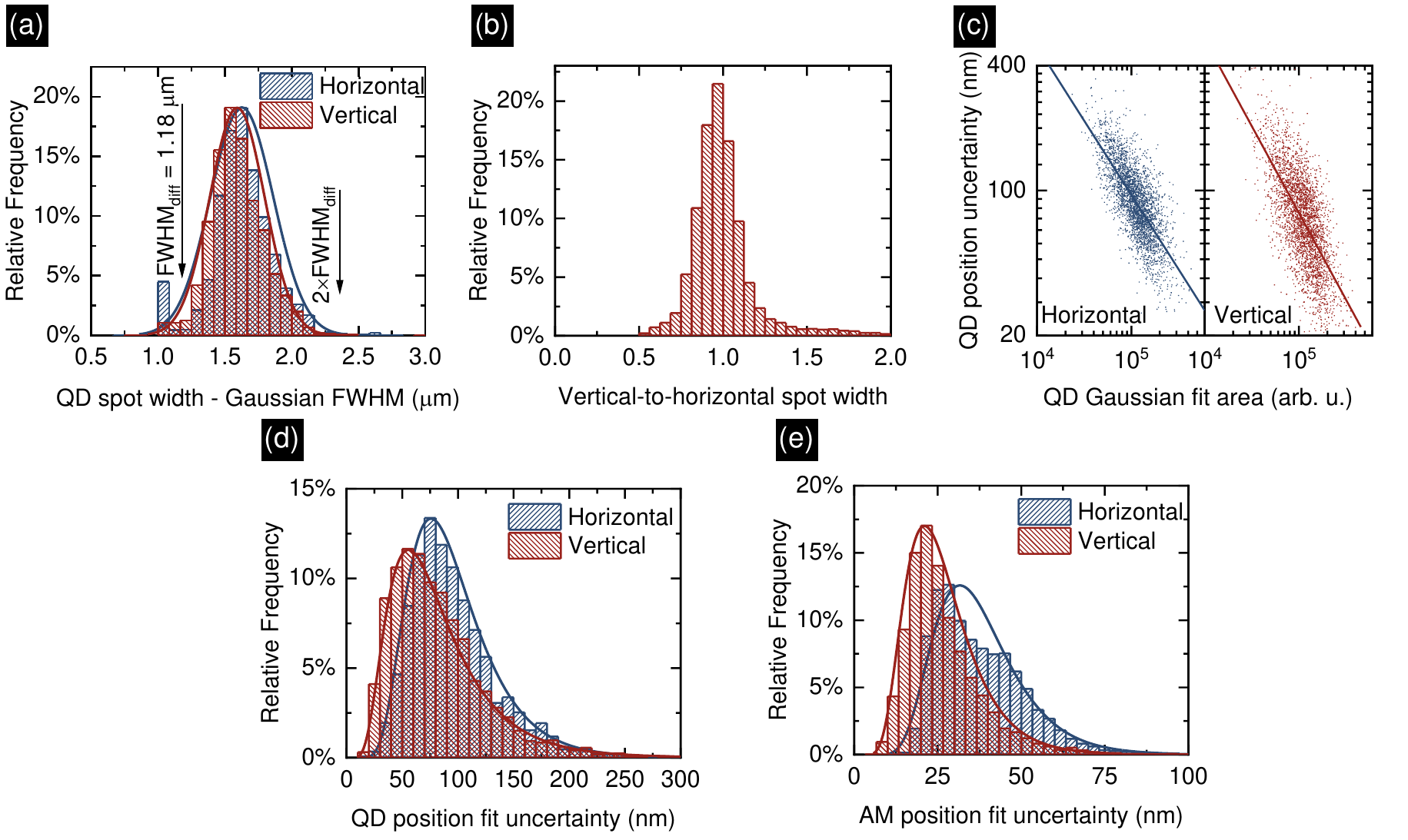} 
	    \end{center}
		\caption{\label{fig:Accuracy}
		Parameters of fitting the microphotoluminescence map cross-sections, shown separately for horizontal and vertical directions.
		\textbf{a},~the histogram of full width at half maximum (FWHM) values for the fits used for the determination of quantum dot (QD) position $Q$ compared with the calculated diffraction-limited value FWHM$_{\mathrm{diff}} = \SI{1.18}{\micro\meter}$ [\eqnref{FHWM-diff}] and fitted with the normal distribution curves,
		\textbf{b},~histogram of the ratio between the widths of the fitted horizontal and vertical cross-sections of the QD Gaussian profiles, based on the panel (a),
		\textbf{c},~QD position fit uncertainty $\Delta\mathbb{Q}/P$ as a function of the QD fit area,
        \textbf{d},~histogram of the QD position fit uncertainty $\Delta\mathbb{Q}/P$,
		\textbf{e},~alignment mark (AM) position fit uncertainty $\Delta\mathbb{M}/P$.
		Histograms in panels (d) and (e) are fitted with log-normal distributions.}
	\end{figure}

We take the calculated FWHM$_{\mathrm{diff}}=\SI{1.18}{\micro\meter}$ (see previous section) as the diffraction-limited spot size and compare it with the histogram of all FWHM values for registered QD spots in \Supplsubfigref{Accuracy}{a}.
The medians for the registered FWHM values are similar for horizontal and vertical cross-sections, $\SI{1.62}{\micro\meter}$ and $\SI{1.58}{\micro\meter}$, respectively, which is $\SI{\sim35}{\percent}$ more than the calculated FWHM$_{\mathrm{diff}}$.
Almost all registered FWHM values are lower than $2\times$FWHM$_{\mathrm{diff}}$.
Therefore, our imaging setup operates close to the diffraction limit with residual broadening originating most probably from the cryostat window between the QDs and microscope objective~\cite{Liu2017}.
The similarity between horizontal and vertical spot widths is evidenced by the histograms of ratios between the widths of the fitted horizontal and vertical cross-sections of the QD Gaussian profiles (Vertical-to-horizontal spot width), presented in \Supplsubfigref{Accuracy}{b}.
The average ratio is determined as $\SI{99.78}{\percent}$, in other words, it differs only by $\SI{0.22}{\percent}$ from the ideal case unity ratio.
The standard sample deviation for the ratio values distribution is calculated to be $\sigma=\SI{19.24}{\percent}$.

As expected, the fitting uncertainty $\Delta\mathbb{Q}_i/P$ is correlated with the QD brightness, here estimated by the Gaussian fit area, and the corresponding plot is shown in \Supplsubfigref{Accuracy}{c}.
We find a strong negative correlation between considered fit parameters (Pearson correlation coefficient of $-0.68$ and $-0.73$ for horizontal and vertical cross-sections, respectively).

\SupplsubfigrefL{Accuracy}{d} presents the fitting uncertainty for QD positions $\Delta\mathbb{Q}_i/P$, separately for horizontal and vertical cross-sections.
The uncertainty is defined as the standard error of the fitting procedure for estimation of the center of the Gaussian profile and in all analyzed cases, the obtained values form positively skewed distributions that follow the log-normal distribution, as shown with fit curves.
The medians for $\Delta\mathbb{Q}_i/P$ are $\SI{88.6}{\nano\meter}$ and $\SI{73.6}{\nano\meter}$ for horizontal and vertical cross-sections, with $\SI{24}{\percent}$ of vertical uncertainties being below $\SI{50}{\nano\meter}$.

Finally, we analyze the fitting uncertainty for AMs positions $\Delta \mathbb{M}_i/P$, as it influences the uncertainty of QD position $\Delta Q_i$ determination via \eqnref{DQi}.
The medians for $\Delta \mathbb{M}_i/P$ distributions, shown in \Supplsubfigref{Accuracy}{e}, are $\SI{24.1}{\nano\meter}$ and $\SI{35.1}{\nano\meter}$ for horizontal and vertical cross-sections.
Importantly, $\SI{83}{\percent}$ of horizontal and $\SI{95}{\percent}$ of vertical cross-section uncertainties $\Delta \mathbb{M}_i/P$ are below $\SI{50}{\nano\meter}$.

Based on these considerations and the correlation between QD spot brightness and uncertainty of QD position fit $\Delta\mathbb{Q}_i/P$, we estimate that working with only the brightest QD spot in each imaging field (due to the average number of $N_{\mathrm{F}}\approx10$ spots detected per field, this amounts to limiting cavity fabrication to $\SI{10}{\percent}$ of the registered spots), our fitting accuracy would be $\Delta\mathbb{Q}_i/P<\SI{53.2}{\nano\meter}$ and $\Delta\mathbb{Q}_i/P<\SI{37}{\nano\meter}$ for horizontal and vertical cross-sections (10th percentiles for the uncertainties distributions).
Taking medians for the fitting uncertainty of AMs positions $\Delta \mathbb{M}_i/P$, we find the uncertainty of QD localization $\Delta Q<\SI{80.1}{\nano\meter}$, cf. \eqnref{DQ}.
Including the electron beam lithography uncertainty $\Delta C = \SI{40}{\nano\meter}$, we find the total uncertainty of cavity placement $\Delta R<\SI{90.3}{\nano\meter}$, according to \eqnref{DR} (see \Suppltabref{uncertainties-brightest}).
Here, we assume $\left(Q_{i} - M_{i,l}\right)=\SI{25}{\micro\meter}$ (center of the field) and for $\Delta P$ we take the median value of standard error of mean $\hat{\sigma}_{\bar{P_i}}$ calculated for 84 exemplary $\upmu$PL maps.

    \begin{table}[htb]
    \caption{\label{tab:uncertainties-brightest} Uncertainties involved in the determination of the accuracy of cavity positioning $\Delta R$ for an example of a quantum dot (QD) with low uncertainty of fitting the QD positions $\Delta\mathbb{Q}_i/P$ -- corresponding to a QD being in the $\SI{10}{\percent}$ of the brightest spots.
    $\Delta \mathbb{Q}_i/P$ -- 1D fit uncertainty for the QD peak center determination,
    $\Delta \mathbb{M}_i/P$ -- 1D fit uncertainty for the alignment mark (AM) center determination,
    $\Delta Q_i$ -- 1D uncertainty of QD localization,
    $\Delta Q$ -- uncertainty of QD localization in 2D.}
    \begin{tabular}{c c c c c c c}
    \toprule\toprule
    \thead{Orientation} & \thead{$\Delta \mathbb{Q}_i/P$} & \thead{$\Delta \mathbb{M}_i/P$} & \thead{$\left(\mathbb{Q}_{i} - \mathbb{M}_{i,l}\right) \Delta P/P^2$} & \thead{$\Delta Q_i$} & \thead{$\Delta Q$} & \thead{$\Delta R$} \\
    \midrule
    Horizontal & $\leq\SI{53.2}{\nano\meter}$ & $\SI{24.1}{\nano\meter}$ & \mr{$\SI{15.1}{\nano\meter}$} & $<\SI{61.1}{\nano\meter}$  & \mr{$<\SI{80.1}{\nano\meter}$} & \mr{$<\SI{90.3}{\nano\meter}$} \\
    Vertical & $\leq\SI{37}{\nano\meter}$ & $\SI{35.1}{\nano\meter}$ &                               & $<\SI{53.2}{\nano\meter}$  &                                &                                \\\addlinespace
    \bottomrule
    \end{tabular}
    \end{table}

\section{Optical setups}\label{sec:OptExp}

\begin{figure}[htb] %
		\begin{center} %
    	\includegraphics[width=0.8\columnwidth]{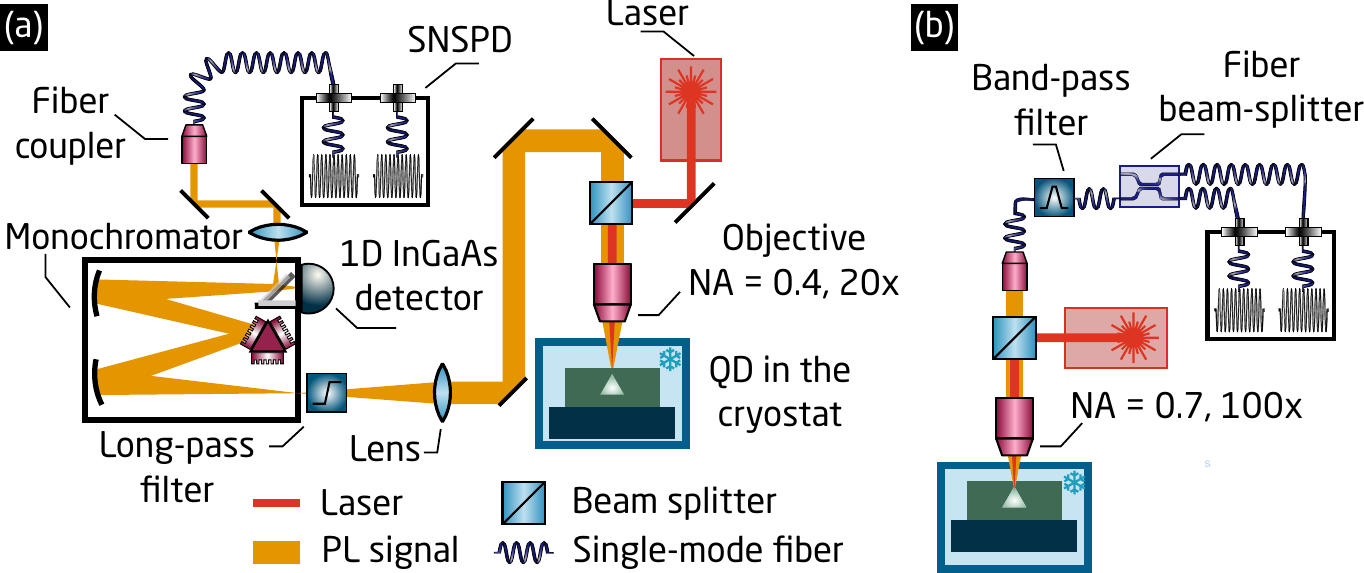} 
	    \end{center}
		\caption{\label{fig:Setup}
		Simplified schemes of the experimental setups used for the device characterization.
		\textbf{a},~the optical setup used for the determination of the photon extraction efficiency and time-resolved microphotoluminescence,
		\textbf{b},~setup used for autocorrelation experiments.
        SNSPD -- superconducting nanowire single-photon detector.}
	\end{figure}

\Supplsubfigref{Setup}{a} presents the setup used for spectroscopy studies of fabricated devices ($\upmu$PL, extraction efficiency determination, time-resolved $\upmu$PL), while the above-band autocorrelation histograms are recorded in a setup shown in \Supplsubfigref{Setup}{b}.
The \methods{} section gives details of the equipment used.

\section{Determination of the photon extraction efficiency}\label{Determination-of-PEE}
To determine the value of photon extraction efficiency $\eta$, we calibrate the optical setup shown in \Supplsubfigref{Setup}{a} and calculate its efficiency $\eta_{\mathrm{Setup}}$ as a multiplication of the transmission of all optical elements and of the efficiency of fiber in-coupling and superconducting nanowire single-photon detector (SNSPD) quantum efficiency.
We start with reflecting the laser tuned to $\SI{1.55}{\micro\meter}$ off a silver mirror placed in the setup instead of the structure.
The signal emitted from the sample passes through the elements enumerated in \Suppltabref{Setup-transmission}, given with their measured transmission efficiencies.
For SNSPD, we take nominal efficiency.
We obtain $\eta_{\mathrm{Setup}}=\SI{1.10(17)}{\percent}$, where the uncertainty is calculated by propagating the assumed uncertainties of transmission of the consecutive elements.

    \begin{table}[htb]
    \caption{\label{tab:Setup-transmission} The transmission of the optical components in the setup used for determination of photon extraction efficiency, as shown in \Supplsubfigref{Setup}{a}.
    SNSPD -- superconducting nanowire single-photon detector.}
    \begin{tabular}{c c}
    \toprule\toprule
    \thead{Element} & \thead{Transmission/Efficiency} \\
    \midrule
    cryostat window         & $\SI{90(2)}{\percent}$\\
    microscope objective    & $\SI{55(3)}{\percent}$\\
    beam splitter           & $\SI{38(2)}{\percent}$\\
    a set of mirrors        & $\SI{85(5)}{\percent}$\\
    focusing lens and long-pass filter & $\SI{93(2)}{\percent}$\\
    monochromator           & $\SI{27(5)}{\percent}$\\
    mirrors for signal coupling & $\SI{96(2)}{\percent}$\\
    fiber in-coupling       & $\SI{41(10)}{\percent}$\\
    fibers and their connections    & $\SI{80(10)}{\percent}$\\
    SNSPD efficiency         & $\SI{87(3)}{\percent}$\\
    \midrule
    Total setup efficiency $\eta_{\mathrm{Setup}}$  & $\SI{1.10(17)}{\percent}$\\
    \bottomrule
    \end{tabular}
    \end{table}

Then, we excite the QDs off-resonantly with a pulsed laser diode with $f_{\mathrm{rep}}=\SI{80}{\mega\hertz}$ repetition rate at the saturation power for each QD.
We collect the emission with the microscope objective ($\mathrm{NA}=0.4$) and take the SNSPD count rate $n_{\mathrm{QD}}$ for the most intense QD lines, and correct them by $\eta_{\mathrm{Setup}}$ and $f_{\mathrm{rep}}$ according to the formula 
\begin{equation}
    \eta=\frac{n_{\mathrm{QD}}}{f_{\mathrm{rep}}\times\eta_{\mathrm{Setup}}}.
\end{equation}
The uncertainty of the photon extraction efficiency for QD-CBG \#1 and \#2 are calculated by propagating the uncertainties as 
\begin{equation}
    \Delta \eta = \sqrt{\left(\frac{\Delta n_{\mathrm{QD}}}{\eta_{\mathrm{Setup}}}\right)^2 + 
    \left(\frac{n_{\mathrm{QD}}\Delta\eta_{\mathrm{Setup}}}{\eta_{\mathrm{Setup}}^2}\right)^2},
\end{equation}
with $\Delta n_{\mathrm{QD}}=1000$ and $\Delta\eta_{\mathrm{Setup}}=\SI{0.17}{\percent}$.
This method assumes unity internal quantum efficiency of QDs ($\eta_{\mathrm{int}}=\SI{100}{\percent}$), so that the QD photon emission rate equals $f_{\mathrm{rep}}$.
It is however difficult to determine experimentally the contribution of non-radiative recombination and hence the real value of $\eta_{\mathrm{int}}$.
As a result, a discrepancy arises between the calculated photon extraction efficiency (Fig. 1b) and measured $\eta$.
The reason for lowered $\eta_{\mathrm{int}}$ (and hence lowered $\eta$) can be attributed to the non-radiative recombination channels introduced to the QDs due to structural defects propagating from the InP substrate or defect states at the side walls of the CBG central mesa, which are introduced during dry etching.
These defects most likely cause additional exciton energy relaxation channels.
The assumption of $\eta_{\mathrm{int}}=\SI{100}{\percent}$ thus sets a lower limit of $\eta$ due to a possible overestimation of the total number of photons emitted by the QD.
	
\section{Supporting microphotoluminescence data for QD-CBG devices}

\subsection{Quantum dot linewidths}	%
\begin{figure}[htb] %
		\begin{center} %
    	\includegraphics[width=0.4\columnwidth]{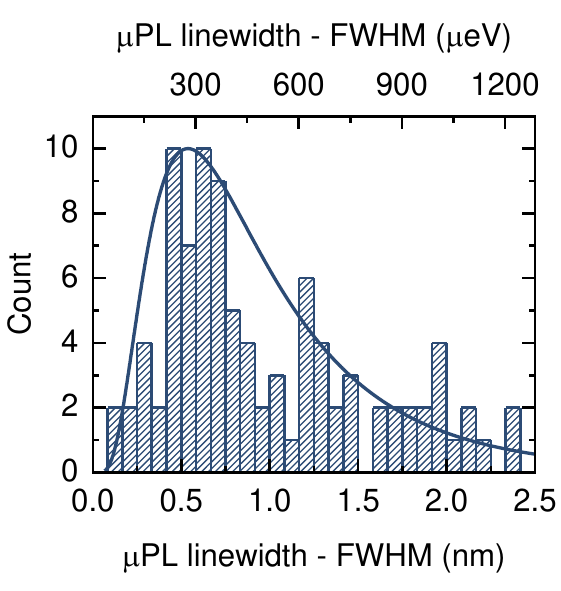} 
	    \end{center}
		\caption{\label{fig:Linewidths}
		The histogram of linewidths for quantum dots in circular Bragg gratings (full width at half maximum, FWHM, of fitted Gaussian profiles).}
	\end{figure}

\SupplfigrefL{Linewidths} presents the histogram of linewidths for all $102$ found QDs emitting in the CBGs.
We use the FWHM of the fitted Gaussian profiles to describe the linewidth and find that the median linewidth is $\SI{0.76}{\nano\meter}$, $1^\mathrm{st}$ quartile $\SI{0.52}{\nano\meter}$, and the minimal value $\SI{0.14}{\nano\meter}$.
Additionally, the linewidths for QD-CBGs \#1--\#3 are shown in \Suppltabref{Linewidths}.

    \begin{table}[htb]
    \caption{\label{tab:Linewidths}  Microphotoluminescence ($\upmu$PL) linewidths of devices \#1--\#3.}
    \begin{tabular}{c c}
    \toprule\toprule
    \thead{Device} & \thead{QD linewidth in $\upmu$PL} \\
    \midrule
    QD-CBG \#1 & $\SI{0.47}{\nano\meter}$\\
    QD-CBG \#2 & $\SI{0.17}{\nano\meter}$\\
    QD-CBG \#3 & $\SI{0.19}{\nano\meter}$\\
    \bottomrule     
    \end{tabular}
    \end{table}

\subsection{Time-resolved microphotoluminescence}
	\begin{figure}[htb] %
		\begin{center} %
    	\includegraphics[width=1\columnwidth]{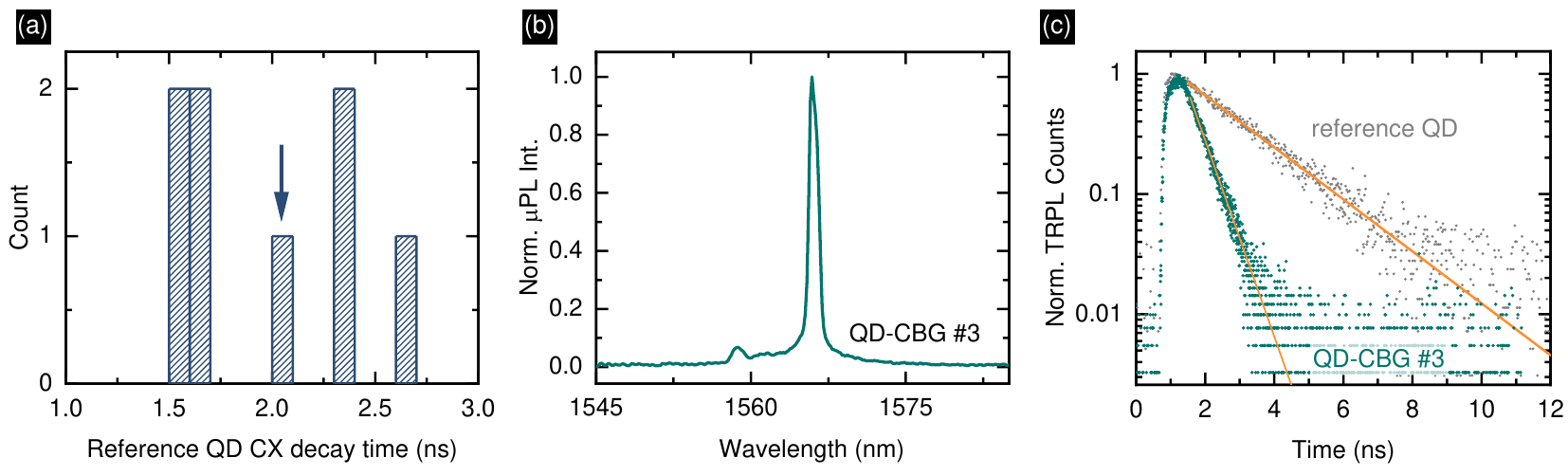} 
	    \end{center}
		\caption{\label{fig:QD-B-and-C-TRPL}
		Additional microphotoluminescence ($\upmu$PL) data.
		\textbf{a},~Histogram of recorded decay times for reference QD trion (CX) lines; arrow points to the quantum dot (QD) shown in Fig.~3b in the article and in panel (c),
		\textbf{b},~$\upmu$PL spectrum for QD-CBG \#3,
		\textbf{c},~time-resolved $\upmu$PL time traces for QD-CBG \#3 and the reference QD.}
	\end{figure}

\SupplfigrefL{QD-B-and-C-TRPL} shows the $\upmu$PL data supporting the determination of \fp{}.
We focus on an additional QD-CBG \#3 with a well-isolated transition line, analogous to QD-CBGs \#1 and \#2 presented in the article, Fig.~3.

\Supplsubfigref{QD-B-and-C-TRPL}{a} shows a histogram of recorded decay times for the reference QD trion lines.
The time-resolved $\upmu$PL results for eight QDs found in the area outside the fabricated cavities with intense trion lines indicate a considerable scattering of the decay times, with minimal and maximal registered times $\SI{1.59}{\nano\second}$ and $\SI{2.69}{\nano\second}$ and the average decay time $\tau_{\mathrm{ref}} = \SI{1.99}{\nano\second}$.
The standard deviation of the fitted normal distribution is $\SI{0.44}{\nano\second}$ and the standard error of mean is $\SI{0.16}{\nano\second}$.
We take the average decay time $\tau_{\mathrm{ref}} = \SI{1.99(16)}{\nano\second}$ to calculate the \fp{} according to formula $F_{\mathrm{P}}=\tau_{\mathrm{ref}}/\tau_{\mathrm{cav}}$.

\Supplsubfigref{QD-B-and-C-TRPL}{b} shows the spectrum of QD-CBG \#3, and \Supplsubfigref{QD-B-and-C-TRPL}{c} presents time-resolved $\upmu$PL decay traces in analogy to Fig.~3b of the article, where the same reference time trace is presented.
Decay time for QD-CBG \#3 is the same as for QD-CBG \#2 within the fitting accuracy, $\tau_{\mathrm{\#3}} = \tau_{\mathrm{\#2}} =\SI{0.53(1)}{\nano\second}$, what translates to $F_{\mathrm{P}} = \SI{3.75(30)}{}$.

\subsection{Temperature-dependent $\upmu$PL}	%
	\begin{figure}[htb] %
		\begin{center} %
    	\includegraphics[width=1\columnwidth]{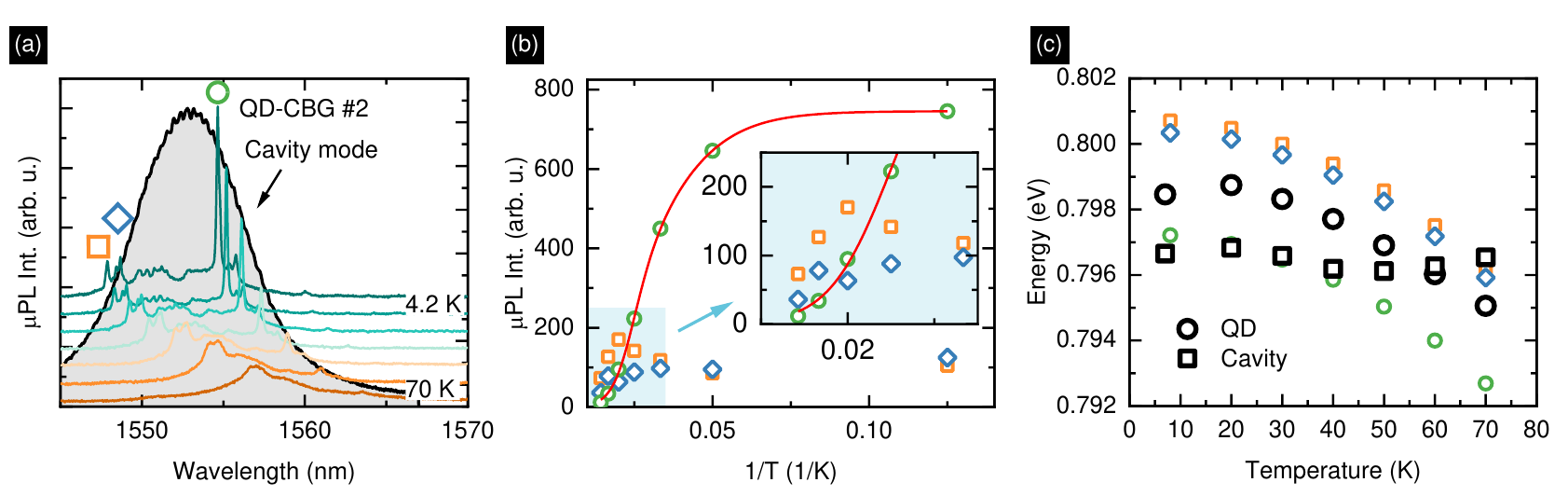} 
	    \end{center}
		\caption{\label{fig:QD-B-and-C-Temp}
		Analysis of the temperature-dependent microphotoluminescence spectra for quantum dot-circular Bragg grating (QD-CBG) device \#2.
		\textbf{a},~Temperature-dependent $\upmu$PL spectra for QD-CBG \#2, overlapped with the cavity mode. 
		\textbf{b},~Arrhenius plot for the integrated $\upmu$PL intensity for three QD lines, indicated with the same symbols in panel~(a).
		Inset: close-up of the marked region.
		\textbf{c},~$\upmu$PL emission energy for the same QD lines.
        Black symbols stand for another QD-CBG device, with QD and cavity mode energies tracked up to $T = \SI{70}{\kelvin}$.
        The mode crossing characteristic for the weak QD-cavity coupling regime is visible.}
	\end{figure}

We record the temperature-dependent $\upmu$PL signal for QD-CBG \#2 in the temperature range of \lowT{} to $\SI{70}{\kelvin}$ under the continuous-wave (CW) excitation and present the stacked spectra in \Supplsubfigref{QD-B-and-C-Temp}{a}.
Three emission lines, marked with rings, can be identified and their emission intensity and energy tracked as the temperature is increased.
We plot also the mode profile for reference.
Note the $\sim\SI{2}{\nano\meter}$ redshift of the most intense line from the central wavelength of the cavity.

The $\upmu$PL intensities change differently for lines with emission energy lower vs. higher compared to the mode profile.
The short-wavelength lines, marked with red and blue rings, are tuned across the mode profile as the temperature raises.
Their $\upmu$PL intensity quench is greatly suppressed, in fact, the intensity of the line marked with a red circle has a maximum at $T=\SI{50}{\kelvin}$.
The behavior for the long-wavelength line, marked with a green ring, is opposite and its intensity quenches fast.

The temperature-dependent $\upmu$PL intensity for the most intense line marked with the green ring is fitted with a standard formula assuming two activation processes~\cite{LambkinAPL1990}:	%
	\begin{equation}\label{ArrhEq}
	\centering
		I\left( T \right) = \frac{I_0}{1+B_{\mathrm{1}} \exp\left(-E_{\mathrm{a,1}}/k_{\mathrm{B}}T\right)+
		B_{\mathrm{2}} \exp\left( -E_{\mathrm{a,2}}/k_{\mathrm{B}}T\right)},
	\end{equation}
where $I_0$ is the $\upmu$PL intensity for $T\to 0$, $E_{\mathrm{a,1}}$ and $E_{\mathrm{a,2}}$ are activation energies, and $B_{\mathrm{1}}$ and $B_{\mathrm{2}}$ are relative rates corresponding to the efficiency of involved processes.
We achieve the activation energies of $E_{\mathrm{a,1}} = \SI{6.9(8)}{\milli\electronvolt}$ and $E_{\mathrm{a,2}} =\SI{27.9(31)}{\milli\electronvolt}$.
We note that $E_{\mathrm{a,2}}$ is similar to the activation energy $E_{\mathrm{Ref}} =\SI{23.6(8)}{\milli\electronvolt}$ found for a CX line from analogous QDs~\cite{Holewa2022Mirror}.
Accordingly to Ref.~\cite{Holewa2020PRB}, this activation energy can be attributed to the charge transfer to higher orbital states, based on the band structure calculations within the 8-band $k\cdot p$ framework.
Hence, another mechanism is responsible for the identification of $E_{\mathrm{a,1}}$ energy which is absent in the case of high-energy lines (blue and red circles, see \Supplsubfigref{QD-B-and-C-Temp}{b}).
This can potentially be associated with the enhancement of the QD emission rate when the overlap between the line and the mode energy is maximized.

In \SupplsubfigrefL{QD-B-and-C-Temp}{c} we plot the temperature dependence of the emission energy for lines analyzed in \Supplsubfigref{QD-B-and-C-Temp}{b} and, additionally, for another investigated QD-CBG device, where the observation of the crossing of the temperature-tuned QD line and the cavity mode under high power off-resonant CW excitation was possible due to the QD energy being higher than the mode energy at \lowT{}.
The temperature-induced evolution of cavity mode energy (black open squares) can be compared with the QD lines, evidencing a weaker temperature dependence, as expected for the cavity mode.

\section{Quantum optics experiments}
\subsection{Off-resonant autocorrelation data for QD-CBG \#2}

\begin{figure}[htb] %
		\begin{center} %
        \includegraphics[width=1\columnwidth]{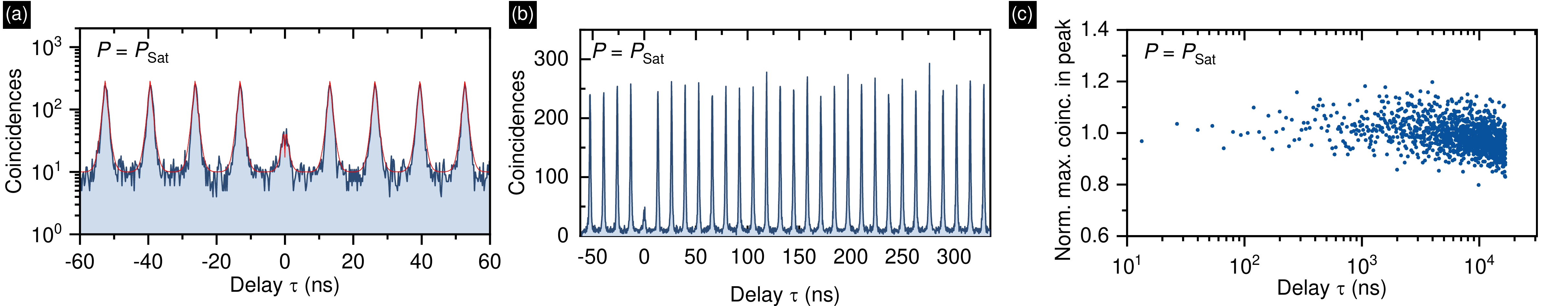} 
	    \end{center}
		\caption{\label{fig:supp-g2-non-res}
        Analysis of the second-order autocorrelation function $\gtwo{\tau}$ of the photons emitted from quantum dot-circular Bragg grating (QD-CBG) \#2 under pulsed off-resonant excitation for laser excitation power $P$ corresponding to saturation of the trion (CX) line.
		\textbf{a},~\textbf{b},~The recorded histogram 
        \textbf{c},~Normalized maximum number of coincidences for all 1260 registered peaks for positive delays of histogram shown in (b).}
	\end{figure}

\SupplfigrefL{supp-g2-non-res} shows the autocorrelation histogram recorded under pulsed off-resonant excitation for QD-CBG~\#2.
We do not observe blinking there, as the normalized coincidences in consecutive peaks (maximum number of coincidences without fitting) in the $\gtwo{\tau}$ histogram is at a constant level (\Supplsubfigref{supp-g2-non-res}{c}).
We show data for all registered 1260 peaks, i.\,e. up to $\tau=\SI{16.7}{\micro\second}$ delay.
Fluctuation in the data originates in the discretization of the $\gtwo{\tau}$ due to finite time binning in the experiment.
The blinking would indicate the occupation of a meta-stable QD state~\cite{Santori2001}, originating, e.\,g., in the emission wavelength fluctuations due to nearby defects~\cite{Pistol1999} or trapped charges~\cite{Robinson2000}.

We fit the histograms with the function~\cite{Dalgarno2008,Miyazawa2016}:
    \begin{widetext}
    \begin{equation}\label{eq:g2-pulsed-Eq}
    C(\tau)=
    B+A\left[\exp\left(-|\tau|/\tau_{\mathrm{dec}}\right)-\exp\left(-|\tau|/\tau_{\mathrm{cap}}\right)\right]\\
    +H \sum_{n\neq 0}\exp\left(-|\tau-n\tau_0|/\tau_{\mathrm{dec}}\right),
    \end{equation}
    \end{widetext}
where $B$ is the level of background coincidences, $A$ is a scaling parameter related to secondary photon emission, $n \neq 0$ is the peak number, $\tau_0$ is the laser pulse period, and $H$ the average height of the peaks at $\tau_n = n \tau_0$.
The second-order correlation function $\gtwo{\tau}$ is then obtained by normalizing $C(\tau)$ with $(H+B)$.

The time-independent level of background coincidences $B$ originates from the detector dark counts and uncorrelated photons contributing to the registered histograms.
Our approach allows taking into account only coincidences caused by the QD signal.
We define the purity as the ratio between the QD emission coincidences registered at $\tau_0$ peak (area of this peak) to the average number of coincidences (peak area) registered at $\tau_{n\neq0}$ peaks.
Then, the $\gtwo{0}$ value is calculated by first subtracting the background contribution $B$, integrating the areas under the central peak and under the non-zero peaks, and dividing these two integrals, according to the formula:
\begin{equation}\label{eq:def-g2-fit-off-resonant}
\begin{split}
    g^{(2)}(0)_{\mathrm{fit}} = \frac{\int_{-\tau_0/2}^{\tau_0/2} A\left[\exp\left(-|\tau|/\tau_{\mathrm{dec}}\right)-\exp\left(-|\tau|/\tau_{\mathrm{cap}}\right)\right]\mathrm{d}\tau}
    {\int_{-\tau_0/2}^{\tau_0/2} H\exp\left(-|\tau|/\tau_{\mathrm{dec}}\right)\mathrm{d}\tau}.
\end{split}
\end{equation}

\SuppltabrefL{g2_off-resonant-fit-parameters} summarizes the fit parameters for off-resonant autocorrelation histograms obtained for $0.5\times P_{\mathrm{sat}}$ and $P_{\mathrm{sat}}$ excitation power.
The uncertainties are determined from the fit uncertainties, and the $g^{(2)}(0)_{\mathrm{fit}}$ is calculated from the fit parameters.

    \begin{table}[htb]
    \caption{\label{tab:g2_off-resonant-fit-parameters} Fitting parameters for off-resonant autocorrelation measurement.}
    \begin{tabular}{c c c}\toprule
    & \multicolumn{2}{c}{Off-resonant excitation power}  \\\cmidrule{2-3}
    Fit parameter & $0.5\times P_{\mathrm{sat}}$ & $P_{\mathrm{sat}}$ \\\midrule
    Background contribution, $B$ & $\SI{14.4(20)}{}$ & $\SI{10.2(07)}{}$ \\
    Scaling of the center peak, $A$ & $\SI{54(15)}{}$ & $\SI{67(9)}{}$\\
    Height of $\tau_{n\neq0}$ peaks, $H$ & $\SI{187(4)}{}$ & $\SI{286(2)}{}$ \\
    Laser period, $\tau_0$ &  $\SI{13.14(1)}{\nano\second}$ & $\SI{13.15(1)}{\nano\second}$ \\
    Decay time, $\tau_{\mathrm{dec}}$ & $\SI{668(29)}{\pico\second}$ & $\SI{714(25)}{\pico\second}$ \\
    Recapture time, $\tau_{\mathrm{cap}}$ & $\SI{514(73)}{\pico\second}$ & $\SI{192(42)}{\pico\second}$ \\    
    $g^{(2)}(0)_{\mathrm{fit}}$, \eqnref{def-g2-fit-off-resonant}   & $\SI{0.05(2)}{}$ & $\SI{0.17(3)}{}$\\
    \bottomrule
    \end{tabular}
    \end{table}

\subsection{Microphotoluminescence excitation spectroscopy of QD-CBG \#2}
\begin{figure}[htb] %
		\begin{center} %
    	\includegraphics[width=0.9\columnwidth]{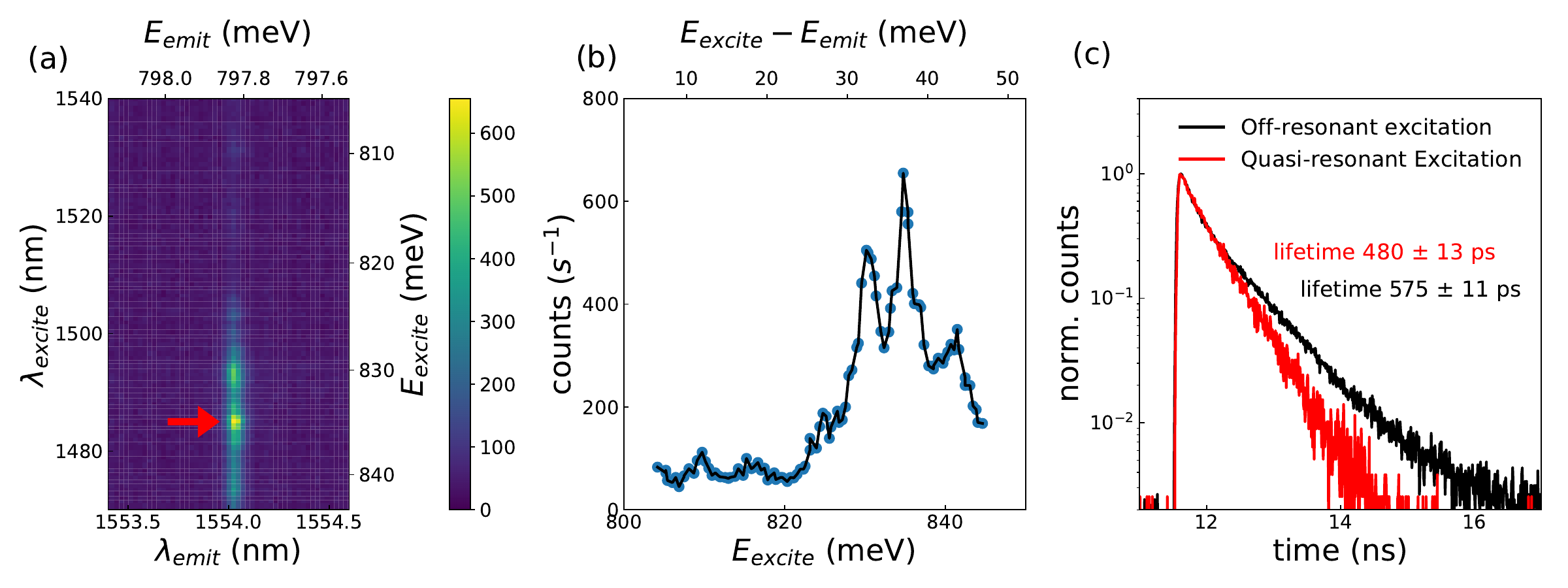} 
	    \end{center}
		\caption{\label{fig:PLE_map}
		Quasi-resonant excitation of quantum dot-circular Bragg grating (QD-CBG) \#2.
        \textbf{a},~Microphotoluminescence excitation map of (QD-CBG) \#2.
        \textbf{b},~Cross-section of the map taken at the center of the QD-CBG \#2 emission line, revealing the quasi-resonant excitation energy of $\sim\SI{0.835}{\electronvolt}$ used to excite the QD for the indistinguishability measurements.
        The energy difference corresponds to the detuning of $\sim\SI{37}{\milli\electronvolt}$ from the emission energy.
        \textbf{c},~Time-resolved $\upmu$PL data for line in QD-CBG \#2 for quasi- and off-resonant excitation.}
	\end{figure}

The two-photon interference experiments were carried out under quasi-resonant excitation conditions.
The exact excitation energy was determined based on the microphotoluminescence excitation ($\upmu$PLE) experiment conducted with a pulsed tunable laser with $\SI{5}{\pico\second}$-long pulses and $\SI{80}{\mega\hertz}$ repetition rate.
The excitation laser wavelength was varied in the range of $\SIrange{1470}{1540}{\nano\meter}$ at constant average excitation power of $\SI{25}{\micro\watt}$ measured in front of the cryostat window.
In the $\upmu$PLE map, shown in \Supplsubfigref{PLE_map}{a}, a clear maximum is visible at the wavelength $\SI{1484.2}{\nano\meter}$, corresponding to $\SI{835.37}{\milli\electronvolt}$ photon energy which was used for all experiments described in this section.
The energy difference of $\SI{37.57}{\milli\electronvolt}$ (see the map cross-section in \Supplsubfigref{PLE_map}{b}) coincides reasonably well with the LO phonon energy of InP of $\SI{43.4}{\milli\electronvolt}$ at low temperature~\cite{Irmer1996}.
Simultaneously, the measured energy difference is far above the calculated trion \textit{p}-shell splitting of about $\sim\SI{20}{\milli\electronvolt}$~\cite{Holewa2020PRB} for QDs very similar in size and chemical composition (P admixture) to the ones investigated here.
Therefore, we assume that the applied quasi-resonant excitation of the QD is LO-phonon-assisted.

The comparison of the time-resolved $\upmu$PL time traces for the trion line in QD-CBG \#2 for quasi- and off-resonant excitation, shown in \Supplsubfigref{PLE_map}{c}, confirms the accelerated relaxation of the excited state by the reduced decay time under for the quasi-resonant excitation.

\subsection{Quasi-resonant autocorrelation data for QD-CBG \#2}
The single-photon purity was verified in autocorrelation measurements taken at various excitation powers to identify the contribution of multi-photon emission and unsuppressed reflected laser light which lead to erroneous coincidences in HOM measurements.
The lowest $\gtwo{0}$ value was found for the excitation power of $\SI{1.5}{\micro\watt}$, and the corresponding autocorrelation data is shown in Fig.~4a of the main text.
In this configuration, the autocorrelation histogram was measured directly through the HOM setup to make sure that no reflected laser obscures the HOM experiment.
Increasing the excitation power is favored by increasing the signal-to-background ratio as long as the QD is in the linear response regime.
On the other hand, increasing the excitation power once the QD emission intensity is saturated results mostly in the increase of the background counts due to the cavity being fed by other sources.
They mostly originate in the low-energy tail of the wetting layer emission or radiative defects present in the sample.

\begin{figure}[htb] %
    \begin{center} %
    \includegraphics[width=1\columnwidth]{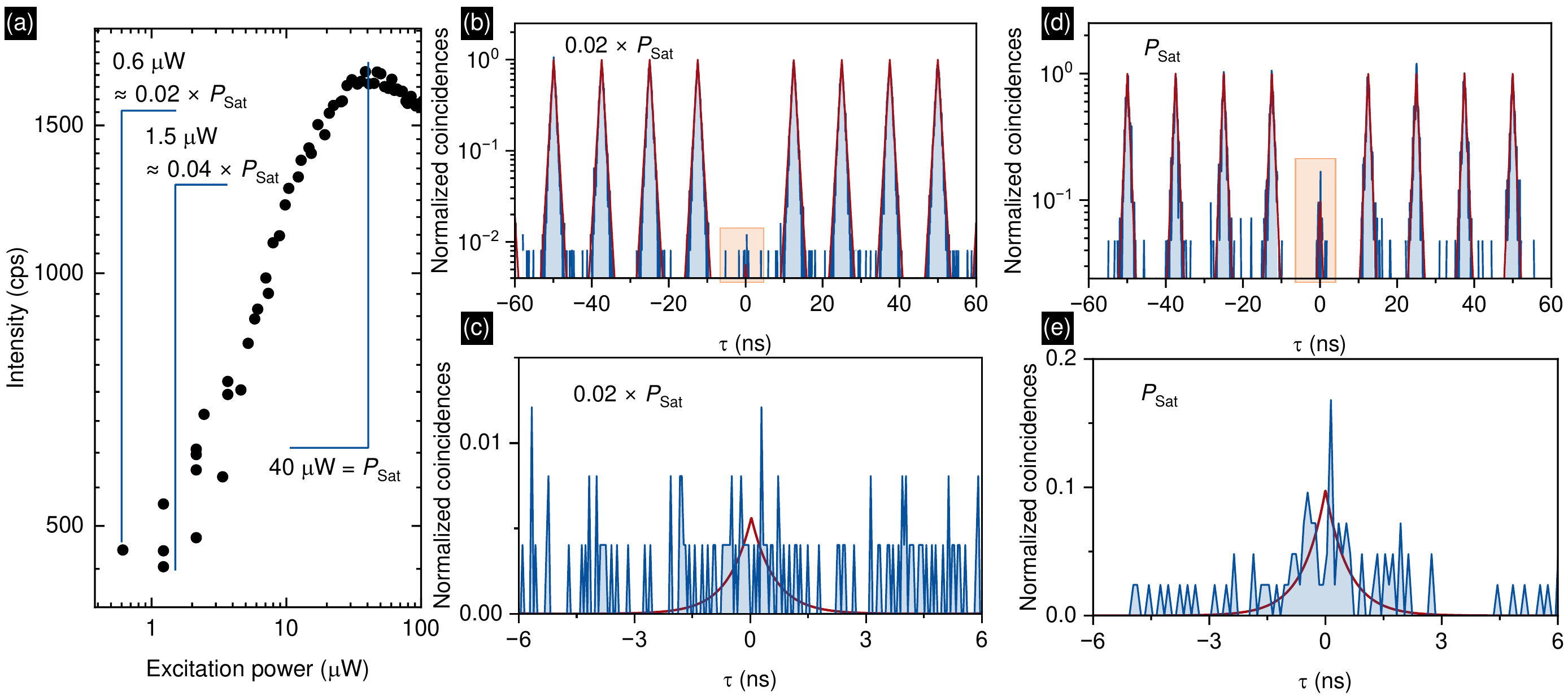} 
    \end{center}
    \caption{\label{fig:hom_power}
    Quasi-resonant autocorrelation data for quantum dot-circular Bragg grating (QD-CBG) \#2.
    \textbf{a},~Excitation-power dependent microphotoluminescence ($\upmu$PL) intensity of the line in QD-CBG \#2 under quasi-resonant excitation with the powers chosen for Hong-Ou-Mandel (HOM) measurements marked.
    \textbf{b}--\textbf{e}~Autocorrelation histograms for QD-CBG \#2 under (b),~(c)~$P_{\mathrm{exc}}=\SI{0.6}{\micro\watt}$ ($0.02\times P_{\mathrm{sat}}$), (d),~(e)~$P_{\mathrm{exc}}=\SI{40}{\micro\watt}$ ($P_{\mathrm{sat}}$).
    Orange boxes mark the region that is zoomed in at the bottom.}
\end{figure}

The low power autocorrelation measurement taken using the HOM configuration leads to an integrated $g^{(2)}(0)_{\mathrm{raw}} = \SI{3.2(6)E-3}{}$ which confirms the suppression of the excitation laser in preparation for the HOM experiment.

To fit the autocorrelation histograms we modify \eqnref{g2-pulsed-Eq} used for fitting off-resonant data, as we do not observe background counts ($B=0$) and no carrier recapture so that the new formula reads 
    \begin{equation}
    C(\tau)=c \cdot \left(g^{(2)}(0)_{\mathrm{fit}} \exp(-|\tau|/\tau_{\mathrm{dec}})+ \sum_{n \neq 0} \exp(-\left| \tau-n \tau_0 \right| / \tau_{\mathrm{dec}}) \right),
    \end{equation}
where $c$ is a global normalization factor.

The extracted fit parameters for the quasi-resonantly excited autocorrelation data shown in the main text and in \Supplsubfigsref{hom_power}{b}{c} are shown in \Suppltabref{g2_fit_parameters}.
Note that in real applications temporal filtering is not always possible which is why we state also the integrated value.
For that, we integrate all coincidences in a window of $\pm \SI{6}{\nano\second}$ around each peak in the histogram. 
Then, we divide the sum of coincidences in the center window by the average of the sums in all side windows, as no blinking is present.
The uncertainty is based on the variance of integrated side peak areas.

    \begin{table}[htb]
    \caption{\label{tab:g2_fit_parameters} Fitting parameters for quasi-resonant autocorrelation measurement.}
    \begin{tabular}{c c c c}\toprule
    & \multicolumn{3}{c}{Quasi-resonant excitation power}  \\\cmidrule{2-4}
    Fit parameter & $P_{\mathrm{exc}}=\SI{0.6}{\micro\watt}$ ($0.02\times P_{\mathrm{sat}}$), \SupplsubfigrefS{hom_power}{b} &  $P_{\mathrm{exc}}=\SI{1.5}{\micro\watt}$ ($0.04\times P_{\mathrm{sat}}$), Fig.~4a & $P_{\mathrm{exc}}=\SI{40}{\micro\watt}$ ($P_{\mathrm{sat}}$), \SupplsubfigrefS{hom_power}{c}  \\\midrule
    Decay time, $\tau_{\mathrm{dec}}$ & $\SI{606(3)}{\pico\second}$ & $\SI{584(3)}{\pico\second}$   & $\SI{591(7)}{\pico\second}$ \\
    Laser period, $\tau_0$ & $\SI{12.49(1)}{\nano\second}$ & $\SI{12.49(1)}{\nano\second}$ & $\SI{12.49(1)}{\nano\second}$ \\
    $g^{(2)}(0)_{\mathrm{fit}}$   & $\SI{5.6(50)E-3}{}$ & $\SI{4.7(26)E-3}{}$               & $\SI{9.81(194)E-2}{}$\\
     $g^{(2)}(0)_{\mathrm{raw}}$  &  $\SI{4.2(2)E-3}{}$ & $\SI{3.2(6)E-3}{}$               & $\SI{8.75(488)E-2}{}$ \\
    \bottomrule
    \end{tabular}
    \end{table}

\subsection{Indistinguishability measurements and data analysis}
\begin{figure}[htb] %
    \begin{center} %
    \includegraphics[width=0.85\columnwidth]{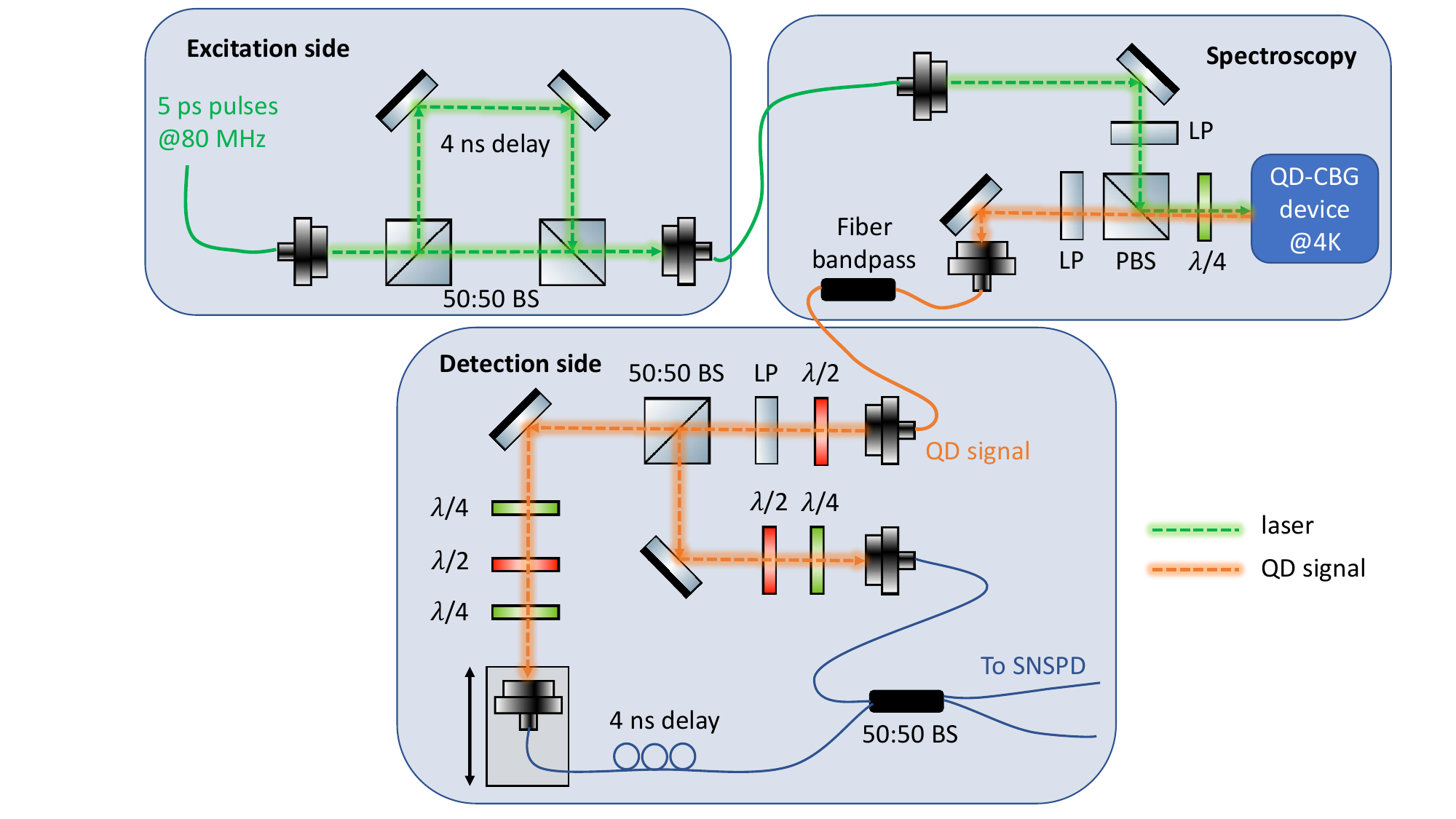} 
    \end{center}
    \caption{\label{fig:HOM_setup}
    The Hong-Ou-Mandel (HOM) experimental setup. BS -- beam splitter, $\lambda/2$, $\lambda/4$ -- half and quarter wave-plate, SNSPD -- superconducting nanowire single-photon detector.}
\end{figure}

\Supplfigref{HOM_setup} presents the configuration of the experimental setup applied to record the HOM histograms.
The $\SI{4}{\nano\second}$ excitation delay is compensated on the detection side.
A cross-polarization setup suppresses the reflected laser light and a $\SI{0.4}{\nano\meter}$ fiber bandpass spectrally filters the emission.
The HOM setup consists of a 50:50 free-space beam splitter and a 50:50 fiber beam splitter in which the interference takes place.
The fiber in-coupling can be translated for optimization of the temporal matching and the polarization is set in free space via waveplates and confirmed with a polarimeter.

The indistinguishability measurements were performed with a $\SI{4}{\nano\second}$ delay.
To do so, the train of laser pulses arriving every $\SI{12.5}{\nano\second}$ was split on the excitation side into two pulses separated by a delay of $\SI{4}{\nano\second}$ which was compensated on the detection side, to interfere subsequently emitted photons in a fiber beam splitter.
The laser was tuned to excite the QD-CBG quasi-resonantly for three excitation powers, $P_{\mathrm{exc}}=\SI{0.6}{\micro\watt}$ ($0.02\times P_{\mathrm{sat}}$),  $P_{\mathrm{exc}}=\SI{1.5}{\micro\watt}$ ($0.04\times P_{\mathrm{sat}}$), and $P_{\mathrm{exc}}=\SI{40}{\micro\watt}$ ($P_{\mathrm{sat}}$).

The obtained histograms are composed of a characteristic pattern of 5 peaks repeated every $\SI{12.5}{\nano\second}$ corresponding to the $\SI{80}{\mega\hertz}$ repetition rate of the excitation laser.
They are related to coincidences resulting from consecutive photons taking different paths in the imbalanced Mach-Zehnder interferometer (MZI).
There are 5 possible final delay combinations leading to the observed pattern~\cite{Thoma2016}.
For the Poissonian statistics of the emission, the intensity ratio of the 5 non-central peaks is expected to be 1:4:6:4:1, whereas for the center peak (produced by coincidences originating in the pair of laser pulses separated by $\SI{4}{\nano\second}$), indistinguishable single photons produce the combination of coincidences 1:2:0:2:1 in contrast to 1:2:2:2:1 for completely distinguishable photons. 

The figure of merit for the photon indistinguishability can be extracted from HOM measurements in different ways.
Typically, the amount of coincidences in the case of expected indistinguishability is compared to the number of coincidences for expected maximum distinguishability, either at cross-polarized interference or from different laser pulses (photons that have not interfered).
As the second approach with side peaks is more susceptible to blinking and imperfect setups, we determine the indistinguishability from the comparison of co- and cross-polarized HOM measurements and extract the visibility as the ratio of the central peak areas via
\begin{equation}\label{eq:V-co-cross}
    V = 1 -A_{\mathrm{Co}}/A_{\mathrm{Cross}},
\end{equation}
while we still mention the values obtained from the side peak method for completeness at the end of this section.
In order to extract also other physical quantities and to compensate for statistical fluctuations the data is fitted according to the model described below and the areas $A_{\mathrm{Co}}=A_3$ for the co-polarized case and $A_{\mathrm{Cross}}=A_3$ for the cross-polarized case are obtained.

\begin{figure}[htb] %
    \begin{center} %
    \includegraphics[width=0.7\columnwidth]{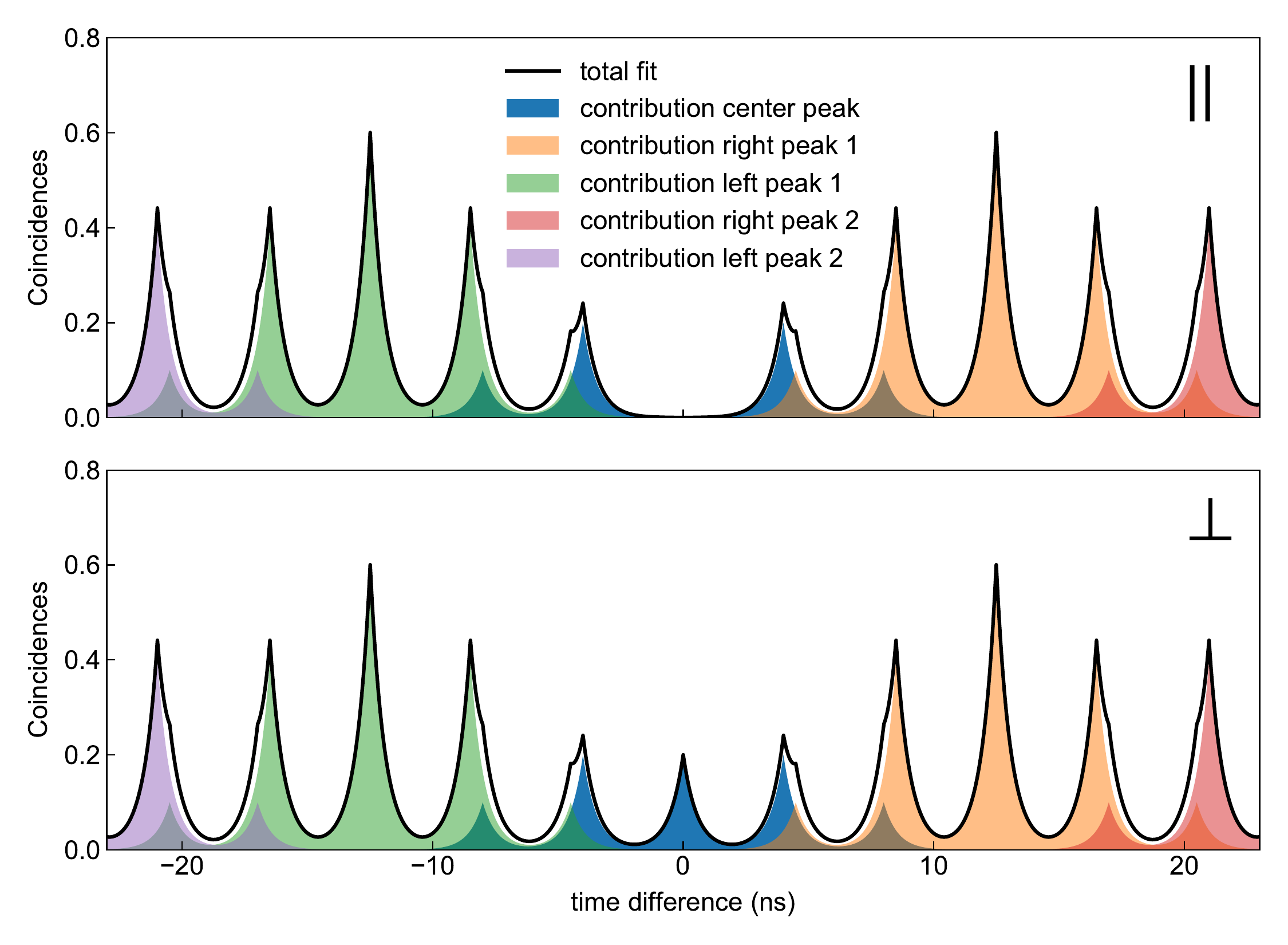} 
    \end{center}
    \caption{\label{fig:hom_expected}
    Expected Hong-Ou-Mandel (HOM) histograms of coincidences with individual peak contributions to the five peak patterns formed by $\SI{4}{\nano\second}$ delay and laser repetition of $\tau_0=\SI{12.5}{\nano\second}$,
    \textbf{a},~for perfectly indistinguishable photons, and 
    \textbf{b},~for maximally distinguishable photons.
    The black lines represent the fits according to \eqnsref{hom-pulsed-Co}{hom-pulsed-Cross} that include all contributions, plotted for $\tau_1=\SI{550}{\pico\second}$ lifetime.
    }
\end{figure}

To properly fit the entire HOM coincidence histograms one has to keep in mind that the areas of the $\pm \SI{4}{\nano\second}$ side peaks belonging to a laser pulse at the time delay $\tau = \SI{0}{\nano\second}$ already overlap with the areas from the side peaks originating from the $\pm \SI{8}{\nano\second}$ laser excitation at $\tau = \pm \SI{12.5}{\nano\second}$ (as they are present in the histogram at $\pm \SI{4.5}{\nano\second}$), see \Supplfigref{hom_expected}.
Consequently, the individual contributions can be extracted by fitting the data by the sum of all contributions according to the formula.
For the HOM histogram recorded for the co-polarized case, the formula reads

\begin{eqnarray}\label{eq:hom-pulsed-Co}
    C_{\mathrm{HOM,Co}}\left(\tau,\left[\tau_1, T_2, \Delta t, \tau_0, \vec{A}, \vec{B}\right]\right)&=
    &A_3 \exp(-|\tau| / \tau_1)\left(1-V_{\mathrm{PS}} \cdot e^{-|\tau| / T_2}\right)+\sum_{\mathrm{i}=\{1,2,4,5\}} A_i \exp(-\left|\tau+\Delta t_i\right| / \tau_1) \\ 
    \notag
    &+& \sum_{n=-10, n \neq 0}^{10}\left[\sum_{\mathrm{i}=\{1,2,3,4,5\}} B_i \exp(-\left|\tau+\Delta t_i+n \cdot \tau_0\right| / \tau_1)\right],
\end{eqnarray}
and for the cross-polarized data
\begin{eqnarray}\label{eq:hom-pulsed-Cross}
    C_{\mathrm{HOM,Cross}}\left(\tau,\left[\tau_1, T_2, \Delta t, \tau_0, \vec{A}, \vec{B}\right]\right)&=
    &A_3 \exp(-|\tau| / \tau_1) +\sum_{\mathrm{i}=\{1,2,4,5\}} A_i \exp(-\left|\tau+\Delta t_i\right| / \tau_1) \\ 
    \notag
    &+& \sum_{n=-10, n \neq 0}^{10}\left[\sum_{\mathrm{i}=\{1,2,3,4,5\}} B_i \exp(-\left|\tau+\Delta t_i+n \cdot \tau_0\right| / \tau_1)\right] 
    \hspace{0.2mm}.
\end{eqnarray}
These formulas for the normalized coincidences as a function of the detection time difference $\tau$ includes the $\SI{4}{\nano\second}$ delay between two interfering photons $\Delta t_{1-5} = \{-8,-4,0,4,8 \} \,$ns, the $\SI{12.5}{\nano\second}$ initial laser pulse delay $\tau_0$, the photoluminescence decay time $\tau_1$, the coherence time $T_2$, the post-selected visibility $V_{\mathrm{PS}}$, the respective peak heights of the center 5-peak-structure $A_{\mathrm{1-5}}$, and the averaged peak heights of all peaks at higher delays $B_{\mathrm{1-5}}$.
The two fits for co- and cross-polarized cases differ only by the existence of the volcano-shaped dip in the central peak of the co-polarized data, being the fingerprint of the two-photon interference.
The post-selected visibility is the value one obtains also when comparing the center peak contribution of the co- and the cross-case at $\tau = 0$ as then 
\begin{equation}\label{eq:V-postselected-new}
V_{\mathrm{PS}} =
1 - \frac{C_{\mathrm{HOM,Co}}(\tau=0)}{C_{\mathrm{HOM,Cross}}(\tau=0)} =
1 - \frac{A_3 (1-V_{\mathrm{PS}})}{A_3}.
\end{equation}
The simulated histograms for our experimental parameters expected from this model are shown in \Supplfigref{hom_expected} top and bottom for perfectly indistinguishable and distinguishable photons, respectively.

\begin{figure}[htb] %
		\begin{center} %
    	\includegraphics[width=0.85\columnwidth]{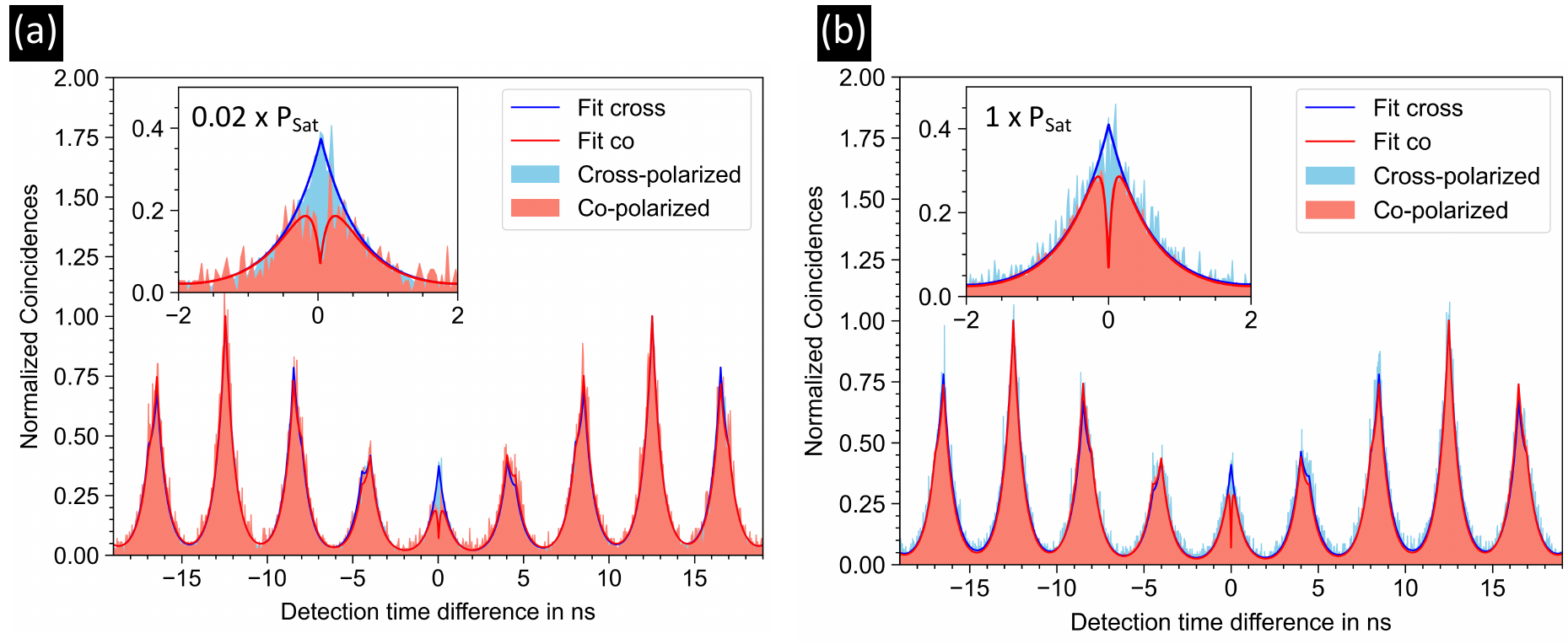} 
	    \end{center}
		\caption{\label{fig:hom_both}
        Comparison of Hong-Ou-Mandel (HOM) histograms taken at 
		\textbf{a},~$0.02\times P_{\mathrm{sat}}$ and
        \textbf{b},~$P_{\mathrm{sat}}$ excitation power.
        HOM measurement recorded for $0.02\times P_{\mathrm{sat}}$ shows higher indistinguishability and longer coherence time than the one taken at $P_{\mathrm{sat}}$, see \Suppltabref{hom_fit_parameters}.}
\end{figure}

While the fitted histogram for $0.04\times P_{\mathrm{sat}}$ is presented in the article, Fig.~4b, the fitted histograms for $0.02\times P_{\mathrm{sat}}$ and $P_{\mathrm{sat}}$ excitation powers are shown in \Supplfigref{hom_both} and the extracted parameters in \Suppltabref{hom_fit_parameters}.
All fits were done on the unbinned raw data without correcting for finite $\gtwo{0}$, as the obtained $\gtwo{0}$ values are small.
Additionally, the fits do not include corrections for the detector time response or subtracting a fixed background.
The uncertainties are determined from the fit errors and propagated whenever a quantity was calculated from the fit parameters.

\begin{table}[htb]
    \caption{Extracted fitting parameters for recorded Hong-Ou-Mandel (HOM) histograms. MFR -- mean fit residuals, TPI -- two-photon interference, pol -- polarized.} 
    \label{tab:hom_fit_parameters}
    \begin{tabular}{c c c c}\toprule
    & \multicolumn{3}{c}{Quasi-resonant excitation power} \\\cmidrule{2-4}
    Fit parameter & $0.02\times P_{\mathrm{sat}}$, shown in \SupplsubfigrefS{hom_both}{a} & $0.04\times P_{\mathrm{sat}}$, shown in Fig.~4c & $P_{\mathrm{sat}}$, shown in \SupplsubfigrefS{hom_both}{b} \\\midrule
    Co-pol areas $\vec{A}$, $\vec{B}$ & $\begin{pmatrix} 1.00 \pm 0.11 \\ 2.08 \pm 0.11 \\ 1.77  \pm 0.21 \\ 2.14 \pm 0.11 \\ 1.38 \pm 0.11 \end{pmatrix}$,
    $\begin{pmatrix} 1.00 \pm 0.06 \\ 4.23 \pm 0.05 \\ 6.66  \pm 0.04 \\ 4.55 \pm 0.05 \\ 1.27 \pm 0.06 \end{pmatrix}$ &
    $\begin{pmatrix} 1.00 \pm 0.04 \\ 1.84 \pm 0.04 \\ 1.68  \pm 0.05 \\ 1.95 \pm 0.04 \\ 1.03 \pm 0.04 \end{pmatrix}$,
    $\begin{pmatrix} 1.00 \pm 0.02 \\ 3.76 \pm 0.02 \\ 5.69  \pm 0.02 \\ 3.75 \pm 0.02 \\ 1.03 \pm 0.02\end{pmatrix}$  &
    $\begin{pmatrix} 1.00 \pm 0.02 \\ 1.96 \pm 0.02 \\ 2.02  \pm 0.03 \\ 1.92 \pm 0.02 \\ 0.95 \pm 0.02 \end{pmatrix}$,
    $\begin{pmatrix} 1.00 \pm 0.01 \\ 3.78 \pm 0.01 \\ 5.68  \pm 0.01 \\ 3.75 \pm 0.01 \\ 0.98 \pm 0.01 \end{pmatrix}$ 
    \\
    Cross-pol area $A_3$, & $2.28 \pm 0.04$  & $2.08 \pm 0.04$ & $2.27 \pm 0.06$
    \\
    Lifetime, $\tau_1$ & $\SI{559(4)}{\pico\second}$ & $\SI{553(2)}{\pico\second}$ & $\SI{563(7)}{\pico\second}$ \\
    Coherence time, $T_2$ & $\SI{176(9)}{\pico\second}$ & $\SI{103(13)}{\pico\second}$ & $\SI{74(6)}{\pico\second}$ \\
    $V_{\mathrm{PS}}$         & $\SI{80(13)}{\percent}$ & $\SI{99(6)}{\percent}$ & $\SI{84(3)}{\percent}$ \\
    MFR, Co-pol fit      & 0.0812            & 0.0286 & 0.0158    \\
    MFR, Cross-pol fit      & 0.0280           & 0.0280 & 0.0526   \\\midrule
    TPI visibility, $V$      & $\SI{22.1(89)}{\percent}$            &   $\SI{19.3(26)}{\percent}$ &  $\SI{11.3(23)}{\percent}$ \\
    \bottomrule
    \end{tabular}
\end{table}

One can clearly see from the graph that the photon indistinguishability is reduced at higher power, as evidenced by the larger central peak area. That is confirmed also by the extracted visibilities of $\SI{22.1(89)}{\%}$, $\SI{19.3(26)}{\%}$ and $\SI{11.3(23)}{\%}$ for low to high power.
That is partly due to a worse single-photon purity at $P_{\mathrm{sat}}$ (see \Supplsubfigref{hom_power}{c}) but mainly due to a reduced coherence, as indicated by the narrower central dip at higher excitation power.
We find that the coherence time $T_2$ is reduced from $\SI{176(9)}{\pico\second}$ over $\SI{103(13)}{\pico\second}$ to $\SI{74(6)}{\pico\second}$ with the increasing excitation power.
The variation between the extracted post-selected values of $V_{\mathrm{PS}} = \SI{80(13)}{\percent}$, $\SI{99(6)}{\percent}$ and $\SI{84(3)}{\percent}$ for different excitation powers results from the finite temporal resolution of our setup ($\SI{57}{\pico\second}$ FWHM jitter that is especially critical when resolving short coherence times), limited statistics due to reduced count rates at lower excitation powers, also evidenced by the larger error and the increased multi-photon contributions at higher power.
Correcting for the $\gtwo{0}$ increase at higher powers, the extracted post-selected visibility values agree within the standard errors.
However, to be in line with future real-world applications, we state the uncorrected values.
The larger uncertainties for the fit parameters for the HOM histogram recorded at $0.02\times P_{\mathrm{sat}}$ are caused by the lower gathered statistics, as also indicated by the larger mean fit residuals (MFR) for the fit of the co-polarized case ($0.0812$ vs. $0.0280$).
Note that for the same reason, the deviation from the expected peak ratios is also larger in this case.
The remaining deviations from the expected 5-peak-ratios ($\vec{A}$, $\vec{B}$) can also be caused by unequal transmissions in the two arms of the MZIs.

\begin{figure}[htb] %
		\begin{center} %
    	\includegraphics[width=0.85\columnwidth]{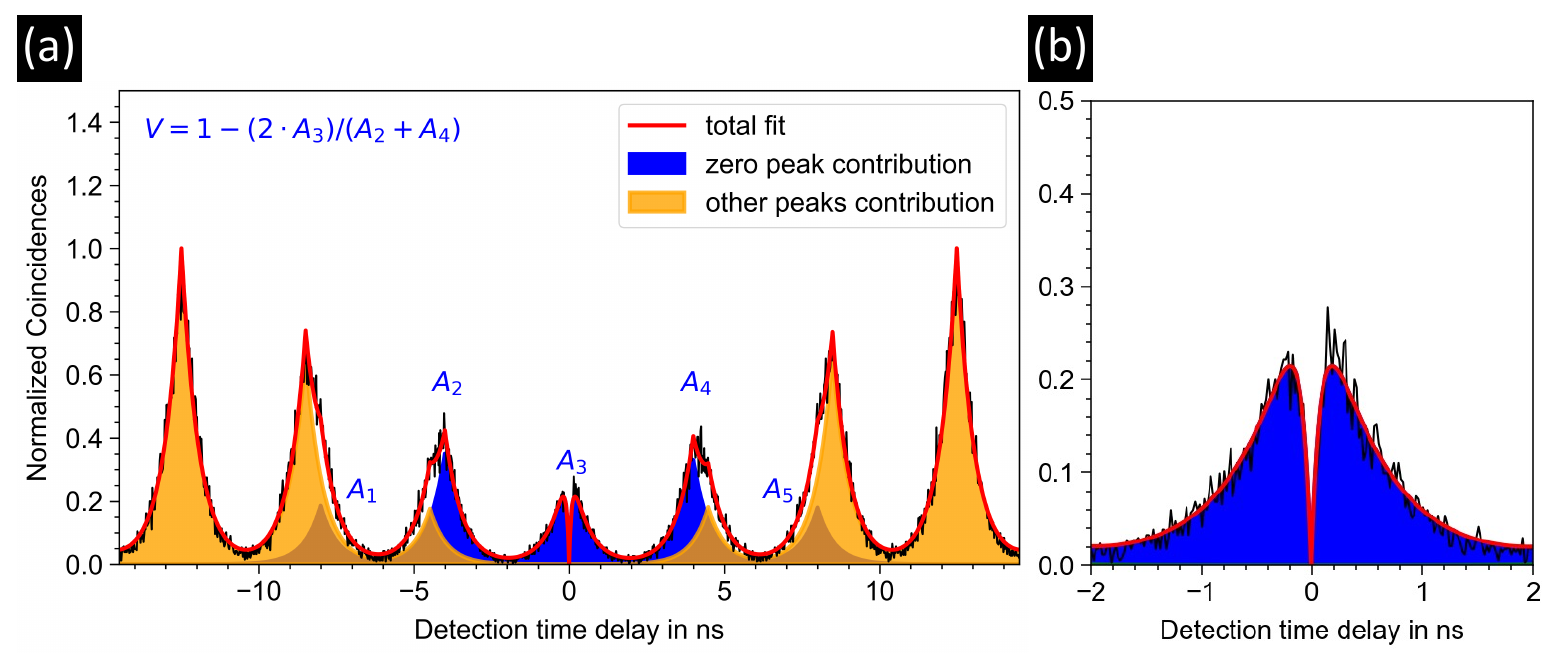} 
	    \end{center}
		\caption{\label{fig:hom_extract-areas}
        Data analysis of the low-power Hong-Ou-Mandel (HOM) histogram.
		\textbf{a},~Individual peak contributions extracted as part of the total fit according to \eqnref{hom-pulsed-Co} for the lowest power HOM measurement ($0.02\times P_{\mathrm{sat}}$).
        \textbf{b},~Close-up of the central peak.}
	\end{figure}

As mentioned at the beginning of this section, in addition to comparing the co- and cross-polarized central peak areas, the indistinguishability can also be extracted by comparing the areas of the central to the outer peaks for only the co-polarized data.
This analysis of the low-power HOM histogram is presented in \Supplfigref{hom_extract-areas}.
Evaluating the visibility from the extracted center $A_3$ and side peak areas $A_2, A_4$, and calculating the visibility as $V_{\mathrm{side\, peaks}}=1-2A_3/(A_2+A_4)$ leads to $\SI{15.9(99)}{\percent}$ for the low excitation power ($0.02\times P_{\mathrm{sat}}$), which agrees with the visibility result reported above obtained from the co-cross-comparison.
The uncertainty is however larger, as the fitting relies on successfully separating the overlapping peak contributions.

Importantly, the obtained visibilities and the post-selected values compare favorably with the reports for GaAs-based QDs emitting at C-band~\cite{Nawrath2019,Nawrath2021,Nawrath2023}.
For planar QDs, a visibility of $\SI{14.4(15)}{\percent}$ has been reported under pulsed resonant excitation~\cite{Nawrath2021} while the raw visibility value of $\SI{71(15)}{\percent}$ was obtained under two-photon-resonant CW excitation~\cite{Nawrath2019}.
Values obtained under CW excitation can be related to the post-selected values determined from pulsed excitation.
For QDs placed non-deterministically in CBGs, the only reported visibility so far is $\SI{8.1(34)}{\percent}$ and the post-selected on the order of $\SI{60}{\percent}$~\cite{Nawrath2023}.

\subsection{Coherence measurements}
\begin{figure}[htb] %
		\begin{center} %
    	\includegraphics[width=0.85\columnwidth]{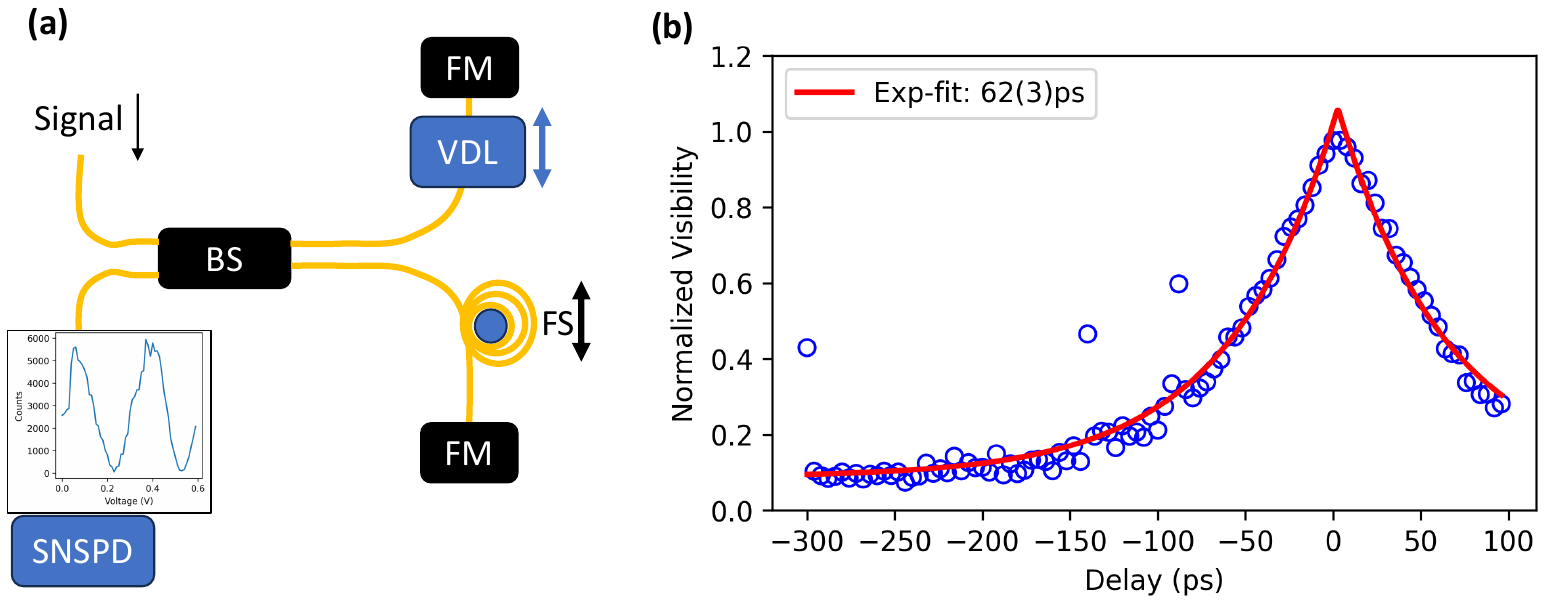} 
	    \end{center}
		\caption{
        \textbf{a},~Schematic of the all-fiber Michelson interferometer for $\SI{1550}{\nano\meter}$.
        The single photon signal is split in a fiber beam splitter (BS) and back-reflected at two Faraday mirrors (FM), while the coarse delay on one arm is controlled with a variable optical delay line (VDL) and the fine scan of the relative path difference is done via a piezo fiber stretcher (FS).
        The counts after the interference are detected on a superconducting nanowire single photon detector (SNSPD).
        Inset: Example of the FS scan for $\SI{0}{\pico\second}$ delay indicating constructive and destructive interference.
        \textbf{b},~Extracted visibility for different delay positions allows the determination of the $T_2$ time using an exponential fit.
        This is data under $\SI{980}{\nano\meter}$ CW excitation for the QD-CBG \#2 from the main text.
        \label{fig:MI_principle}}
	\end{figure}

To access the coherence time $T_2$ directly, we performed measurements using a Michelson interferometer (MI) under above-barrier CW excitation ($\SI{980}{\nano\meter}$) of QDs integrated into CBGs and planar sample regions.
To do so, we employed an all-fiber MI shown in \Supplsubfigref{MI_principle}{a}, where the inset shows a typical interference scan.
While the resulting $T_2$ values are not directly comparable to the values extracted from the HOM experiments, they provide lower bounds on the coherence time (as above-barrier CW excitation results typically in substantially enhanced decoherence effects), and can be used to compare different QD devices.
The highest coherence time observed in the MI measurements was obtained from the QD-CBG \#2 from the main text to be $\SI{62(3)}{\pico\second}$ (c.f. \Supplsubfigref{MI_principle}{b}), extracted from double-exponential fits.
Note the artifacts, which are caused by temporary spectral jumps (telegraphic noise), as also discussed in the main text as one of the factors limiting the indistinguishability.
The extracted $T_2$ value is in line with coherence times of $\sim\SI{50}{\pico\second}$ reported in the literature for InAs/InP QDs grown in the Stranski-Krastanov mode \cite{Anderson2021}.

\begin{figure}[htb] %
		\begin{center} %
    	\includegraphics[width=0.85\columnwidth]{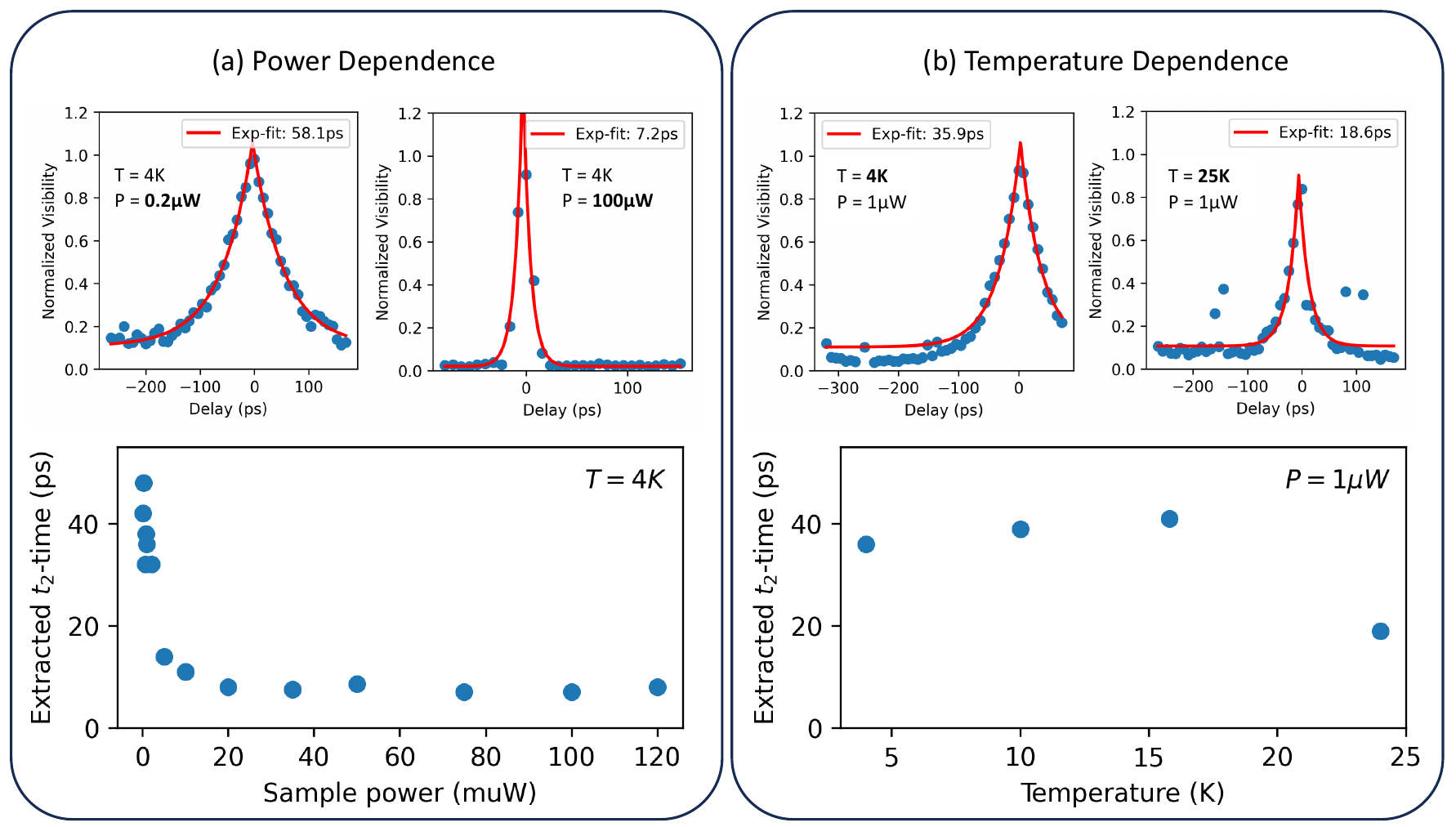} 
	    \end{center}
		\caption{
        \textbf{a},~Measured coherence of the QD-CBG device as a function of excitation power shows coherence drop when increasing power.
        \textbf{b},~Coherence as a function of temperature shows a slight increase, potentially due to a spectral jump into a more coherent state, before decreasing as expected.
        \label{fig:MI_CBG2}}
	\end{figure}

To gain further insight, we have measured the $T_2$ time for the QD-CBG~\#2 from the manuscript for different excitation powers (\Supplsubfigref{MI_CBG2}{a}) as well as for different temperatures (\Supplsubfigref{MI_CBG2}{b}).
As expected, the $T_2$ time drops when the temperature or the excitation power is increased.
The slight increase in coherence when temperature increases might be explained by the fact that the trion transition occasionally showed a random telegraphic noise.
Here, the different states may be associated with different coherence times, as confirmed by repeated measurements under the same conditions.
The sudden spectral jumps during a Michelson scan yield sudden changes in signal rate, which leads to outliers for the extracted interference contrast (c.f., four data points in \Supplsubfigref{MI_CBG2}{b}).
 
\begin{figure}[htb] %
		\begin{center} %
    	\includegraphics[width=0.85\columnwidth]{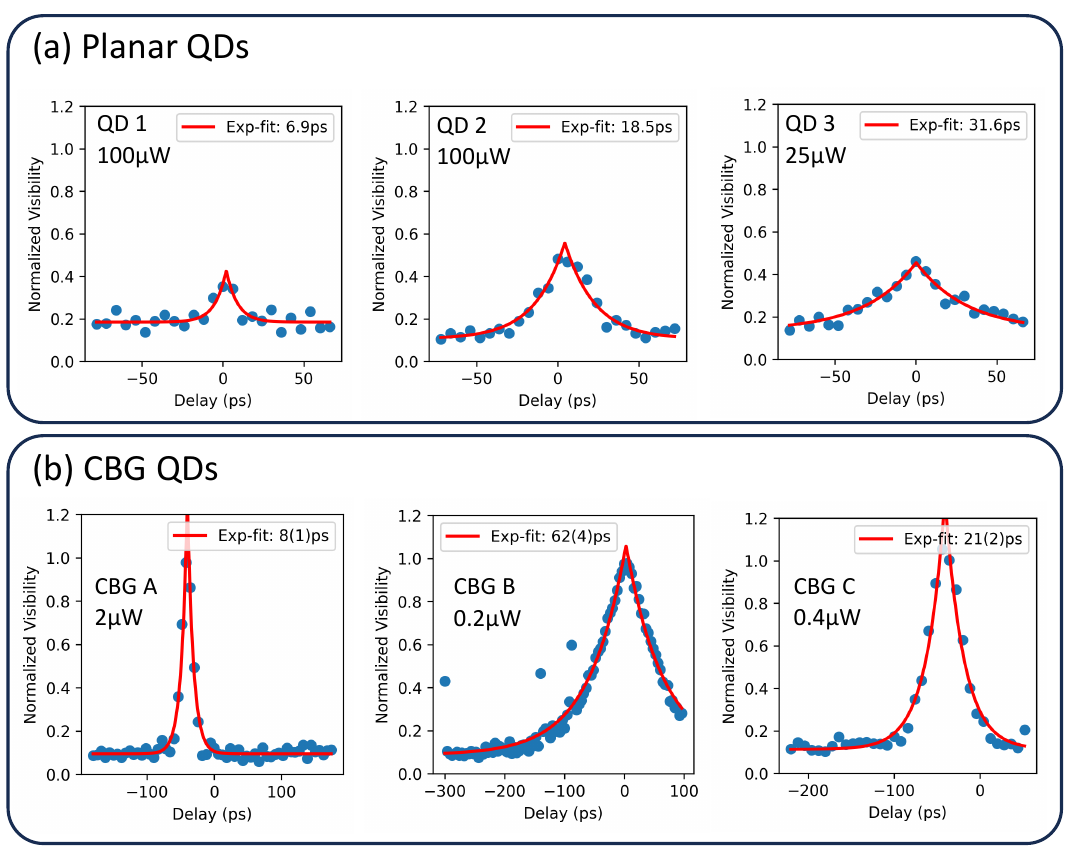}
	    \end{center}
		\caption{Comparison of coherence measurements results. 
        \textbf{a},~Data for the selection of planar QDs,
        \textbf{b},~QDs deterministically integrated into CBG structures.
        While the exact coherence time depends on power, the general trend is that the $T_2$ time is, on average, not reduced when integrating the QD into a photonic structure.
        \label{fig:MI_comparison}   }
	\end{figure}

While the strong dependence of the coherence on excitation power makes a direct comparison between QDs in the planar state and in CBGs difficult, we chose a compromise of evaluating the coherence for 3 exemplary (one high coherence, one medium, and one low) planar and CBG-integrated QDs, respectively.
The results are compared in \Supplfigref{MI_comparison}. 
As the planar QDs do not couple to the excitation laser efficiently, more power was required to get a sufficient signal, yielding an overall reduced coherence.
Also, due to the smaller count rate, the maximum interference contrast is not well resolved.
We do find, however, a spread of measured maximum coherence times between $\SIrange{6}{30}{\pico\second}$ for the set of planar QDs investigated (\Supplsubfigref{MI_comparison}{a}). 

For the QDs deterministically integrated into CBG structures, a lower excitation power is possible. Even though only a small number of QDs were investigated, the spread of coherence times from $\SIrange{18}{62}{\pico\second}$ indicates at least no deterioration of the coherence by integrating it into CBG structures.
If the integration yields a significant Purcell enhancement, thus reducing the $T_1$ time, while not strongly reducing the $T_2$ time, as it seems to be the case here, one can conclude that the CBG cavity brings the emission closer to the Fourier limit of $T_1 = 2 \times T_2$. 
 
\clearpage